\documentclass{aa}
\usepackage{natbib,epsfig,txfonts,graphicx}
\bibpunct{(}{)}{;}{a}{}{,}

\def\Ha {H$_\alpha$\,}
\def\Hbeta {H$_\beta$\,}
\def\Hgama {H$_\gamma$\,}
\def\Hdelta{H$_\delta$\,}

\def\SiII {Si II $\lambda\lambda$4128, 4131\,}

\def\SiIV {Si IV $\lambda$4116\,}

\def\Mdot {$\dot M$\,}
\def\MV {M$_{\rm V}$\,}

\def\Yhe {$Y_{\rm He}$\,} 
\def\kms {km~s$^{\rm -1}$\,} 
\def\Rstar {$R_\star$\,}
\def\Mstar {$M_\star$\,}
\def\Teff {$T_{\rm eff}$\,}
\def\vinf {$v_\infty$\,}
\def\vesc {$v_{\rm esc}$\,}
\def\vsini {$v \sin i$\,}
\def\logg {$\log g$\,}
\def\loggtr {$\log g_{\rm true}$\,}
\def\Vr {$V_{\rm r}$\,}

\def\logl {$\log L/L_\odot$\,}
\def\vmac {$v_{\rm mac}$\,}
\def\vmic {$v_{\rm mic}$\,}
 
\def \logwm {$\log D_{\rm mom}$\,}

\def\Msun {$M_\odot$\,}
\def\Rsun {$R_\odot$\,}

\def \Rstare {R_\star}
\def \Mstare {M_\star}
\def \Teffe {T_{\rm eff}}
\def \vinfe {v_\infty}
\def \vesce {v_{\rm esc}}
\def \Mdote {\dot M}

\def \Rsune{R_\odot}
\def \ao{\frac{1}{\alpha'}}
\def \af{\frac{4}{\alpha'}}
\def \half{\frac{1}{2}}
\def \threehalf{\frac{3}{2}}
\def \Cinf{C_{\infty}}
\def \Dmom {D_{\rm mom}}
\def \Do {D_{0}}
\def \geff {g_{\rm eff}}
\def \beq{\begin{equation}}
\def \eeq{\end{equation}}
\def \ben{\begin{enumerate}} 
\def \een{\end{enumerate}} 
\def \beqa{\begin{eqnarray}}
\def \eeqa{\end{eqnarray}}
\begin{document}
\title{Bright OB  stars in the Galaxy}
\subtitle{IV. Stellar and wind parameters of early to late B supergiants}

\author{N. Markova\inst{1} 
  \and J. Puls\inst{2}} 
	   
\offprints{N. Markova,\\ \email{nmarkova@astro.bas.bg}}

\institute{Institute of Astronomy, 
       National Astronomical Observatory,
       Bulgarian Academy of Sciences, 
	P.O. Box 136, 4700 Smolyan, Bulgaria\\ 
	\email{nmarkova@astro.bas.bg}
	\and Universit\"{a}ts-Sternwarte,
	      Scheinerstrasse 1, D-81679 M\"unchen, Germany\\
	\email{uh101aw@usm.uni-muenchen.de}}
\date{Received; Accepted }

\abstract{B-type supergiants represent an important phase in the evolution 
of massive stars. Reliable estimates of their stellar and wind parameters,
however, are scarce, especially at mid and late spectral subtypes.
}
{We apply the NLTE atmosphere code FASTWIND to perform a spectroscopic study of 
a small sample of Galactic B-supergiants from B0 to B9. By means of the
resulting data and incorporating additional datasets from alternative
studies, we investigate the properties of OB-supergiants and compare our
findings with theoretical predictions.}
{Stellar and wind parameters of our sample stars are determined by line
profile fitting, based on synthetic profiles, a Fourier technique to
investigate the individual contributions of stellar rotation and
``macro-turbulence'' and an adequate approach to determine the Si abundances
in parallel with micro-turbulent velocities.}
{Due to the combined effects of line- and wind-blanketing, the temperature
scale of Galactic B-supergiants needs to be revised downward, by 10 to
20\%, the latter value being appropriate for stronger winds.  Compared to
theoretical predictions, the wind properties of OB-supergiants indicate 
a number of discrepancies. 
In fair accordance with recent results, our sample indicates a gradual
decrease in \vinf over the bi-stability region, where the limits of this
region are located at lower \Teff than those predicted. Introducing
a distance-independent quantity $Q'$ related to wind-strength, we show
that this quantity is a well defined, monotonically increasing function of
\Teff $outside$ this region. $Inside$ and from hot to cool, \Mdot
changes by a factor (in between 0.4 and 2.5) which is much smaller than
the predicted factor of 5.} 
{The decrease in \vinf over the bi-stability region is $not$ over-compensated by 
an increase of \Mdot, as frequently argued,  provided that 
wind-clumping properties on both sides of this region do not differ substantially.
}
   

\titlerunning{Galactic B supergiants}
\authorrunning{N. Markova and J. Puls}

\maketitle

 %

\section{Introduction}

Hot massive stars are key objects for  studying and understanding many exciting 
phenomena in the Universe such as re-ionisation and $\gamma$-ray bursters. Due
to their powerful stellar winds hot massive stars are important contributors
to the chemical and dynamical evolution of galaxies, and in the distant
Universe they dominate the integrated UV radiation in young galaxies.

While the number of Galactic O and early B stars with reliably determined 
stellar and wind parameters has progressively increased during the last few
years (e.g., \citealt{Herrero02, repo, GB04, bouret05, martins05,
crowther06}), mid and late B supergiants (SGs) are currently
under-represented in the sample of stars investigated so far. Given the fact
that B-SGs represent an important phase in the evolutionary sequence of
massive stars, any study aiming to increase our knowledge of these stars
would be highly valuable, since it would allow several important issues to
be addressed (see below).

Compared to O-type stars the B-SG spectra are more complicated due to a
larger variety of atomic species being visible, the most important among
which is Silicon, the main temperature indicator in the optical domain.
Thus, the reproduction of these spectra by methods of {\it quantitative 
spectroscopy} is a real challenge for state-of-the art model atmosphere
codes, since it requires a good knowledge of the physics of these objects,
combined with accurate atomic data. In turn, any discrepancy that might
appear between computed and observed spectral features would help to
validate the physical assumptions underlying the model calculations as well
as the accuracy of the adopted atomic models and data.

Numerical simulations of the non-linear evolution of the line-driven flow
instability (for a review, see \citealt{O94}), with various degrees of
approximation concerning the stabilising diffuse, scattered radiation field
(\citealt{op96, OP99}) as well as more recent simulations concentrating on
the outer wind regions \citep{ro02, ro05}, predict that hot star winds are
not smooth but structured, with clumping properties depending on the
distance to the stellar surface. However, recent observational studies of
clumping in O-SGs have revealed inconsistencies both between results
originating from different wind diagnostics, such as UV resonance lines, 
\Ha and the IR-/radio-excess \citep{FMP06, puls06}, and between theoretical
predictions and observed constraints on the radial stratification of the
clumping factor \citep{bouret05, puls06}. In addition, there are
observational results which imply that clumping might depend on wind
density.  Because of their dense winds, B-SGs might provide additional clues
to clarify these points.

Due to their high luminosities, BA-SGs can be resolved and observed,
both photometrically and spectroscopically, even in rather distant,
extragalactic stellar systems (e.g., \citealt{Kud99, Bresolin02, Urb03,
Bianchi06}). This fact makes them potential standard candles, allowing us to
determine distances by means of purely spectroscopic tools using the 
wind-momentum luminosity relationship (WLR, \citealt{klp95}). Even though 
certain discrepancies between predicted and observed wind momenta of early
B0 to B3 subtypes have been revealed \citep{crowther06}, relevant
information about later subtypes is still missing.  

During the last years, the quantitative analyses of spectra in the far-UV/UV
and optical domains (e.g.,
\citealt{Herrero02,BG02,Crowther02,bouret03,repo,massey04,heap06}) have
unambiguously shown that the inclusion of line-blocking and blanketing and
wind effects (if present) significantly modifies the temperature scale of
O-stars (for a recent calibration at {\it solar} metallicity, see
\citealt{Markova04, martins05}). Regarding B-SGs, particularly of later
subtype, this issue has not been addressed so far, mostly due to lacking 
\Teff estimates.
  
The main goal of this study is to test and to apply the potential of our
NLTE atmosphere code FASTWIND \citep{Puls05} to provide reliable estimates
of stellar and wind parameters of SGs with temperatures ranging from 30 to
11 kK. By means of these data and incorporating additional datasets from 
alternative studies, we will try to resolve the questions outlined above.

In Sects.~\ref{obs} and \ref{sample}, we describe the stellar sample and the
underlying observational material used in this study. In Sect.~4 we outline
our procedure to determine the basic parameters of our targets, highlighting
some problems faced during this process. In Sect.~5 the effects of line
blocking/blanketing on the temperature scale of B-SGs at solar and SMC
metallicities will be addressed, and in Sect.~6 we investigate the wind
properties for Galactic B-SGs (augmented by O-and A-SG data), by comparison
with theoretical predictions. Particular emphasis will be given to the
behaviour of the mass-loss rate over the so-called bi-stability jump.
Sect.~7 gives our summary and implications for future work.  
\begin{table}
\caption{Galactic B-SGs studied in this work, together with
adopted photometric data. For multiple entries, see text.}
\label{log_phot}
\tabcolsep0.8mm
\begin{tabular}{llllllll}
\hline 
\hline
\\
\multicolumn{1}{l}{Object}
&\multicolumn{1}{l}{spectral}
&\multicolumn{1}{c}{member-}
&\multicolumn{1}{c}{d}
&\multicolumn{1}{c}{$V$}
&\multicolumn{1}{l}{$B-V$}
&\multicolumn{1}{c}{$(B-V)_{\rm 0}$}
&\multicolumn{1}{c}{$M_{\rm V}$}\\
\multicolumn{1}{l}{(HD\#)}
&\multicolumn{1}{l}{type}
&\multicolumn{1}{c}{ship}
\\
\hline
185\,859  &B0.5 Ia   &          &             &      &       &       &-7.0*\\                                 
190\,603  &B1.5 Ia+  &        &1.57$^{c}$   &5.62  &0.760  &-0.16  &-8.21\\
      &          &        &   &5.62         &0.54$\pm$0.02& &-7.53\\ 

206\,165  &B2 Ib     &Cep OB2   &0.83$^{a}$   &4.76  &0.246  &-0.19  &-6.19\\
198\,478  &B2.5 Ia   &Cyg OB7   &0.83$^{a}$   &4.81  &0.571  &-0.12  &-6.93\\  
          &           &         &             &4.84  &0.40$\pm$0.01 &&-6.37\\
191\,243  &B5 Ib     &Cyg OB3   &2.29$^{a}$   &6.12  &0.117  &-0.12  &-6.41\\
             &          &          &1.73$^{b}$   &      &       &       &-5.80\\
199\,478  &B8 Iae    &NGC 6991  &1.84$^{d}$   &5.68  &0.408  &-0.03  &-7.00\\
212\,593  &B9 Iab    &          &             &      &       &       &-6.5*\\
202\,850  &B9 Iab    &Cyg OB4   &1.00$^{a}$   &4.22  &0.098  &-0.03  &-6.18\\
\hline
\end{tabular}
\newline
$^{a}$ \citet{Humphreys78}; $^{b}$ \citet{GS}; $^{c}$ \citet{BC}; $^{d}$ \citet{DH}\\
$^{*}$ from calibrations \citep{HM}\\
\end{table}

\section{Observations and data reduction}
\label{obs}

High-quality optical spectra were collected for eight Galactic B-type 
SGs of spectral types B0.5 to B9 using the Coud\'{e} spectrograph of
the NAO 2-m telescope of the Institute of Astronomy, Bulgarian Academy of 
Sciences. The observations
were carried out using a BL632/14.7 grooves \,mm$^{-1}$ grating in first 
order, together with a PHOTOMETRICS CCD (1024 x 1024, 24$\mu$) as a
detector.\footnote{This detector is characterised by an $rms$ read-out
noise of 3.3 electrons per pixel (2.7 ADU with 1.21 electrons per ADU).}
This configuration produces spectra with a reciprocal dispersion of
$\sim$0.2 \AA\ pixel$^{-1}$ and an effective resolution of 
$\sim$~2.0~pixels, resulting in a spectral resolution of $\sim$~15\,000 at
\Ha.

The signal-to-noise (S/N) ratio, averaged over all spectral regions
referring to a given star, has typical values of 200 to 350, being 
lower in the blue than in the red.

We observed the wavelength range between 4\,100 and 4\,900 \AA, where most 
of the strategic lines of H, He and Si ions are located, together with the 
region around H$_\alpha$. Since our spectra sample about 200 \AA, five settings 
were used to cover the ranges of interest. These settings are as follows:
\ben

\item[i)] From 4110 to 4310 \AA\ (covering \SiII,  \SiIV and He~II $\lambda$4200). 

\item[ii)] From 4315 to 4515 \AA\ (He~I~$\lambda\lambda$4387, 4471 and
\Hgama).

\item[iii)] From 4520 to 4720 \AA\ (Si~III~$\lambda\lambda$4553, 4568, 4575, 
He~I~$\lambda$4713 and He~II~$\lambda\lambda$4541, 4686). 

\item[iv)] The region around \Hbeta including  
Si~III~$\lambda\lambda$4813, 4820, 4829 and He~I~$\lambda$4922.

\item[v)] The region around  \Ha  including 
He~I~$\lambda$6678 and He~II $\lambda\lambda$6527, 6683. 
\een
To minimise the effects of temporal spectral variability (if any), all 
spectra referring to a given star were taken one after the other, with a time
interval between consecutive exposures of about half an hour. Thus, we 
expect our results to be only sensitive to temporal variability of less than
2 hours.

The spectra were reduced following standard procedures and using the 
corresponding IRAF\footnote{The IRAF package is distributed by the 
National Optical Astronomy Observatories, which is operated by the 
Association of Universities for Research in Astronomy, Inc., under 
contract with the National Sciences Foundation.} routines.

\section{Sample stars}
\label{sample}

Table~\ref{log_phot} lists our stellar sample, together with corresponding 
spectral and photometric characteristics, as well as association/cluster 
membership and distances, as adopted in the present study. For 
hotter and intermediate temperature stars,  spectral types and luminosity 
classes (Column 2) were taken from the compilation by \citet{Howarth97}, 
while for the remainder, data from $SIMBAD$  have been used. 

Since $HIPPARCOS$ based distances are no longer reliable in the distance
range considered here (e.g., \citealt{Z99, Schroeder04}), we have adopted
photometric distances collected from various sources in the literature
(Column~4). In particular, for stars which are members of OB associations, we
drew mainly from \citet{Humphreys78} but also consulted the lists published
by \citet{GS} and by \citet{BC}. In most cases, good agreement
between the three datasets was found, and only for Cyg~OB3 did the distance
modulus provided by Humphreys turned out to be significantly larger than
that provided by Garmany \& Stencel. In this latter case two entries for $d$
are given in Table~\ref{log_phot}.

Apart from those stars belonging to the OB associations, there are two objects in
our sample which have been recognised as cluster members: HD~190\,603 and
HD~199\,478. The former was previously assigned as a member of Vul OB2
(e.g.  \citealt{Lennon92}), but this assignment has been questioned by
\citet{McE99} who noted that there are three aggregates at approximately 1,
2 and 4 kpc in the direction of HD~190\,603. Since it is not obvious to which
of them (if any) this star belongs, they adopted a somewhat arbitrary
distance of 1.5 kpc. This value is very close to the estimate of 1.57 kpc
derived by \citet{BC}, and it is this latter value which we will use in the
present study. However, in what follows we shall keep in mind that the
distance to HD~190\,603 is highly uncertain. For the second cluster member,
HD~199\,478, a distance modulus to its host cluster as used by \citet{DH}
was adopted.

Visual magnitudes, $V$, and $B-V$ colours (Column~5 and 6) have been taken
from the {\it HIPPARCOS Main Catalogue (I/239)}.  While for the majority of
sample stars the {\it HIPPARCOS} photometric data agree quite well (within
0.01 to 0.04 mag both in $V$ and $B-V$) with those provided by {\it SIMBAD},
for two of them (HD~190\,603 and HD~198\,478) significant differences
between the two sets of $B-V$ values were found. In these latter cases two
entries for $B-V$ are given, where  the second one represents the mean 
value averaged over all measurements listed in {\it SIMBAD}).

Absolute magnitudes, \MV (Column 8), were calculated using the standard 
extinction law with $R = 3.1$ combined with intrinsic colours, $(B-V)_{\rm
0}$, from \citet{FG}  (Column~7) and distances, $V$ and $B-V$ magnitudes as
described above. For the two stars which do not belong to any
cluster/association (HD~185\,859 and HD~212\,593), absolute magnitudes
according to the calibration by \citet{HM} have been adopted.

For the majority of cases, the absolute magnitudes we derived, agree
within $\pm$0.3~mag with those provided by the Humphreys-McElroy
calibration. Thus, we adopted this value as a measure for the uncertainty in
\MV for cluster members (HD~199\,478) and members of spatially more
concentrated OB associations (HD~198\,478 in Cyg~OB7, see
\citealt{crowther06}). For other stars with known membership, a somewhat
larger error of $\pm$0.4~mag was adopted to account for a possible spread in
distance within the host association. Finally, for HD~190\,603 and those two
stars with calibrated M$_{\rm V}$, we assumed a typical uncertainty of $\Delta$\MV =
$\pm$0.5~mag, representative for the spread in \MV of OB stars within a
given spectral type \citep{crowther04}.\footnote{For a hypergiant such
as HD~190\,603 this value might be even higher.}

\section{Determination of stellar and wind parameters}
\label{results}

The analysis presented here was performed with FASTWIND, which
produces spherically symmetric, NLTE, line-blanketed model atmospheres and
corresponding spectra for hot stars with winds.  While detailed information
about the latest version used here can be found in \citet{Puls05}, we
highlight only those points which are important for our analysis of B
stars.
\begin{itemize}
\item[(a)] In addition to H and He, Silicon is used as an explicit element
(i.e., by means of a detailed model atom and using a comoving frame
transport for the bound-bound transitions). All other elements (e.g., C, N,
O, Fe, Ni etc.) are treated as background elements, where the major
difference (compared to the explicit ones) relates to the line transfer, which
is performed within the Sobolev approximation. 
\item[(b)] A detailed description of the Silicon atomic model can be found 
in \citet{Trundle04}.

\item[(c)] Since previous applications of FASTWIND have concentrated 
on O and early B stars, we briefly note that correct
treatment of cooler stars requires sufficiently well described iron group
ions of stages II/III, whose lines are dominating the background
opacities for these objects. Details of the corresponding model atoms and
line-lists (from {\it superstructure}, \citealt{ Eissner74, Nussbaumer78},
augmented by data from \citealt{Kurucz92}) can be found in
\cite{Pauldrach01}. In order to rule out important effects from still
missing data, we have constructed an alternative dataset which uses {\it all} 
Fe/Ni II/III lines from the Kurucz line-list. Corresponding models
(in particular temperature structure and emergent fluxes) turned out to
remain almost unaffected by this alteration, so we are confident 
that our original database is fairly complete, and can be used
for calculations roughly down to 10~kK.

\item[(d)] A consistent temperature stratification utilising 
a flux-correction method in the lower wind and the thermal balance of
electrons in the outer part is calculated, with a transition point between
the two approaches located roughly at a Rossland optical depth of $\tau_{\rm
R} \approx 0.5$ (in dependence of wind density).
\end{itemize}
To allow for an initial assessment of the basic parameters, a coarse grid of
models was generated using this code (appropriate for the considered
targets). The grid involves 270 models covering the temperature range
between 12 and 30~kK (with increments of 2~kK) and including
\logg values from 1.6 to 3.4 (with increments of 0.2 dex). An extended 
range of wind-densities, as combined in the optical depth invariant 
$Q$ (=\Mdot/(\vinf \Rstar)$^{1.5}$, cf. \citealt{Puls96}) 
has been accounted for as well, to allow for both thin and thick winds.

All models have been calculated assuming solar Helium ($Y_{\rm He}$ = 0.10, 
with $Y_{\rm He}=N({\rm He})/N({\rm H})$) and Silicon abundance (log (Si/H)
= -4.45 by number\footnote{According to latest results
\citep{Asplund}, the actual solar value is slightly lower, log (Si/H) = -
4.49, but such a small difference has no effect on the
quality of the line-profile fits.}, cf. \citealt{GS98} and references
therein), and a micro-turbulent velocity, \vmic, of 15 \kms for hotter
and 10 \kms for cooler subtypes, with a border line at 20 kK.  

By means of this model grid, initial estimates on \Teff, \logg and \Mdot
were obtained for each sample star. These estimates were subsequently used
to construct a smaller subgrid, specific for each target, to derive the
final, more exact values of the stellar and wind parameters (including
 Y$_{\rm He}$, log (Si/H) and  \vmic).

\paragraph{Radial velocities}

To compare observed with synthetic profiles, radial velocities and
rotational speeds of all targets have to be known. We started our analysis
with radial velocities taken from the General Catalogue of Mean Radial
Velocities (III/213, \citealt{BB}). These values were then 
modified to obtain better fits to the analysed absorption profiles. In doing
so we gave preference to Silicon rather than to Helium or Hydrogen lines
since the latter might be influenced by (asymmetrical) wind
absorption/emission. The finally adopted \Vr-values which provide the ``best''
fit to most of the Silicon lines are listed in Column~3 of
Table~\ref{para_1}. The accuracy of these estimates is typically $\pm$
2\kms.
\begin{table}
\caption{Radial velocities (from Si), projected rotational velocities,
macro- and micro-turbulent velocities (all in \kms) and Si abundances, given
as log [$N$(Si)/$N$(H] + 12, of the sample stars 
as determined in the present study. The number in brackets refers to the number
of lines used to derive \vsini and \vmac.}
\label{para_1}
\tabcolsep1.5mm
\begin{tabular}{llrllrc}
\hline 
\hline
~\\
\multicolumn{1}{l}{Object}
&\multicolumn{1}{l}{Sp}
&\multicolumn{1}{c}{\Vr}
&\multicolumn{1}{c}{\vsini}
&\multicolumn{1}{c}{\vmac}
&\multicolumn{1}{c}{\vmic}
&\multicolumn{1}{c}{Si abnd}
\\
\hline
HD 185\,859  &B0.5 Ia   &12      &62(5)  &58(3)   &18  &7.51 \\
HD 190\,603  &B1.5 Ia+  &50      &47(8)  &60(3)   &15  &7.46 \\
HD 206\,165  &B2 Ib     &0       &45(7)  &57(3)   &8   &7.58 \\
HD 198\,478  &B2.5 Ia   &8       &39(9)  &53(3)   &8   &7.58 \\
HD 191\,243  &B5 Ib     &25      &38(4)  &37(3)   &8   &7.48 \\
HD 199\,478  &B8 Iae    &-12     &41(4)  &40(3)   &8   &7.55 \\
HD 212\,593  &B9 Iab    &-13     &28(3)  &25(3)   &7   &7.65 \\
HD 202\,850  &B9 Iab    &13      &33(3)  &33(3)   &7   &7.99 \\
\end{tabular}
\end{table}

\subsection {Projected rotational velocities and 
macro-turbulence}

As a first guess for the projected rotational velocities of the sample
stars, \vsini, we used values obtained by means of the Spectral type -
\vsini\ calibration for Galactic B-type SGs provided by
\citet{Abt02}. However, during the fitting procedure it was found 
that 
(i) these values provide poor agreement between observed and synthetic
profiles and (ii) an additional line-broadening agent must be introduced 
to improve the quality of the fits. These findings are consistent with
similar results from earlier investigations claiming that absorption line
spectra of O-type stars and B-type SGs exhibit a significant amount of
broadening in excess to the rotational broadening \citep{Rosenh70, CE77,
LDF93, Howarth97}. Furthermore, although the physical mechanism responsible
for this additional line-broadening is still not understood 
we shall follow \citet{Ryans} and refer to it as ``macro-turbulence''.
\begin{figure*}
\begin{minipage}{8.8cm}
\resizebox{\hsize}{!}
{\includegraphics{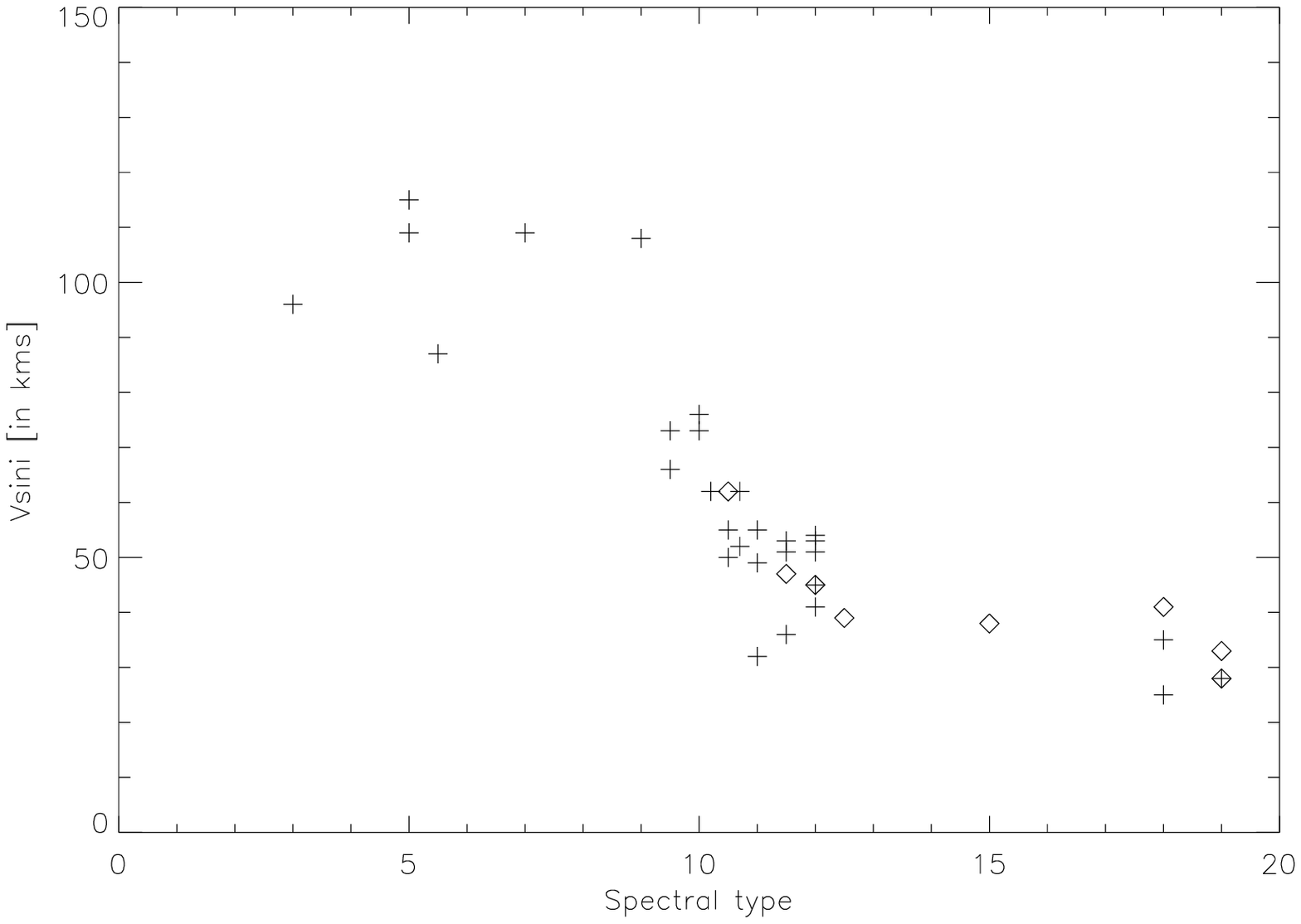}}
\end{minipage}
\hfill
\begin{minipage}{8.8cm}
\resizebox{\hsize}{!}
{\includegraphics{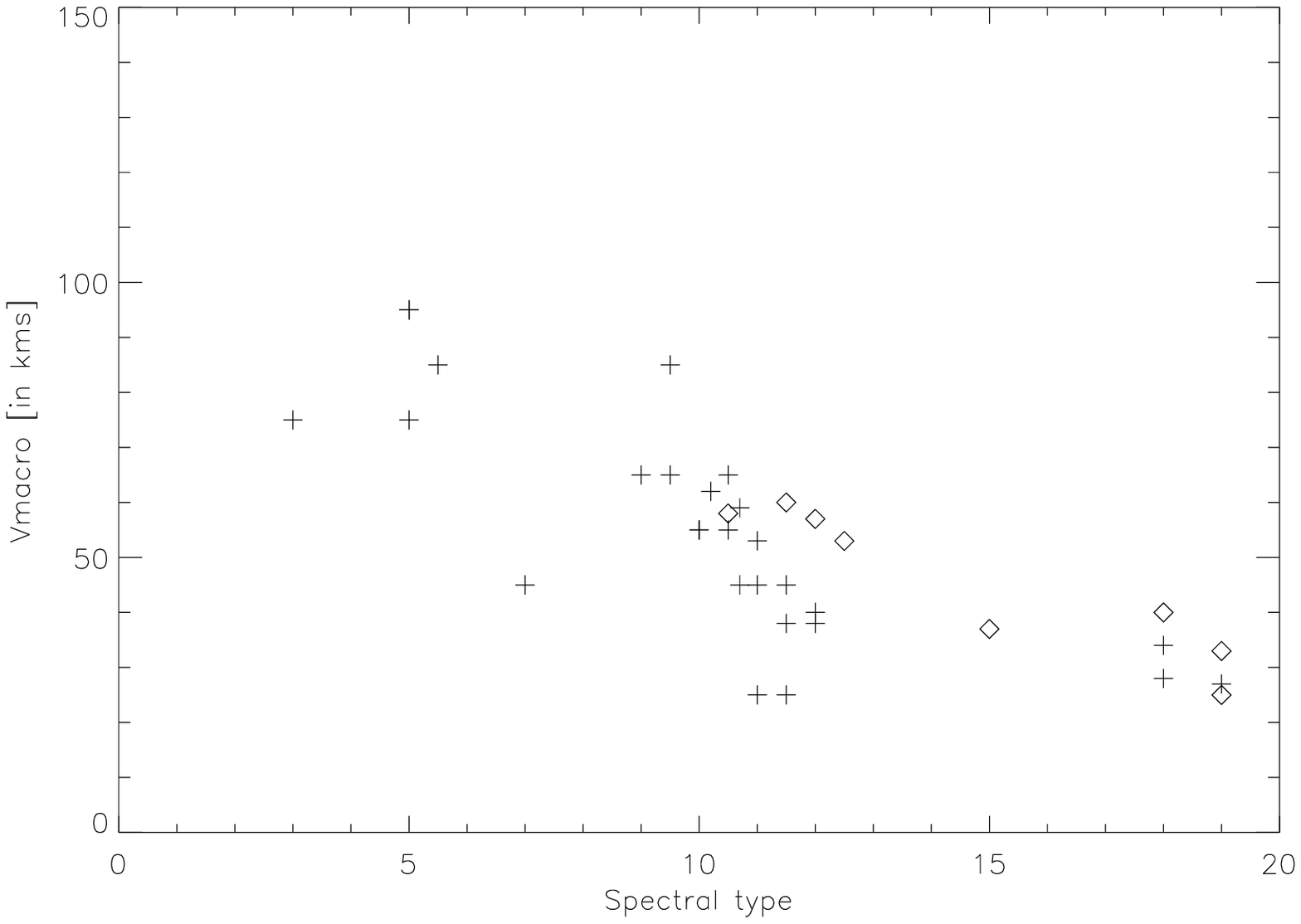}}
\end{minipage}
\caption{Projected rotational (left panel) and macro-turbulent (right panel)
velocities of OB-SGs (spectral types refer to O-stars, i.e., 10
corresponds to B0 and 20 to A0). Data derived in the present study are
marked with diamonds while crosses refer to published data \citep{dft06,simon}.
}
\label{vsini_vmac}
\end{figure*}

Since the effects of macro-turbulence are similar to those caused by axial 
rotation, (i.e., they do not change the line strengths but ``only'' modify 
the profile shapes) and since stellar rotation is a key parameter, such as 
for
stellar evolution calculation (e.g., \citealt{MM00}, \citealt{HMM} and
references therein), it is particularly important to distinguish between the
individual contributions of these two processes. 

There are at least two possibilities to approach this problem: either exploiting
the goodness of the fit between observed and synthetic profiles
\citep{Ryans} or analysing the shape of the Fourier transforms (FT) of
absorption lines \citep{Gray73,Gray75,simon}. Since the second method 
has been proven to provide better constraints \citep{dft06}, we 
followed this approach to separate and measure the relative magnitudes of
rotation and macro-turbulence.

The principal idea of the FT method relates to the fact that in Fourier 
space the convolutions of the ''intrinsic line profile" (which
includes the natural, thermal, collisional/Stark and
microturbulence broadening) with the instrumental, rotational and
macro-turbulent profiles, become simple products of the corresponding Fourier
components, thus allowing the contributions of the latter two processes to
be separated by simply dividing the Fourier components of the observed
profile by the components of the thermal and instrumental profile. 

The first minimum of the Fourier amplitudes of the obtained residual
transform will then fix the value of \vsini\ while the shape of the first
side-lobe of the same transform will constrain $v_{\rm mac}$. 

The major requirements to obtain reliable results from this method is the
presence of {\it high quality} spectra (high S/N ratio and high spectral
resolution) and to analyse only those lines which are free from strong
pressure broadening but are still strong enough to allow for reliable \vsini\
estimates.

For the purpose of the present analysis, we have used the implementation of
the FT technique as developed by \citet{simon} (based on the original method
proposed by \citealt{Gray73, Gray75}) and applied it to a number of
preselected absorption lines fulfilling the above requirements. In
particular, for our sample of {\it early} B subtypes, the Si~III multiplet
around 4553 \AA\ but also lines of O~II and N~II were selected, whereas for
the rest the Si~II doublet around 4130 \AA\ and the Mg~II line at
$\lambda$4481 were used instead. 

The obtained pairs of (\vsini, \vmac), averaged over the measured lines,
were then used as input parameters for the fitting procedure and
subsequently modified to improve the fits.\footnote{Note that in their FT
procedure \citet{simon} have used a Gaussian profile (with EW equal to that
of the observed profile) as ``intrinsic profile''.
Deviations from this shape due to, e.g., natural/collisional broadening are
not accounted for, thus allowing only rough estimates of \vmac to
be derived, which have to be adjusted during the fit
procedure.} 
The finally adopted values of \vsini\ and \vmac\ are listed in Columns~4 and
5 of Table~\ref{para_1}, respectively. Numbers in brackets refer to the
number of lines used for this analysis.  The uncertainty of these estimates
is typically less than $\pm$10~km~s$^{\rm -1}$, being largest for those stars 
with a
relatively low rotational speed, due to the limitations given by the
resolution of our spectra ($\sim$35~\kms). 
Although the sample size is small, the \vsini\ and \vmac\ data listed in
Table~\ref{para_1} indicate that:  

\smallskip
\noindent
$\bullet$
in none of the sample stars is rotation alone  able to reproduce the 
observed line profiles (width and shape). 

\noindent
$\bullet$
both \vsini\ and \vmac\ decrease towards later subtypes (lower 
\Teff), being about a factor of two lower at B9 than at B0.5.

\noindent
$\bullet$
independent of spectral subtype, the size of the
macro-turbulent velocity  is similar to the size of the 
projected rotational velocity. 

\noindent
$\bullet$
also in all cases, \vmac\ is well beyond the speed of sound.

\smallskip
\noindent
Compared to similar data from other investigations for stars in common 
(e.g. \citealt{Rosenh70, Howarth97}), our \vsini\ estimates are always 
smaller, by up to 40\%, which is understandable since these earlier
estimates refer to an interpretation in terms of rotational broadening
alone. 

On the other hand, and within a given spectral subtype, our estimates of
\vsini\ and \vmac\ are consistent with those derived by  \citet{dft06}  
and \citet{simon} (see Figure~\ref{vsini_vmac}).
From these data it is obvious that both \vsini\ and \vmac appear to 
decrease (almost monotonically) in concert, when proceeding from early-O to 
late B-types.

\subsection {Effective temperatures, \Teff}
\label{teff}

For B-type stars the primary temperature diagnostic at optical 
wavelengths is Silicon \citep{BB90, Killian91, McE99, Trundle04} which shows
strong lines from three ionisation stages through all the spectral types: 
Si~III/Si~IV for earlier and Si~II/Si~III for later subtypes, with a 
``short'' overlap at B1.5 - B2. To evaluate \Teff (and \logg), to a large
extent we employed the method of line profile fitting instead of using fit
diagrams (based on EWs), since in the latter case the corresponding
estimates rely on interpolations and furthermore do not account for the
profile shape. Note, however, that for certain tasks (namely the derivation
of the Si-abundance together with the micro-turbulent velocity), EW-methods
have been applied (see below).

In particular, to determine \Teff we used the Si~II features at
$\lambda\lambda$4129, 4131, the Si~III features at $\lambda\lambda$4553,
4568, 4575 and at $\lambda\lambda$4813, 4819, 4828, with a preference on the
first triplet (see 
\begin{figure*}
\begin{minipage}{8.8cm}
\resizebox{\hsize}{!}
{\includegraphics{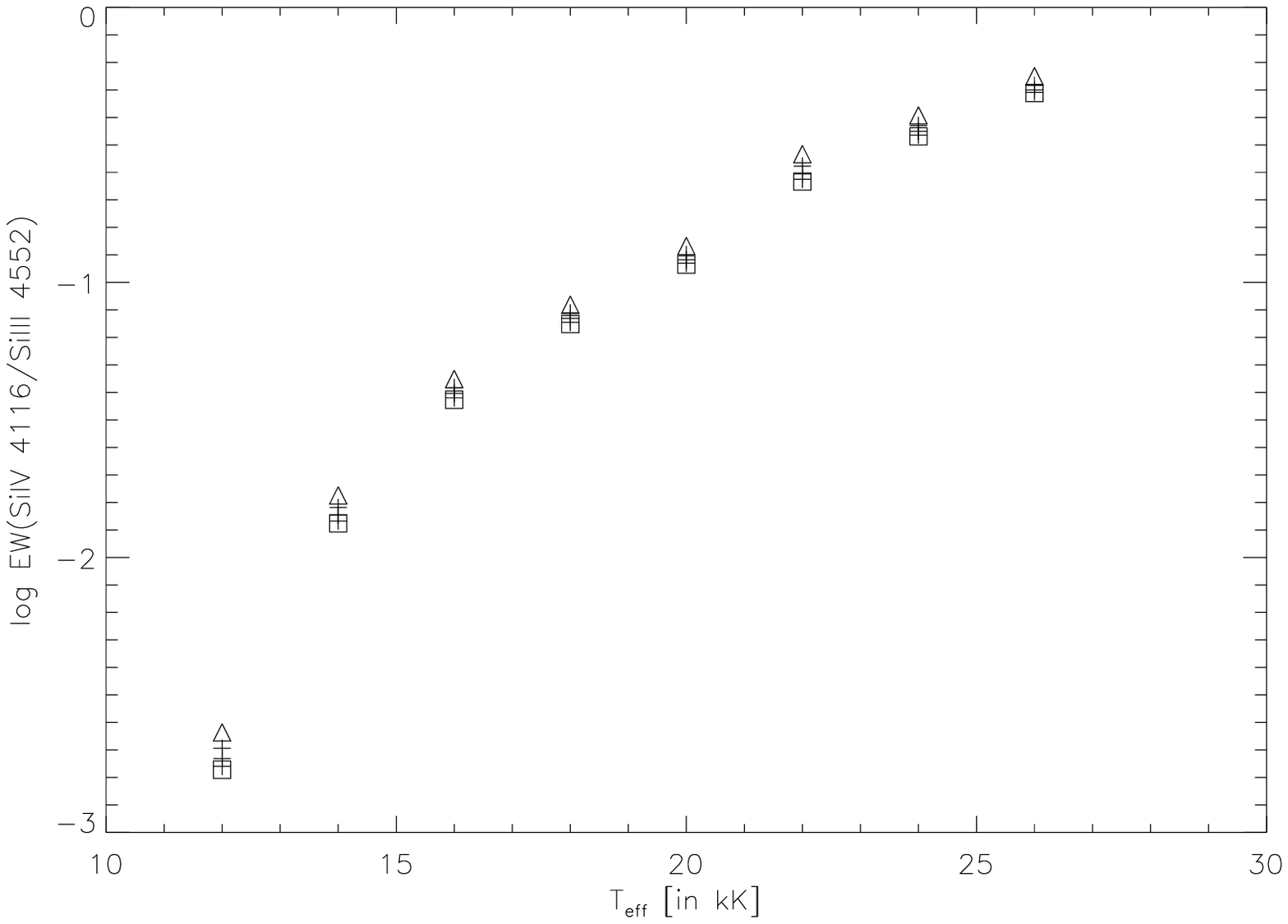}}
\end{minipage}
\hfill
\begin{minipage}{8.8cm}
\resizebox{\hsize}{!}
{\includegraphics{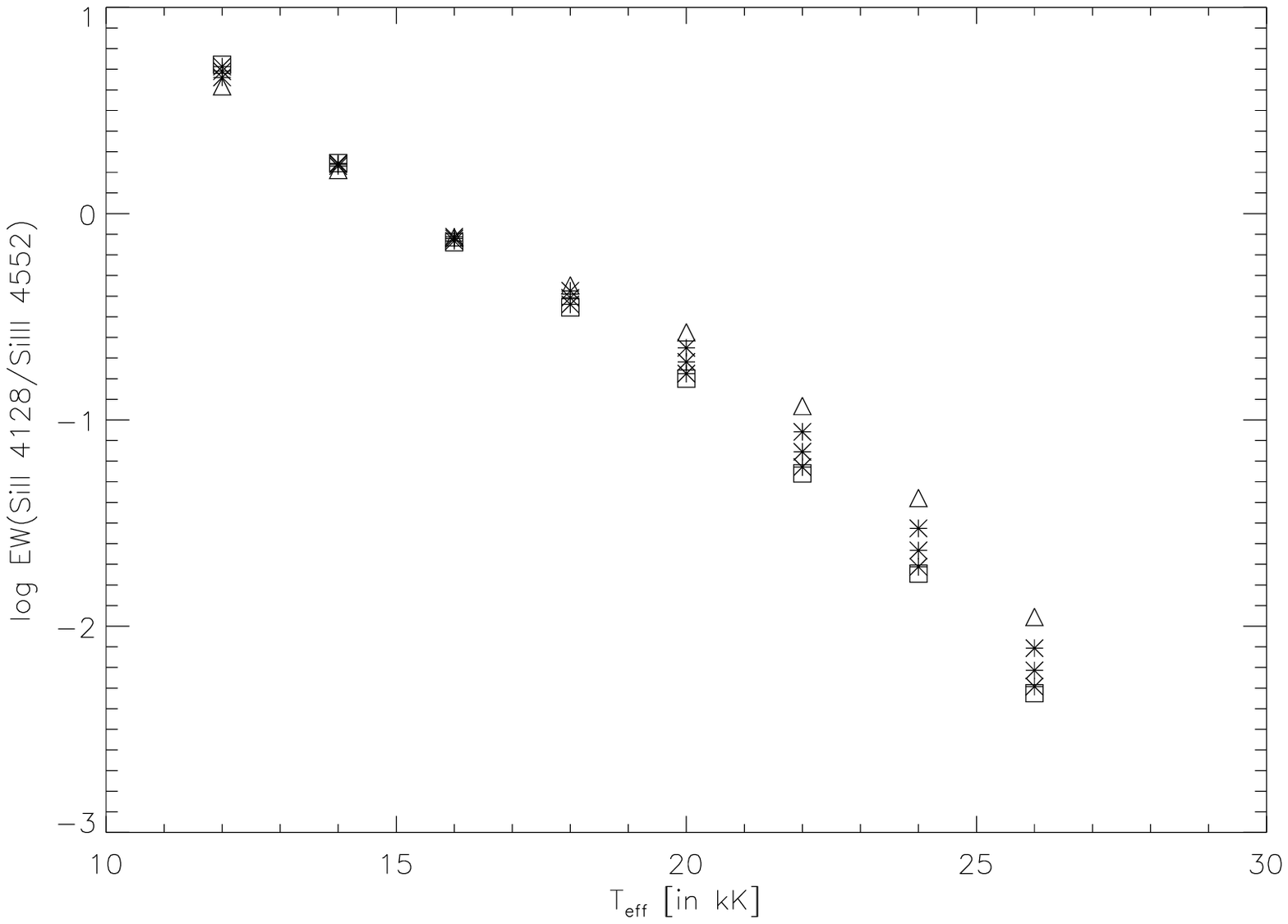}}
\end{minipage}
\caption{Effects of micro-turbulence on the strength of Si~IV~4116/Si~III~4553 
(left panel) and Si~II~4128/Si~III~4553 (right panel) equivalent width 
ratios for B-type SGs. $V_{\rm mic}$ ranging from 4 to 18 \kms,
with increments of 4 \kms. Triangles refer to the lowest value 
of $v_{\rm mic}$, squares to the highest (see text).}
\label{Si_ratio}
\end{figure*}
below) and the Si~IV feature at 
$\lambda$4116.\footnote{Si~IV $\lambda$4089 is unavailable in our
spectra. Given that in early B-SGs this line is strongly blended by O~II
which cannot be synthesised by FASTWIND with our present model atoms, this
fact should not affect the outcome of our analysis.} In addition, for stars
of spectral type B2 and earlier the Helium ionisation balance was exploited
as an additional check on \Teff, involving He~I transitions at
$\lambda\lambda$4471, 4713, 4387, 4922 and He~II transitions at
$\lambda\lambda$4200, 4541, 4686.

\subsubsection{Micro-turbulent velocities  and Si abundances}
\label{vmic}

Though the introduction of a non-vanishing micro-turbulent velocity can
significantly improve the agreement between synthetical profiles and
observations \citep{McE98, SH98}, it is still not completely clear whether
such a mechanism (operating on scales below the photon mean free
path) is really present or whether it is an artefact of some deficiency in
the present-day description of model atmospheres (e.g., \citealt{McE98,
Villamariz} and references therein). 

Since micro-turbulence can strongly affect the strength of Helium and
metal lines, its inclusion into atmospheric models and profile functions
can significantly modify the derived stellar
abundances but also effective temperatures and surface gravities (the latter
two parameters mostly indirectly via its influence on line blanketing:
stronger \vmic implies more blocking/back-scattering, and thus lower \Teff.)

Whereas \citet{Villamariz} showed the effects of \vmic to be relatively
small for O-type stars, for B-type stars this issue has only been 
investigated for early B1-B2 giants (e.g., \citealt{Vrancken}) and a few, 
specific BA supergiants (e.g. \citealt{Urb, Przb06}). Here, we report on the
influence of micro-turbulence on the derived effective
temperatures\footnote{Note that the Balmer lines remain almost unaffected by
\vmic so that a {\it direct} effect of \vmic on the derived \logg is
negligible.} for the complete range of B-type SGs. For this purpose
we used a corresponding sub-grid of FASTWIND models with \vmic
ranging from 4 to 18 \kms (with increments of 4 \kms) and $\log Q$ values
corresponding to the case of relatively weak winds. Based on these models we
studied the behaviour of the SiIV4116/SiIII4553 and SiII4128/SiIII4553 line
ratios and found these ratios to be almost insensitive to variations in
\vmic (Fig.~\ref{Si_ratio}), except for the case of SiII/SiIII beyond 18~kK
where differences of about 0.3 to 0.4~dex can be seen (and are to be expected,
due to the large difference in absolute line-strengths caused by strongly
different ionisation fractions). Within the temperature ranges of interest
(18~$\leq$~\Teff~$\leq$~28 kK for SiIV/SiIII and 12~$\leq$~\Teff~$\leq$~18
kK for SiII/SiIII), however, the differences are relatively small, about
0.15~dex or less, resulting in temperature differences lower than 1\,000~K,
i.e., within the limits of the adopted uncertainties (see below).

Based on these results, we relied on the following strategy to determine
$T_{\rm eff}$, \vmic\ and Si abundances. As a first step, we used the FASTWIND model
grid as described previously (with \vmic=10 and 15 \kms and ``solar'' Si
abundance) to put initial constraints on the stellar and wind parameters of
the sample stars. Then, by varying \Teff\ (but also \logg, \Mdot\ and
velocity-field parameter $\beta$) within the derived limits and by changing
\vmic within $\pm$5 \kms to obtain a satisfactory fit to most of the
strategic Silicon lines, we fixed \Teff/\logg and derived rough estimates of
$v_{\rm mic}$.
Si abundances and final values for \vmic resulted from the following
procedure: for each sample star a grid of 20 FASTWIND models was calculated,
combining four abundances and five values of micro-turbulence (ranging from
10 to 20 \kms or from 4 to 12 km~s$^{\rm -1}$, to cover hot and cool stars,
respectively). By means of this grid, we determined those abundance ranges
which reproduce the observed individual EWs (within the corresponding
errors) of several previously selected Si lines from different ionisation
stages. Subsequently, we sorted out the value of \vmic which provides the
best overlap between these ranges, 
\begin{figure*}
\begin{minipage}{2.9cm}
\resizebox{\hsize}{!}
{\includegraphics{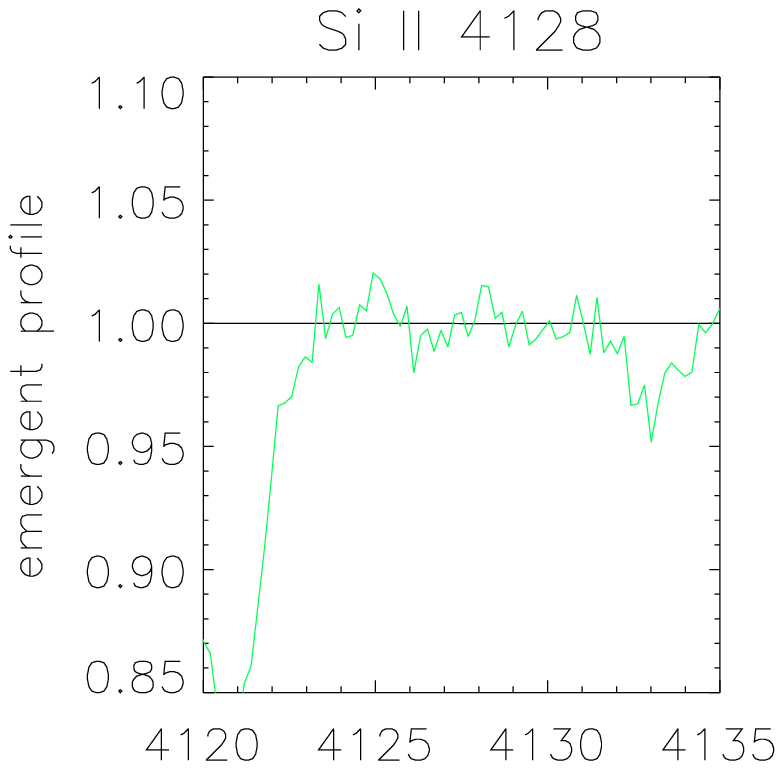}}
\end{minipage}
\hfill
\begin{minipage}{2.9cm}
\resizebox{\hsize}{!}
{\includegraphics{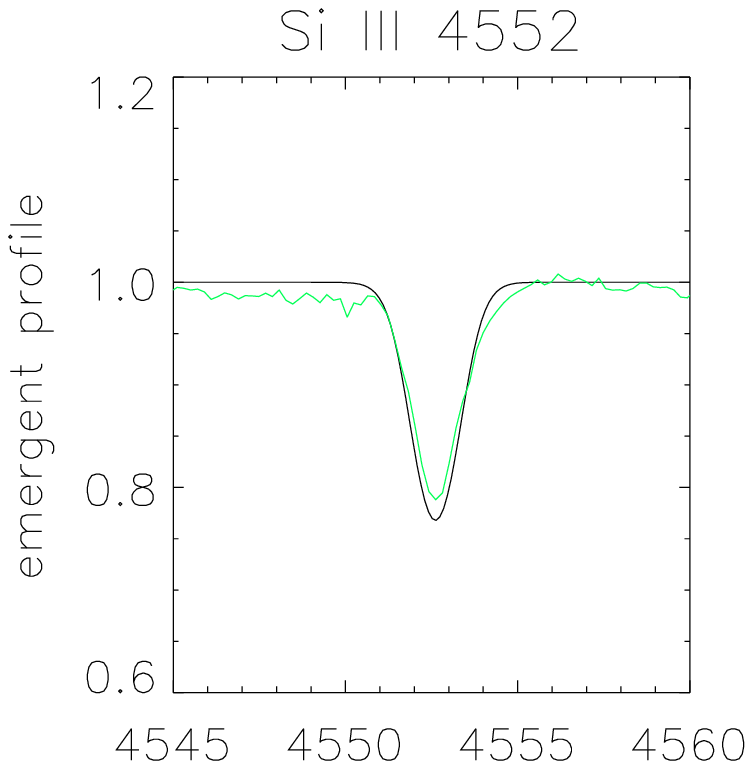}}
\end{minipage}
\hfill
\begin{minipage}{2.9cm}
\resizebox{\hsize}{!}
{\includegraphics{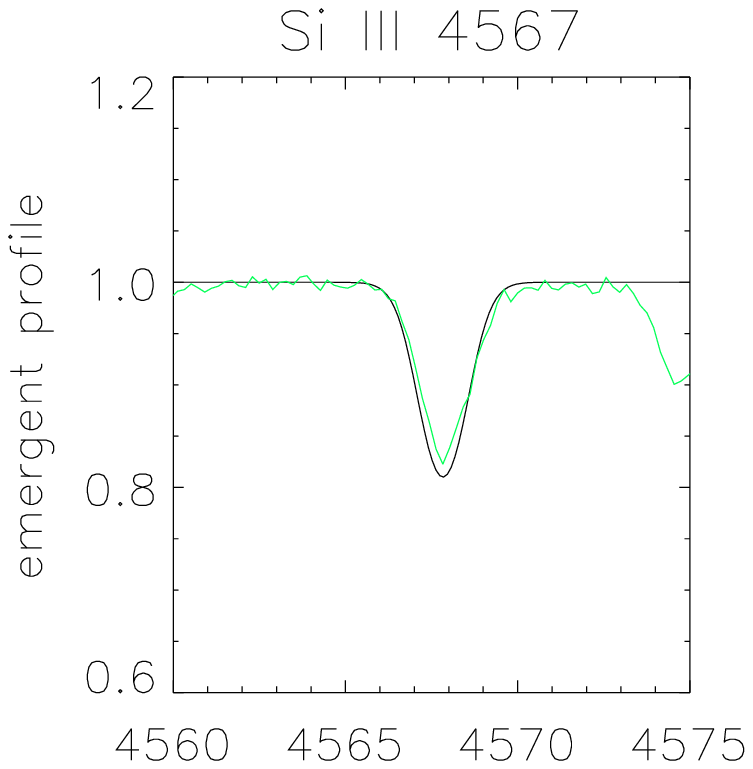}}
\end{minipage}
\hfill
\begin{minipage}{2.9cm}
\resizebox{\hsize}{!}
{\includegraphics{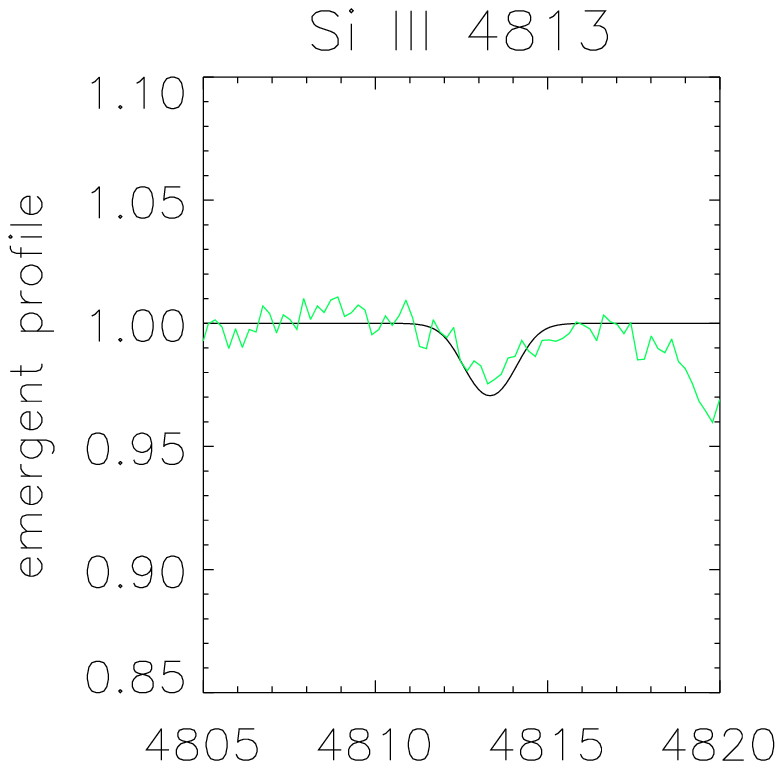}}
\end{minipage}
\hfill
\begin{minipage}{2.9cm}
\resizebox{\hsize}{!}
{\includegraphics{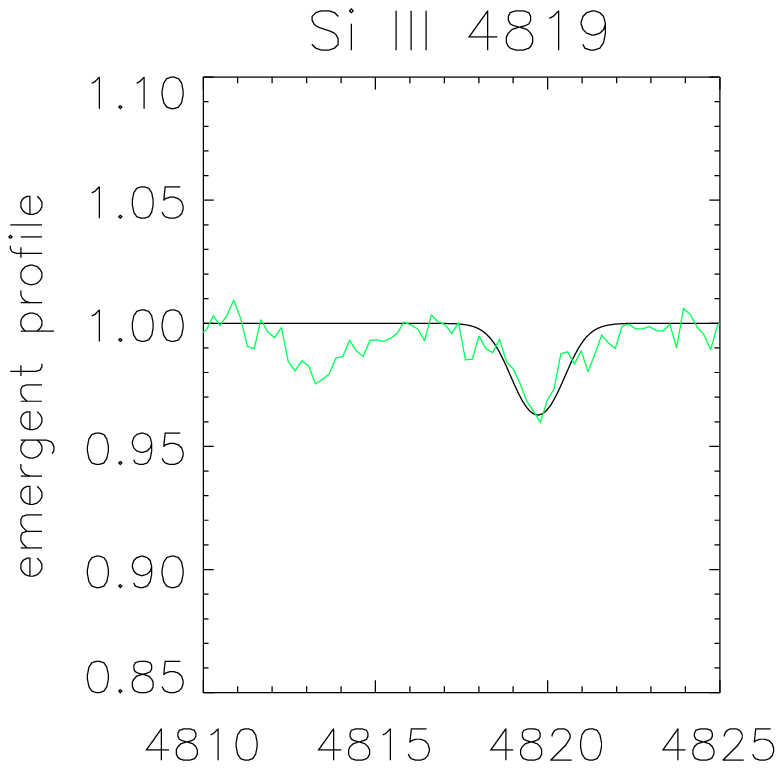}}
\end{minipage}
\hfill
\begin{minipage}{2.9cm}
\resizebox{\hsize}{!}
{\includegraphics{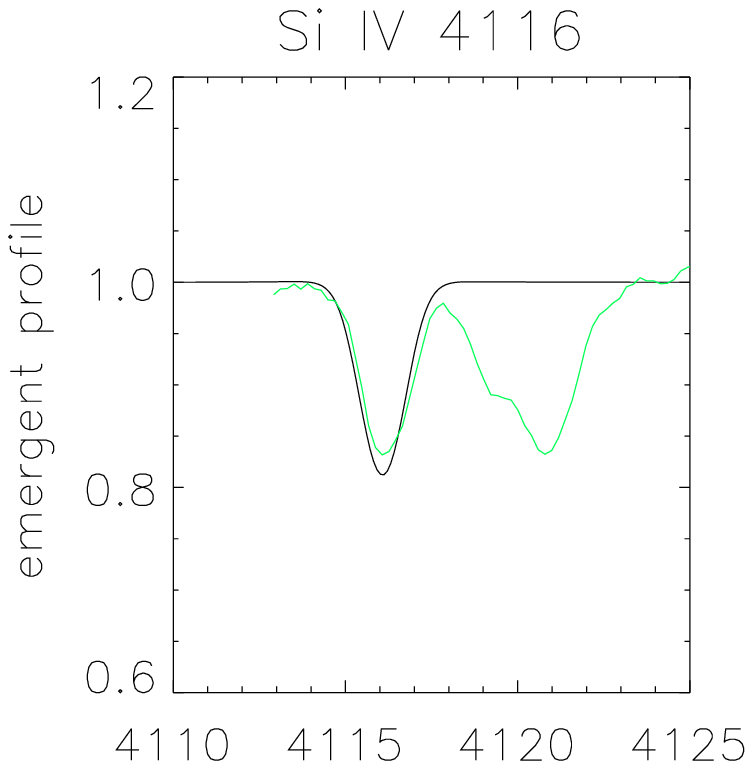}}
\end{minipage}
\\
\begin{minipage}{2.9cm}
\resizebox{\hsize}{!}
{\includegraphics{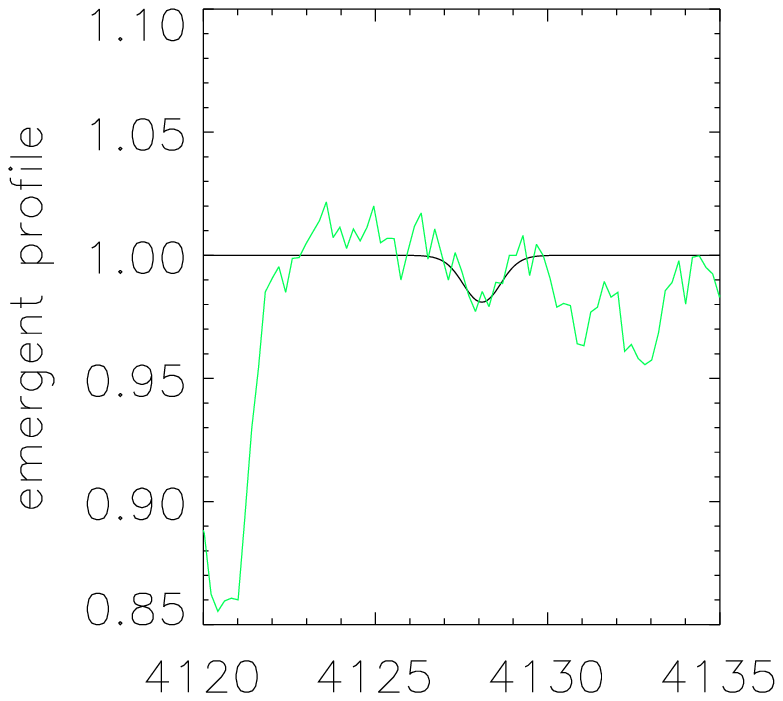}}
\end{minipage}
\hfill
\begin{minipage}{2.9cm}
\resizebox{\hsize}{!}
{\includegraphics{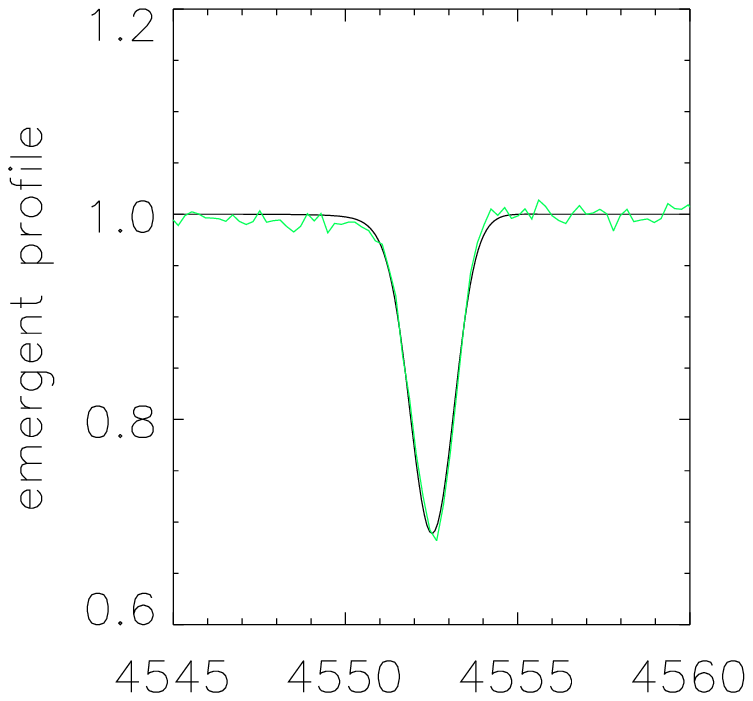}}
\end{minipage}
\hfill
\begin{minipage}{2.9cm}
\resizebox{\hsize}{!}
{\includegraphics{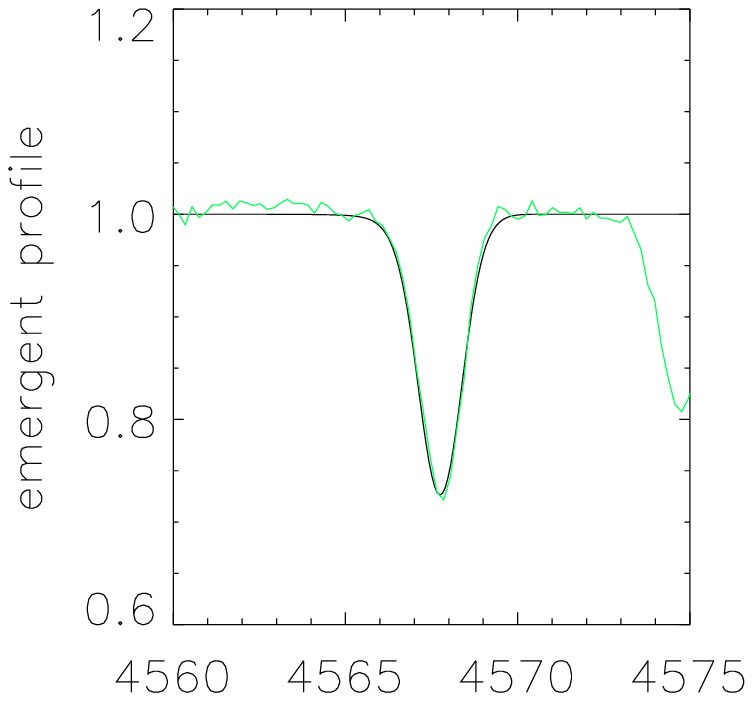}}
\end{minipage}
\hfill
\begin{minipage}{2.9cm}
\resizebox{\hsize}{!}
{\includegraphics{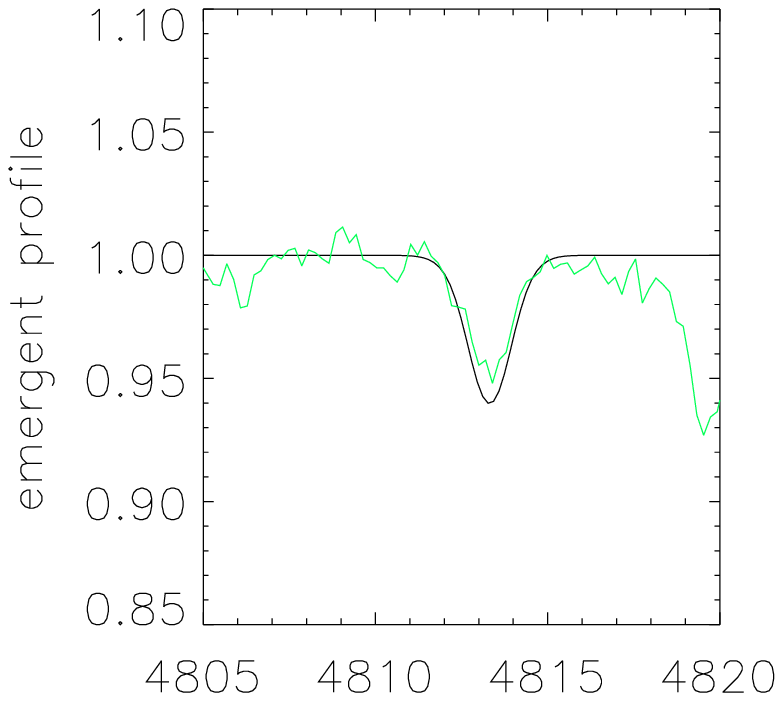}}
\end{minipage}
\hfill
\begin{minipage}{2.9cm}
\resizebox{\hsize}{!}
{\includegraphics{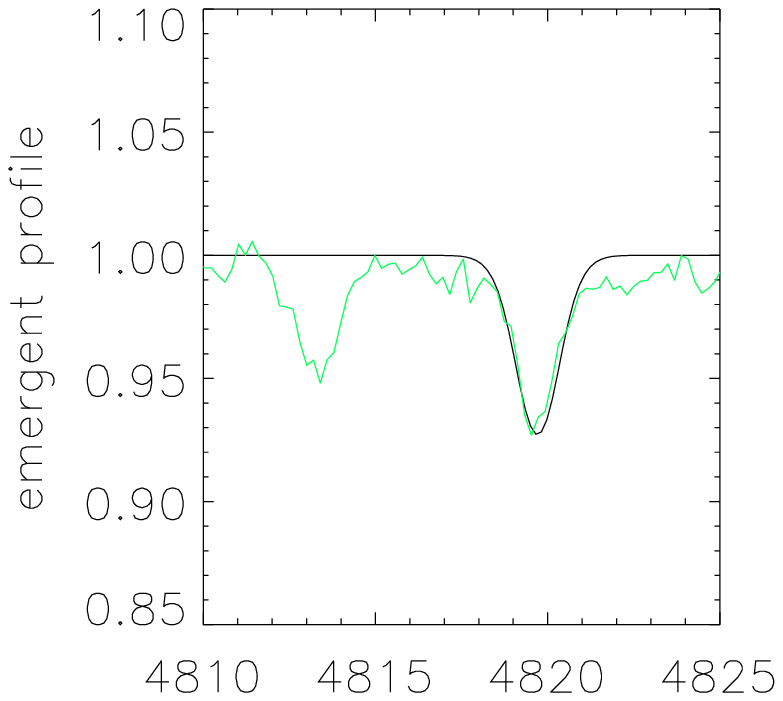}}
\end{minipage}
\hfill
\begin{minipage}{2.9cm}
\resizebox{\hsize}{!}
{\includegraphics{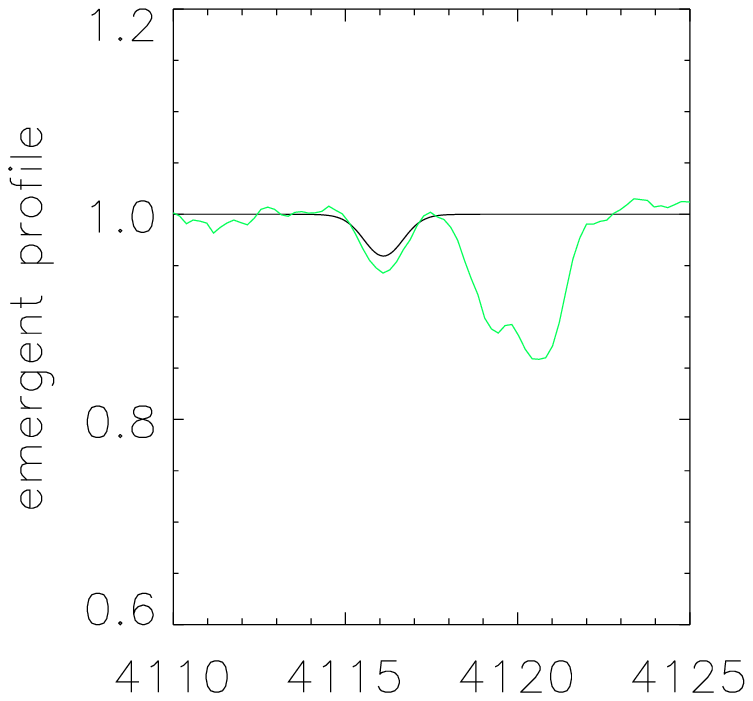}}
\end{minipage}
\\
\begin{minipage}{2.9cm}
\resizebox{\hsize}{!}
{\includegraphics{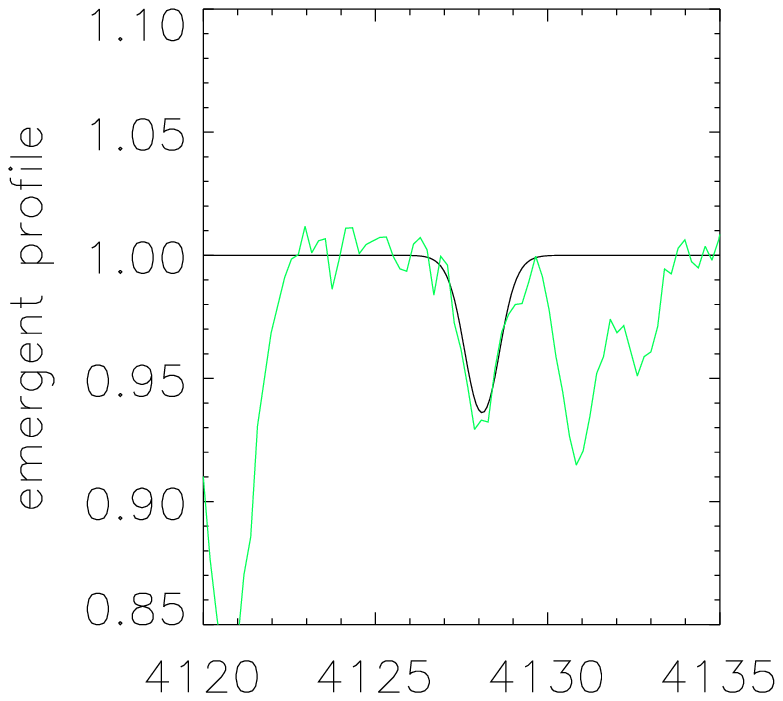}}
\end{minipage}
\hfill
\begin{minipage}{2.9cm}
\resizebox{\hsize}{!}
{\includegraphics{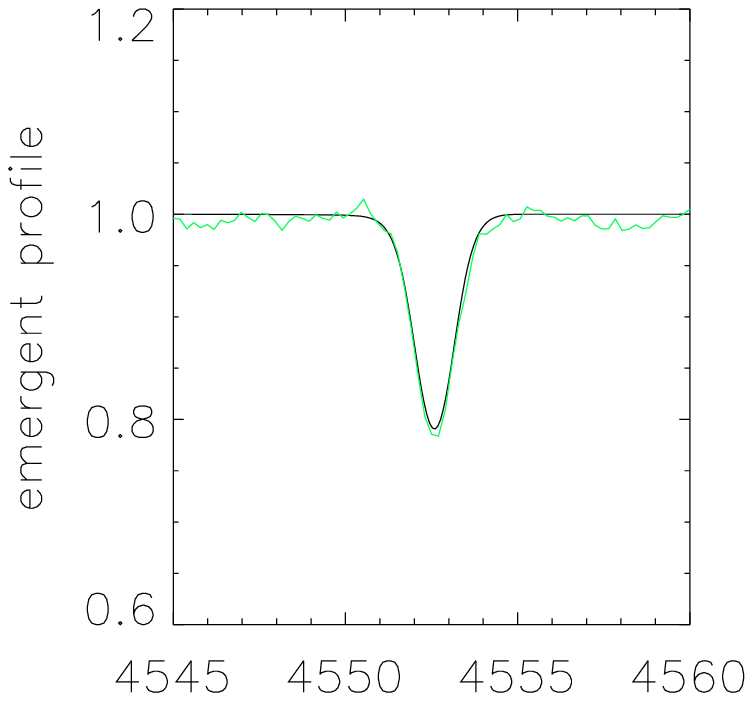}}
\end{minipage}
\hfill
\begin{minipage}{2.9cm}
\resizebox{\hsize}{!}
{\includegraphics{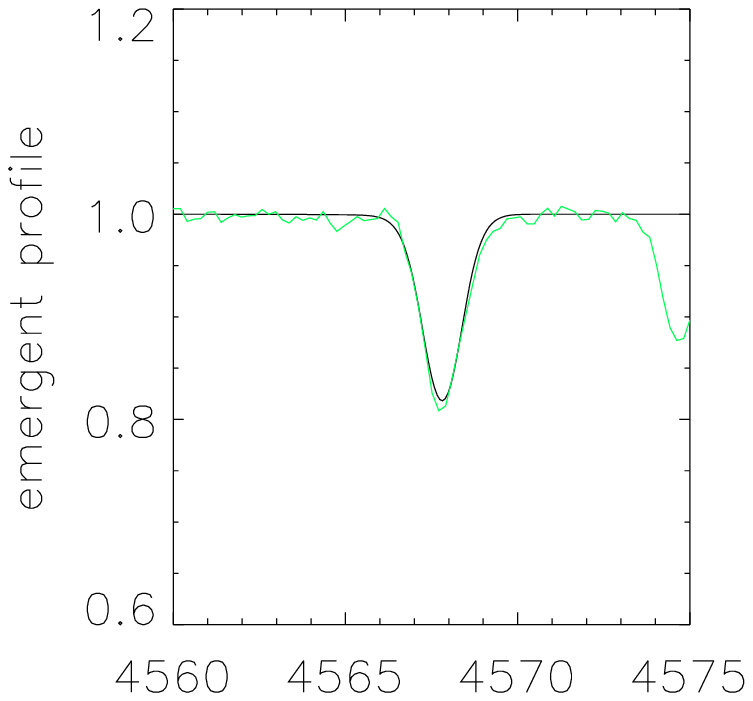}}
\end{minipage}
\hfill
\begin{minipage}{2.9cm}
\resizebox{\hsize}{!}
{\includegraphics{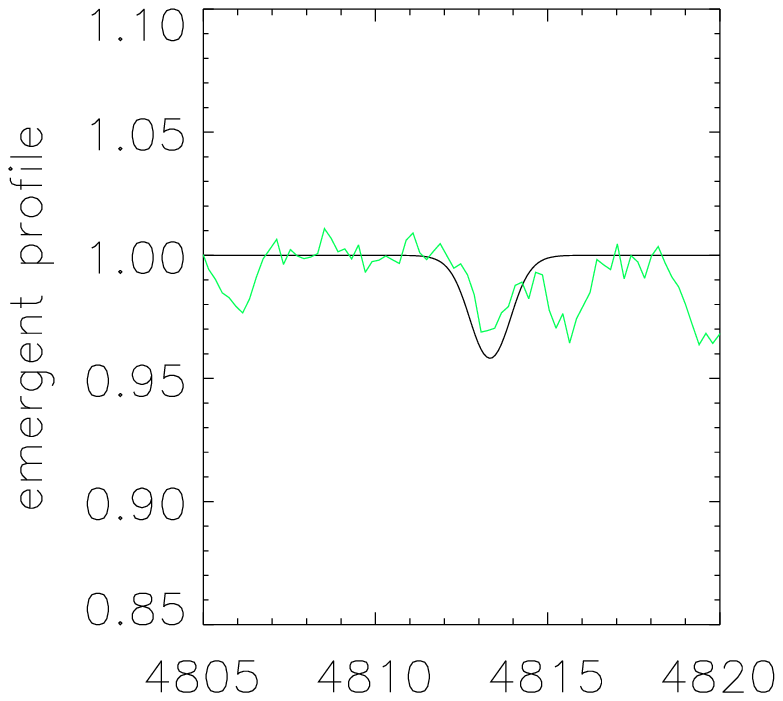}}
\end{minipage}
\hfill
\begin{minipage}{2.9cm}
\resizebox{\hsize}{!}
{\includegraphics{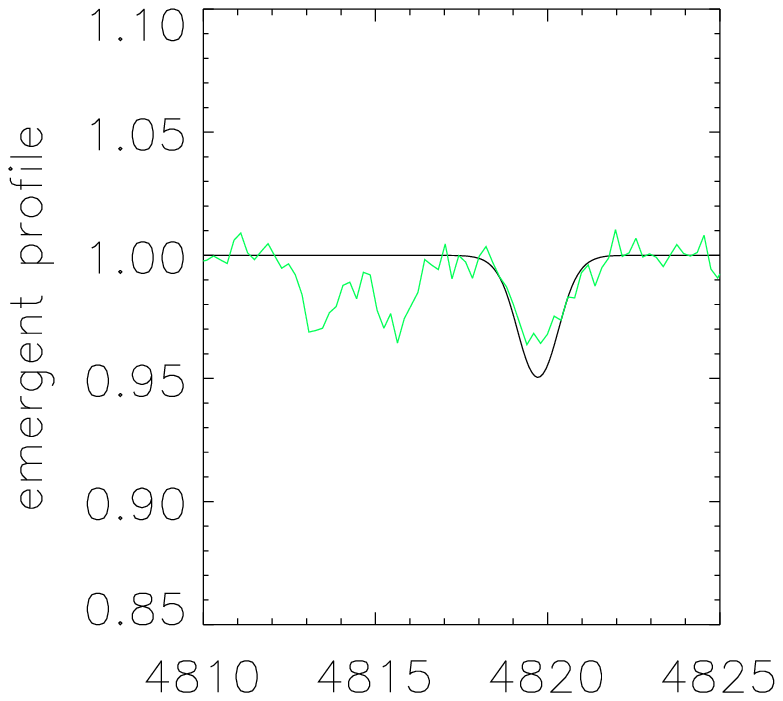}}
\end{minipage}
\hfill
\begin{minipage}{2.9cm}
\resizebox{\hsize}{!}
{\includegraphics{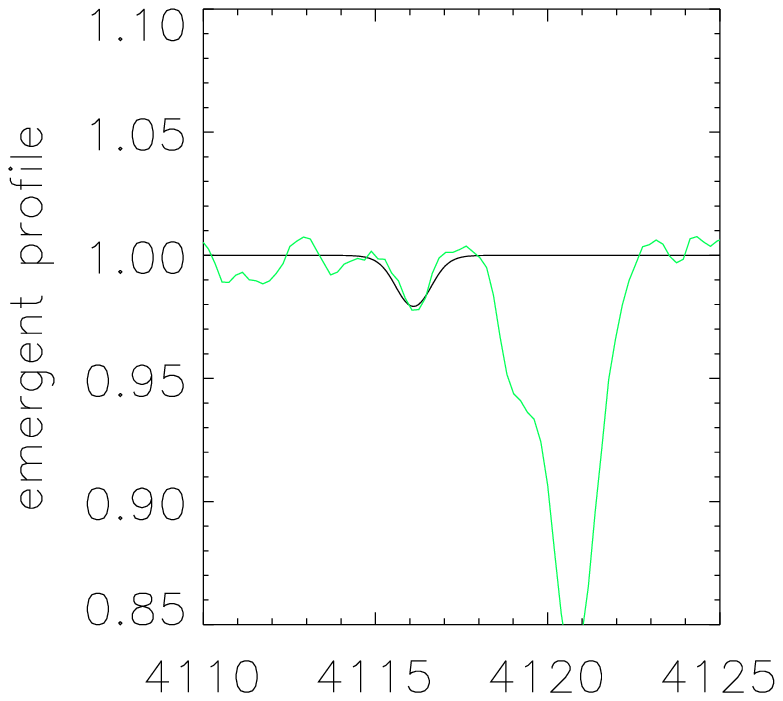}}
\end{minipage}
\\
\begin{minipage}{2.9cm}
\resizebox{\hsize}{!}
{\includegraphics{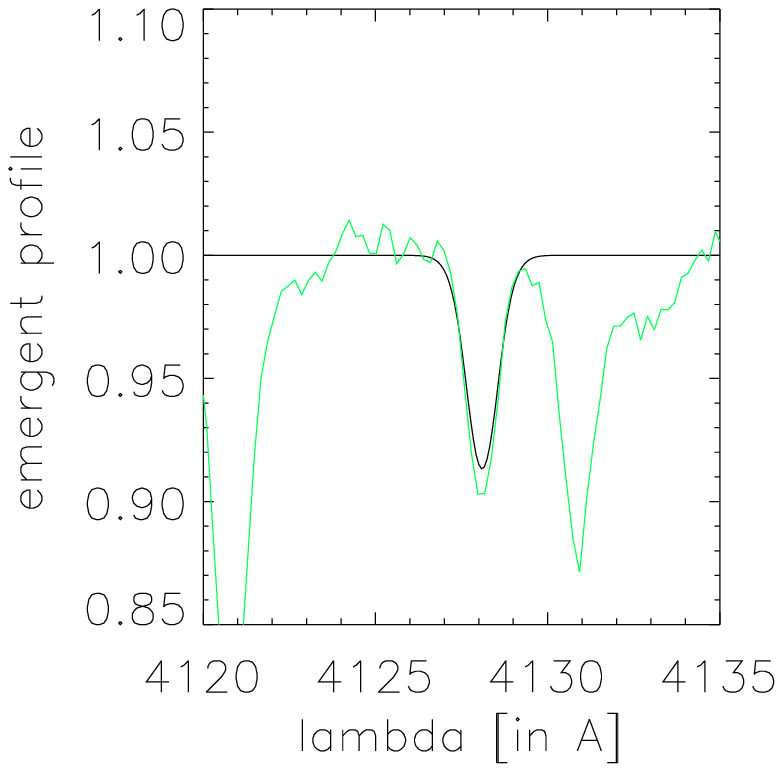}}
\end{minipage}
\hfill
\begin{minipage}{2.9cm}
\resizebox{\hsize}{!}
{\includegraphics{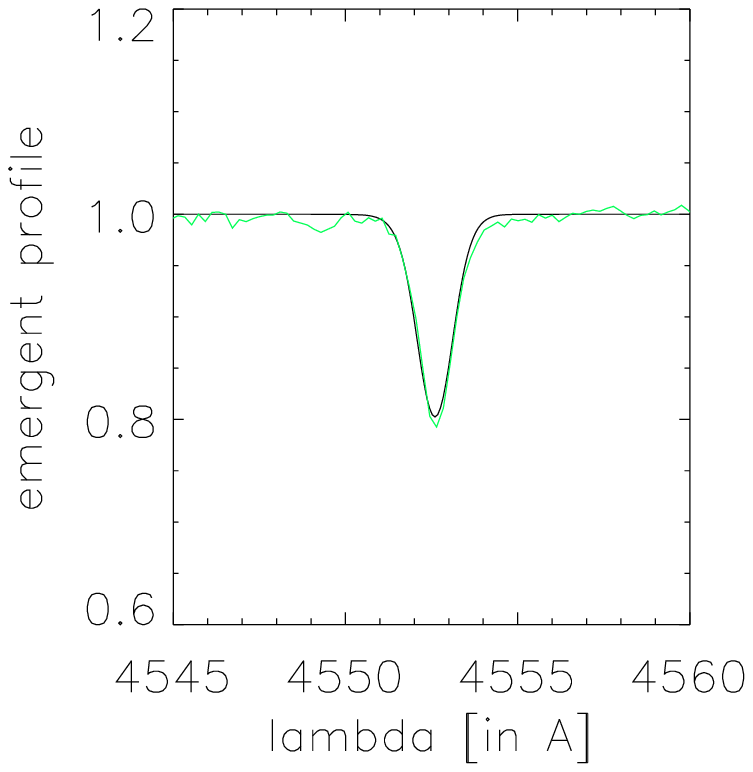}}
\end{minipage}
\hfill
\begin{minipage}{2.9cm}
\resizebox{\hsize}{!}
{\includegraphics{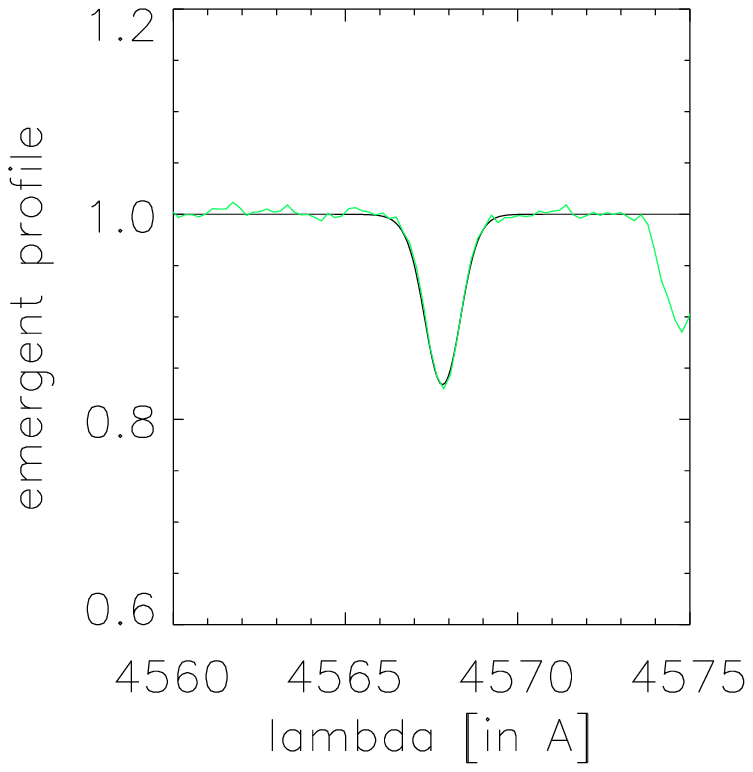}}
\end{minipage}
\hfill
\begin{minipage}{2.9cm}
\resizebox{\hsize}{!}
{\includegraphics{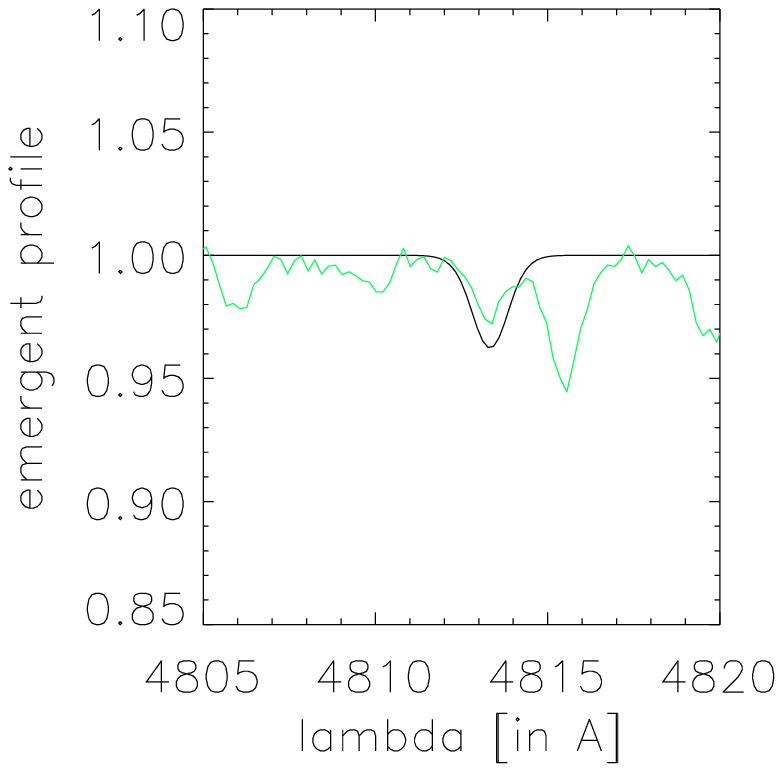}}
\end{minipage}
\hfill
\begin{minipage}{2.9cm}
\resizebox{\hsize}{!}
{\includegraphics{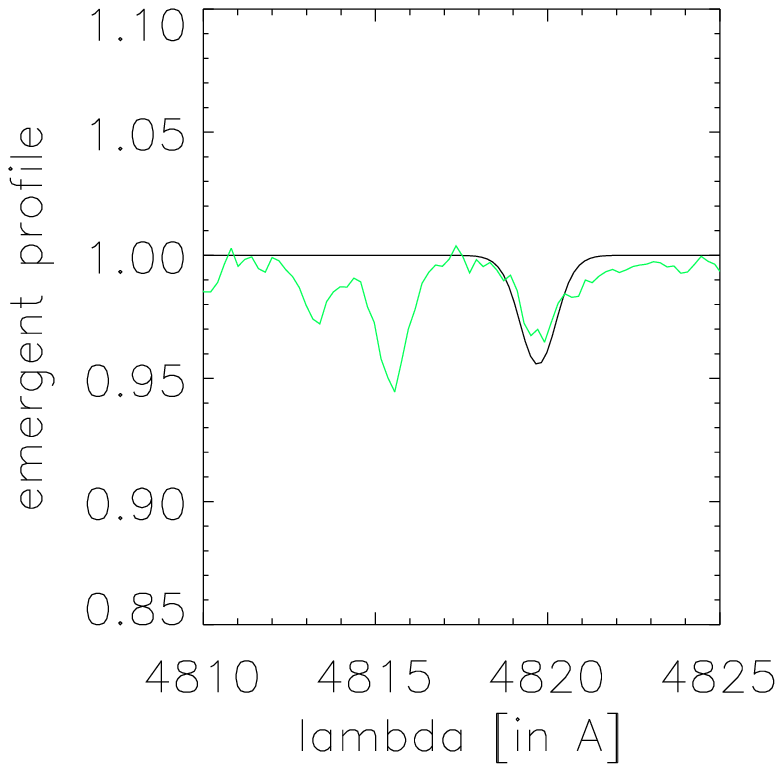}}
\end{minipage}
\hfill
\begin{minipage}{2.9cm}
\resizebox{\hsize}{!}
{\includegraphics{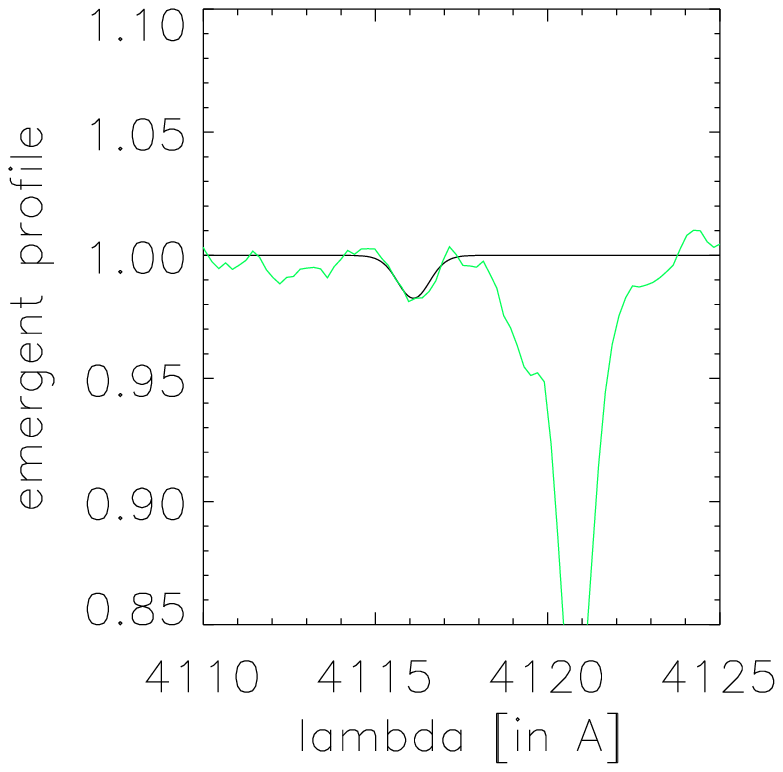}}
\end{minipage}
\caption{Examples for the quality of our final (``best'') Silicon
line-profile fits.  From top to bottom: HD~185\,859 (B0.5Ia), HD~190\,603
(B1.5Ia+), HD~206\,165 (B2Ib) and HD~198\,478 (B2.5Ia). Note the systematic
discrepancy between synthetic and observed profiles of the Si~III lines at
$\lambda\lambda$4813,4819. }
\label{si_prof}
\end{figure*}
i.e., defines a unique abundance together
with appropriate errors (for more details, see e.g. \citealt{Urb, simon1} and
references therein).

Our final results for \vmic were almost identical (within about 
$\pm$1~\kms) to those derived from the ``best'' fit to Silicon. Similarly, 
for all but one star, our final estimates for the Si abundance are
quite similar to the initially adopted ``solar'' one, within 
$\leq\pm$0.1~dex, and only for HD~202\,850 an increase of 0.44 dex was
found.
Given that Si is not involved in CNO nuclear processing, the latter
result is difficult to interpret. On the one hand, fitting/analysis
problems are highly improbable, since no unusual results have been obtained
for the other late B-SG, HD~212\,593. Indeed, the overabundance is almost
``visible'' because the EWs of at least 2 of the 3 strategic Si lines are
significantly larger (by about 20 to 40\%) in HD~202\,850 than in
HD~212\,593.  On the other hand, the possibility that this star is metal
rich seems unlikely given its close proximity to our Sun. Another
possibility might be that HD~202\,850 is a Si star, though its magnetic
field does not seem particularly strong (but exceptions are still known,
e.g. V~828 Her B9sp, EE~Dra B9, \citealt{BBM}). A detailed abundance
analysis may help to solve this puzzling feature.

Finally, we have verified that our newly {\it derived} Si abundances (plus
\vmic-values) do not affect the stellar parameters (which refer to the
initial abundances), by means of corresponding FASTWIND models. Though an 
increase of 0.44 dex (the exceptional case of HD~202\,850) 
makes the Si-lines stronger, this strengthening does not affect
the derived $T_{\rm eff}$, since the latter parameter depends on line {\it ratios}
from different ionisation stages, being thus almost independent of
abundance. In each case, however, the quality of the Silicon line-profile
fits has been improved, as expected.

In Column 6 and 7 of Table~\ref{para_1} we present our final values for
\vmic\ and Si abundance. The error of these estimates depends on the 
accuracy of the measured equivalent widths (about 10\%) and is typically
about $\pm$2~\kms and $\pm$0.15~dex for \vmic\ and the logarithmic Si
abundance, respectively. A closer inspection of these data indicates that
the micro-turbulent velocities of B-type SGs might be closely
related to spectral type  (see also \citealt{McE99}), being highest at
earlier (18~\kms at B0.5) and lowest at later B subtypes (7~\kms at B9).
Interestingly, the latter value is just a bit larger than the typical
values reported for A-SGs (3 to 8~\kms, e.g., \citealt{venn}), thus
implying a possible decline in micro-turbulence towards even later spectral
types.

\subsubsection {Silicon line profiles fits -- a closer inspection}
\label{siIII}

During our fitting procedure, we encountered the problem that the strength
of the Si~III multiplet near 4813 \AA\ was systematically over-predicted
(see Fig.~\ref{si_prof} for some illustrative examples). Though this
discrepancy is not very large (and vanishes if \Teff is modified within
($\pm$500 K), this discrepancy might point to some weaknesses in our model
assumptions or data. Significant difficulties in reproducing the strength of
Si~III multiplets near 4553 and 4813 \AA\ have also been  encountered by
\citet{McE99} and by \citet{BB90}. While in the former study both multiplets
were found to be {\it weaker} in their lowest-gravity models with \Teff\
beyond 22\,500 K, in the latter study the second multiplet was overpredicted,
by a factor of about two. 

The most plausible explanation for the problems encountered by
\citet{McE99} (which are opposite to ours) is the neglect of
line-blocking/blanketing and wind effects in their NLTE model calculations,
as already suggested by the authors themselves. 
The discrepancies reported by \citet{BB90}, on the other hand, are in
qualitative agreement to our findings, but
much more pronounced (a factor of two against 20 to 30\%). Since both
studies use the same Si~III model ion whilst we have updated the
oscillator strengths of the multiplet near 4553 \AA(!) drawing from the
available atomic databases, we suggest that it is these improved oscillator
strengths in conjunction with modern stellar atmospheres which have reduced
the noted discrepancy.

Regardless of these improvements, the remaining discrepancy must have an
origin, and there are at least two possible explanations: (i) too small an
atomic model for Si~III (cutoff effects) and/or (ii) radially stratified
micro-turbulent velocities (erroneous oscillator strengths cannot be
excluded, but are unlikely, since all atomic databases give similar
values).

The first possibility was discussed by \citet{BB90} who concluded that this 
defect cannot be the sole origin of their problem, since the required
corrections are too large and furthermore would affect the other term
populations in an adverse manner. Because the discrepancy found by us has
significantly decreased since then, the possibility of too small an 
atomic model can no longer be ruled out though. Future work on improving the
complete Si atom will clarify this question.

In our analyses, we have used the same value of \vmic\ for {\it all} 
lines in a given spectrum, i.e., assumed that this quantity does
not follow any kind of stratification throughout the atmosphere, whilst the
opposite might actually be true (e.g. \citealt{McE99, Vrancken, Trundle02,
Trundle04, hunter}). If so, a micro-turbulent velocity being a factor of two
lower than inferred from the ``best'' fit to Si~III 4553 and the Si~II
doublet would be needed to reconcile calculated and observed strengths of
the 4813 \AA\ multiplet. Such a number does not seem unlikely, 
given the difference in line strengths, but clearly further investigations
are necessary (after improving upon the atomic model) to clarify this
possibility (see also \citealt{hunter} and references therein).

Considering the findings from above, we decided to follow \citet{BB90}
and to give preference to the Si~III multiplet near 4553 \AA\ 
throughout our analysis. Since this multiplet is observable over the whole B
star temperature range while the other (4813 \AA) disappears at mid-B
subtypes, such a choice has the additional advantage of providing consistent
results for the complete sample, from B0 to B9.
\begin{figure}
\begin{minipage}{2.8cm}
\resizebox{\hsize}{!}
{\includegraphics{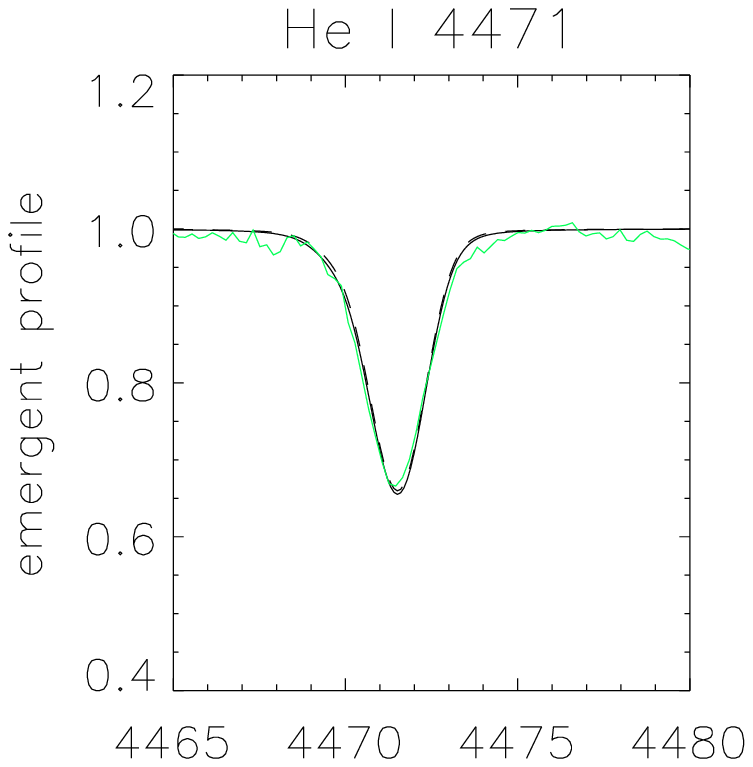}}
\end{minipage}
\hfill
\begin{minipage}{2.8cm}
\resizebox{\hsize}{!}
{\includegraphics{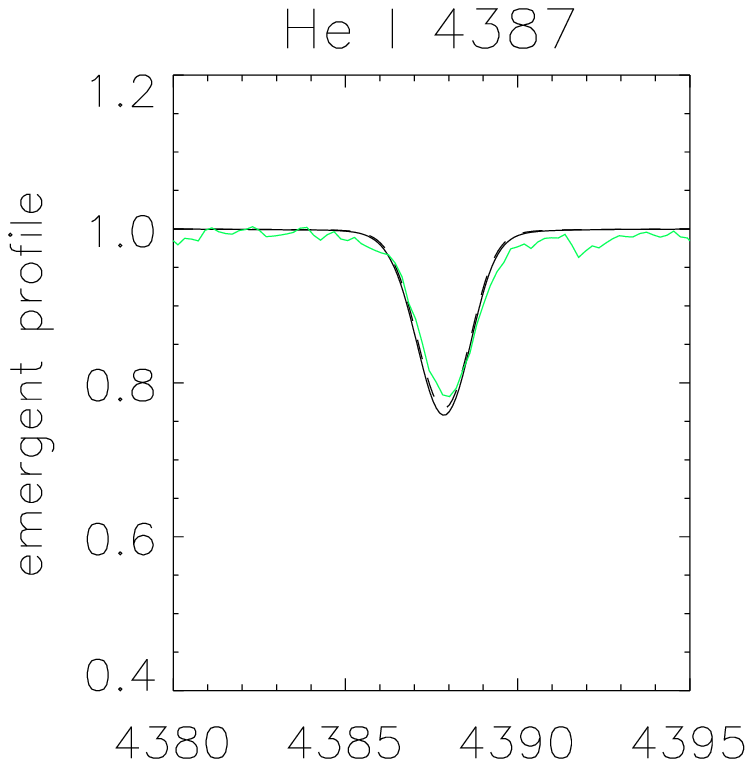}}
\end{minipage}
\hfill
\begin{minipage}{2.8cm}
\resizebox{\hsize}{!}
{\includegraphics{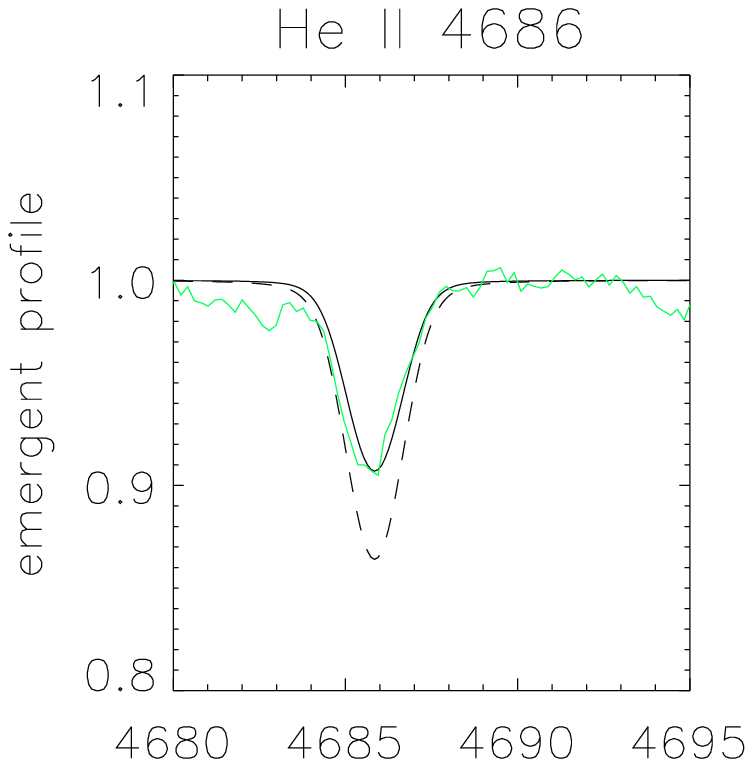}}
\end{minipage}
\\
\begin{minipage}{2.8cm}
\resizebox{\hsize}{!}
{\includegraphics{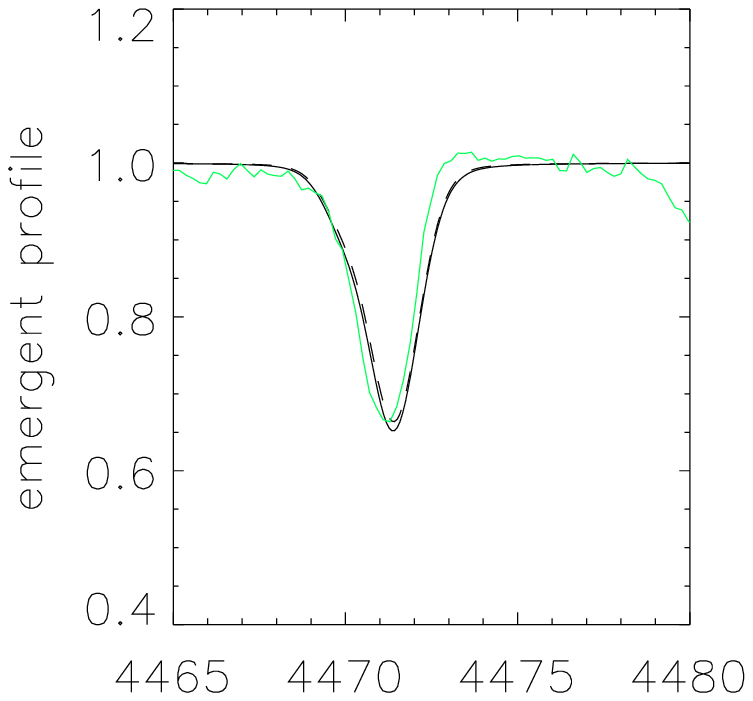}}
\end{minipage}
\hfill
\begin{minipage}{2.8cm}
\resizebox{\hsize}{!}
{\includegraphics{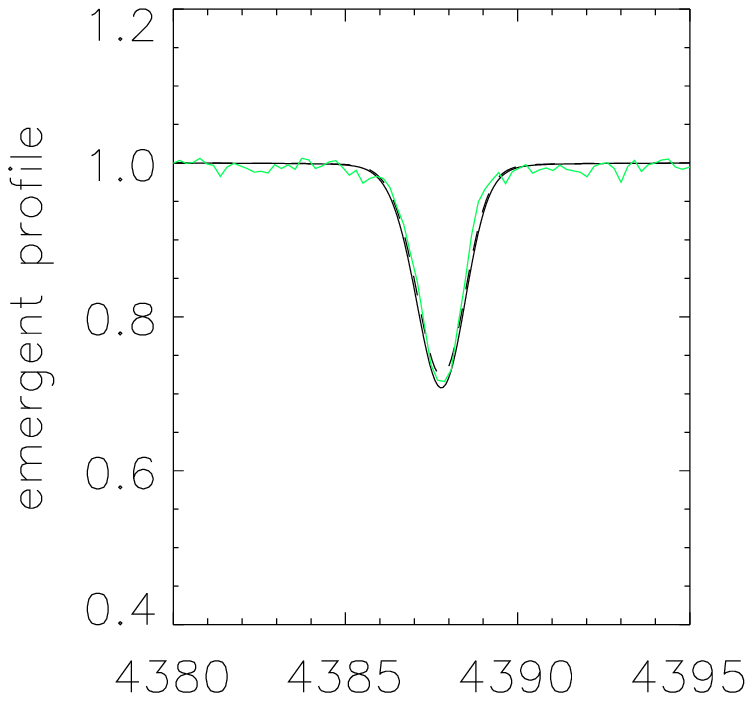}}
\end{minipage}
\hfill
\begin{minipage}{2.8cm}
\resizebox{\hsize}{!}
{\includegraphics{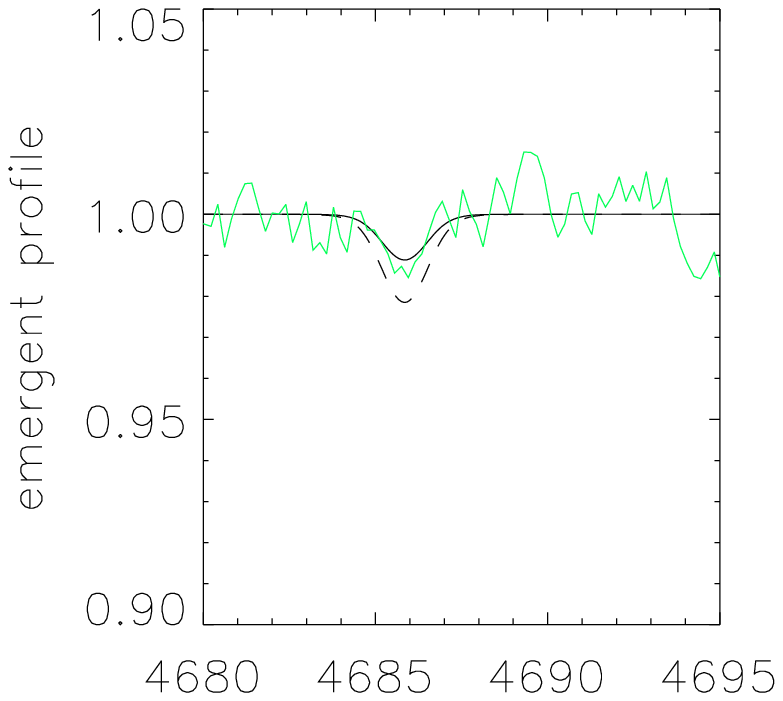}}
\end{minipage}
\\
\begin{minipage}{2.8cm}
\resizebox{\hsize}{!}
{\includegraphics{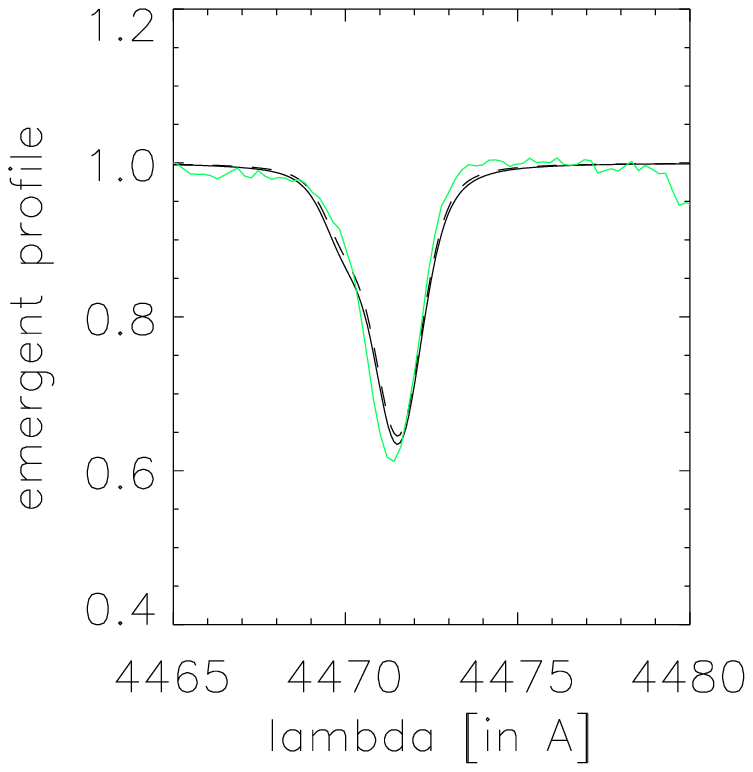}}
\end{minipage}
\hfill
\begin{minipage}{2.8cm}
\resizebox{\hsize}{!}
{\includegraphics{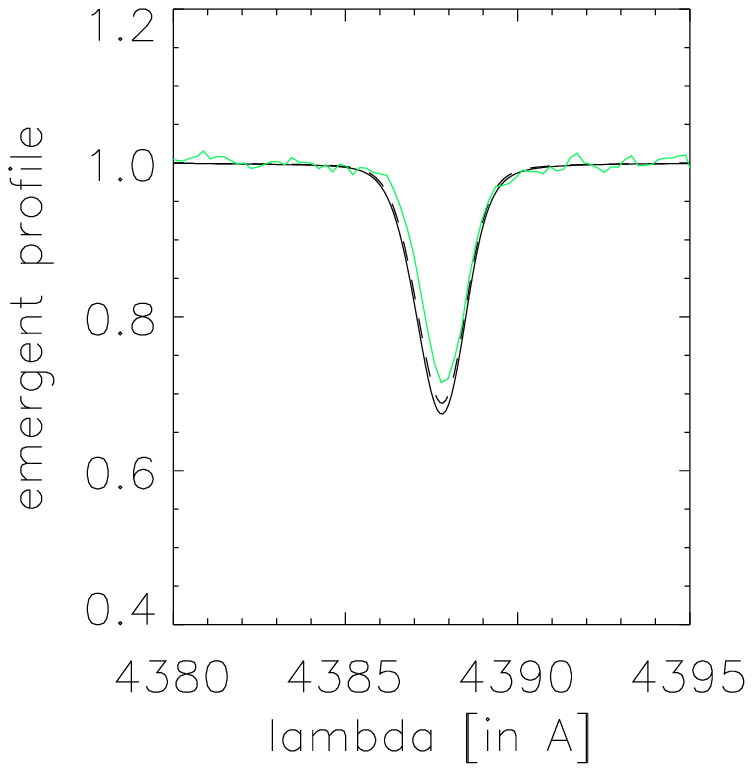}}
\end{minipage}
\hfill
\begin{minipage}{2.8cm}
\resizebox{\hsize}{!}
{\includegraphics{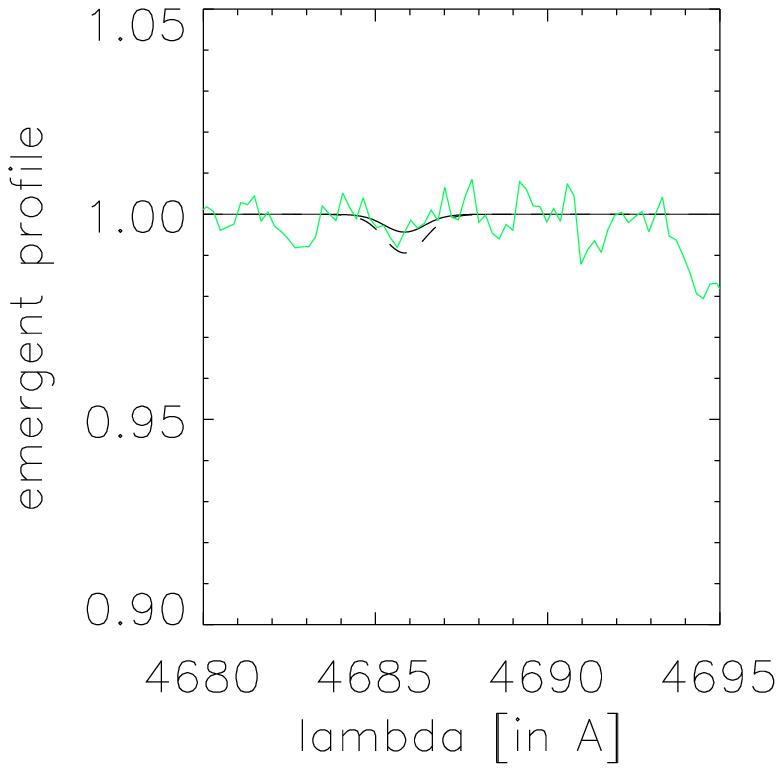}}
\end{minipage}
\caption{Observed and synthetic Helium line-profiles, calculated at the upper 
(\Teff  + $\Delta$\Teff: dashed) and  lower (\Teff - $\Delta$\Teff: solid) 
limit of the corresponding \Teff determined  from the ``best" fits to 
strategic Silicon lines. From top to bottom: HD~185\,859 (B0.5Ia), 
HD~190\,603 (B1.5Ia+) and  HD~206\,165 (B2Ib).}
\label{he_prof_1}
\end{figure}

\subsubsection{Helium line-profiles fits and  Helium abundance}
\label{He}

As pointed out in the beginning of this section, for early-B subtypes,
the Helium ionisation balance can be used to determine $T_{\rm eff}$. 
Consequently, for the three hottest stars in our sample we used 
He~I and He~II lines to derive independent constraints on $T_{\rm eff}$, 
assuming
helium abundances as discussed below.\footnote{Though He~II is 
rather weak at these temperatures, due to the good quality of our 
spectra its strongest features can be well resolved down to B2.}
Interestingly, in all these cases
satisfactory fits to the available strategic Helium lines could be obtained 
in parallel with Si~IV and Si~III (within the adopted uncertainties,
$\Delta$\Teff=$\pm$500K). This result is illustrated in
Fig.~\ref{he_prof_1}, where a comparison between observed and synthetic
Helium profiles is shown, the latter being calculated at the upper and lower
limit of the \Teff range derived from the Silicon ionisation balance. Our
finding contrasts \citet{Urb} who reported differences in the stellar
parameters beyond the typical uncertainties, if either Silicon or Helium was
used independently.  

Whereas no obvious discrepancy between He~I singlets and triplets (``He~I
singlet problem", \citealt{najarro06} and references therein) has been seen
in stars of type B1.5 and earlier, we faced several problems when trying to
fit Helium in parallel with Silicon in stars of mid and late subtypes (B2
and later).\footnote{At these temperatures, only He~I is present, and no
conclusions can be drawn from the ionisation {\it balance}.} 

\begin{figure*}
\begin{minipage}{3.5cm} 
\resizebox{\hsize}{!}
{\includegraphics{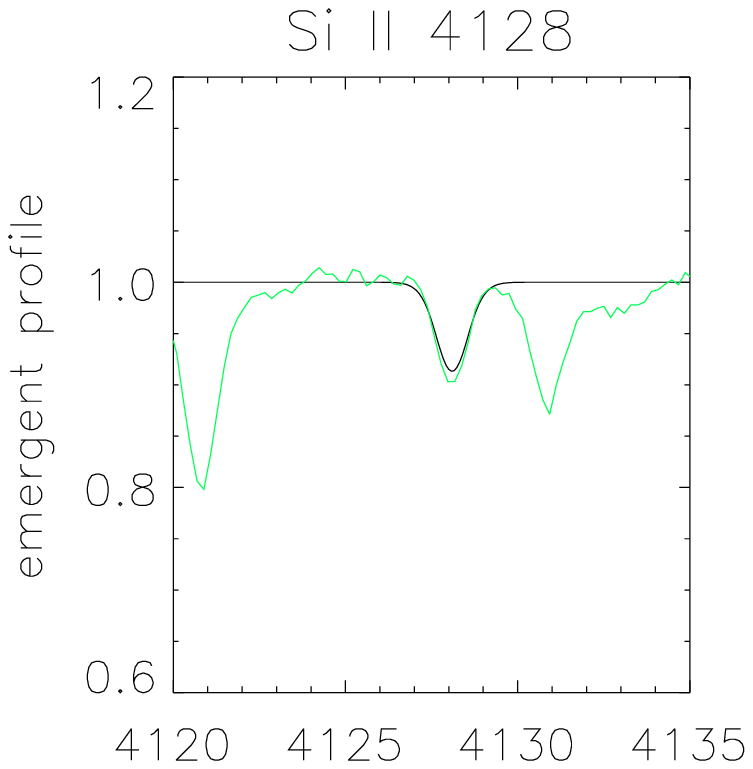}}
\end{minipage}
\hfill
\begin{minipage}{3.5cm} 
\resizebox{\hsize}{!}
{\includegraphics{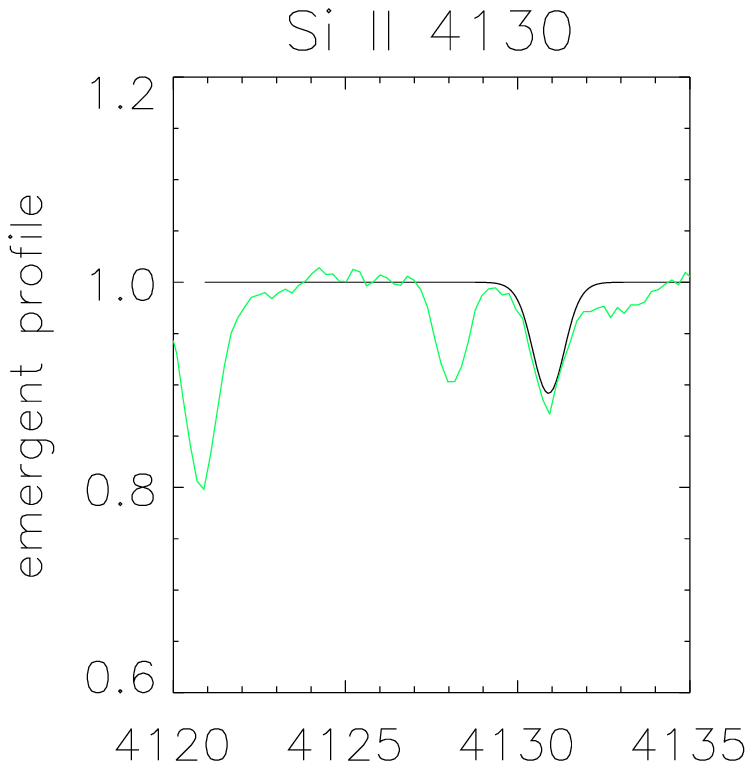}}
\end{minipage}
\hfill
\begin{minipage}{3.5cm} 
\resizebox{\hsize}{!}
{\includegraphics{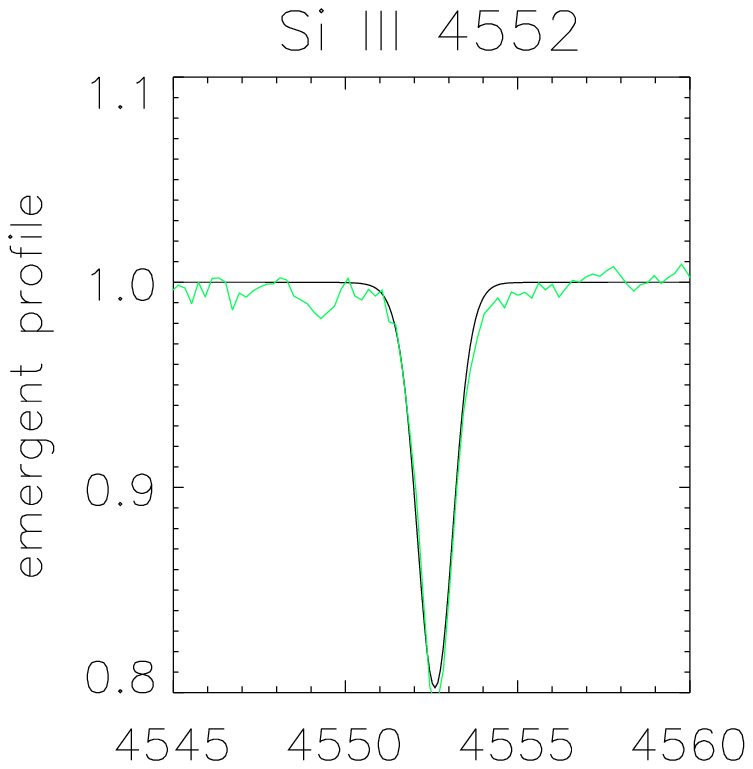}}
\end{minipage}
\hfill
\begin{minipage}{3.5cm}
\resizebox{\hsize}{!}
{\includegraphics{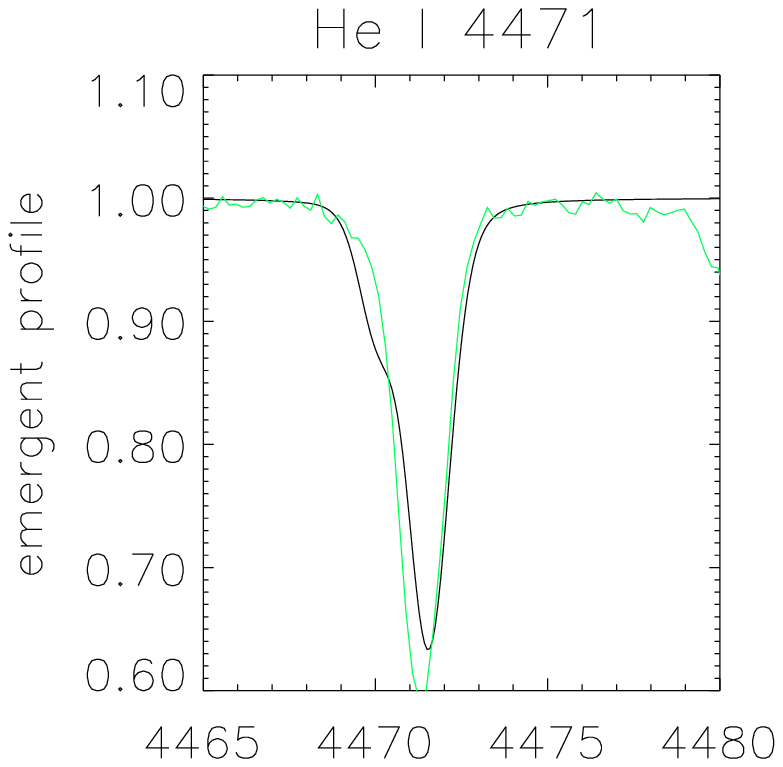}}
\end{minipage}
\hfill
\begin{minipage}{3.5cm}
\resizebox{\hsize}{!}
{\includegraphics{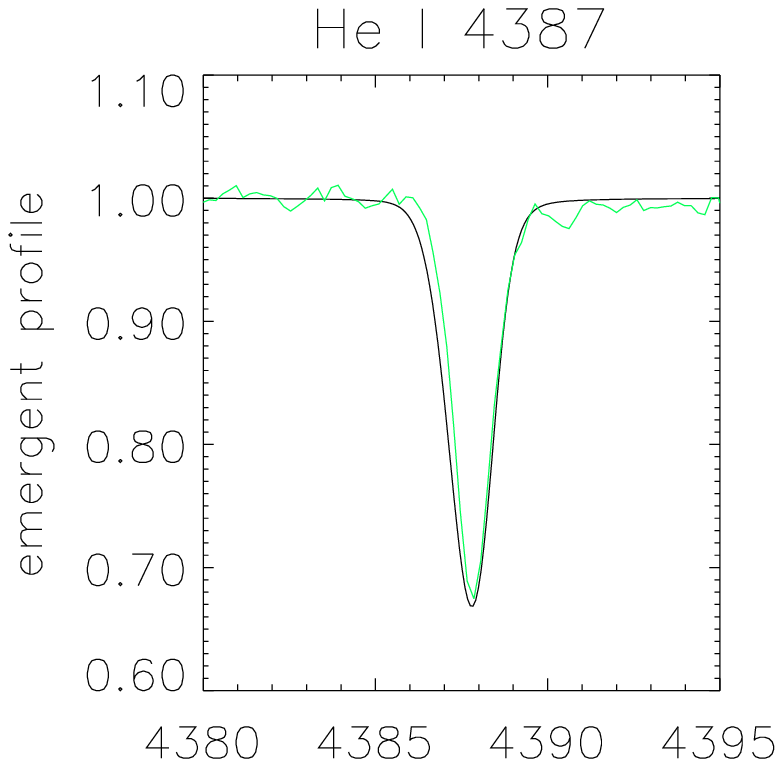}}
\end{minipage}
\begin{minipage}{3.5cm}
\resizebox{\hsize}{!}
{\includegraphics{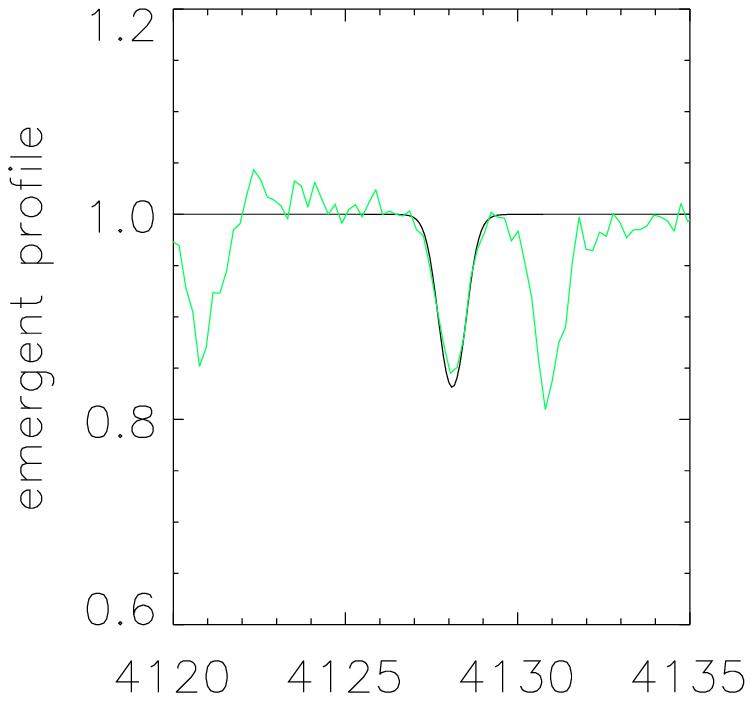}}
\end{minipage}
\hfill
\begin{minipage}{3.5cm}
\resizebox{\hsize}{!}
{\includegraphics{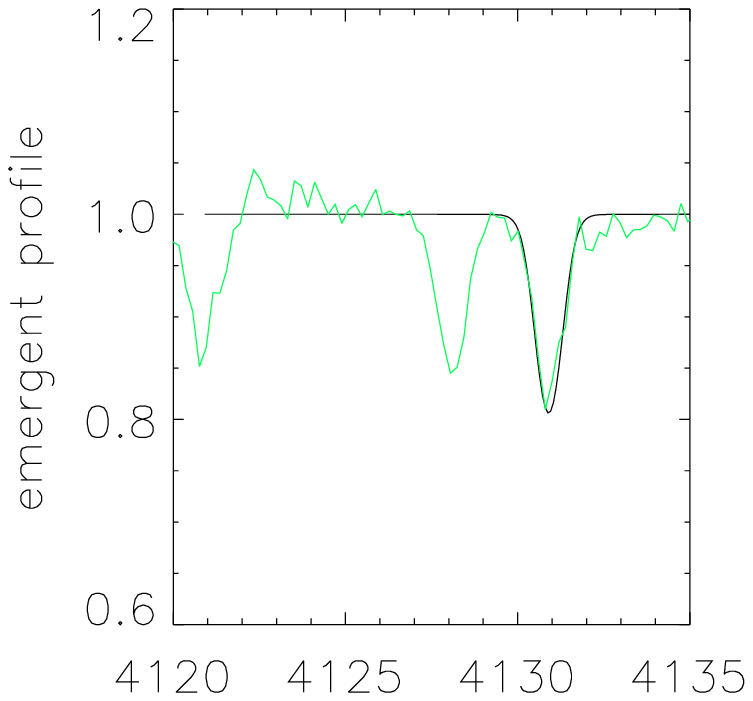}}
\end{minipage}
\hfill
\begin{minipage}{3.5cm}
\resizebox{\hsize}{!}
{\includegraphics{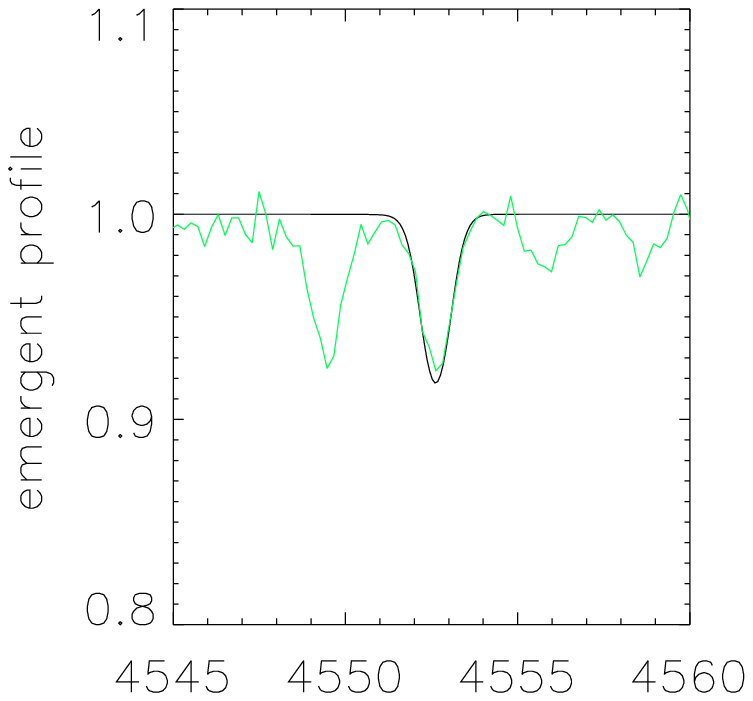}}
\end{minipage}
\hfill
\begin{minipage}{3.5cm}
\resizebox{\hsize}{!}
{\includegraphics{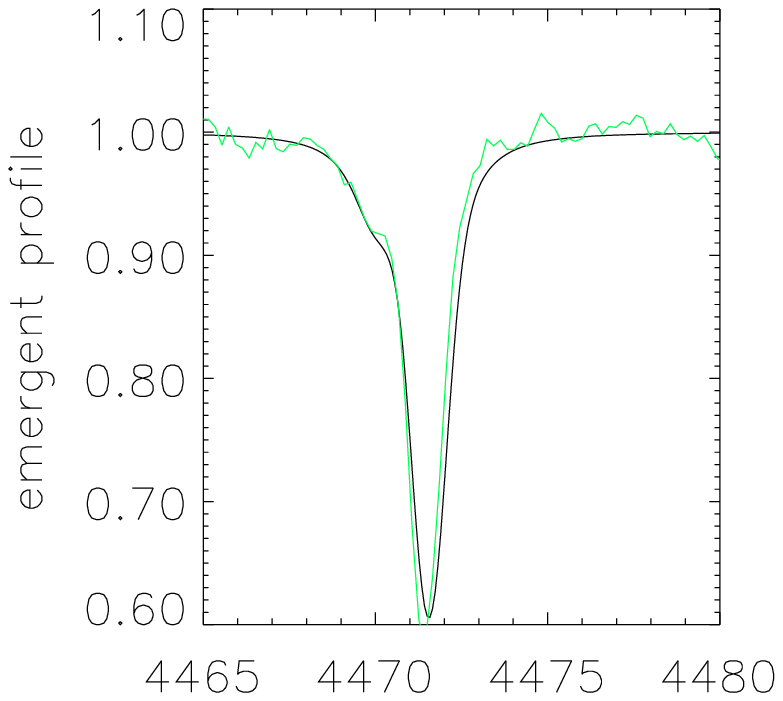}}
\end{minipage}
\hfill
\begin{minipage}{3.5cm}
\resizebox{\hsize}{!}
{\includegraphics{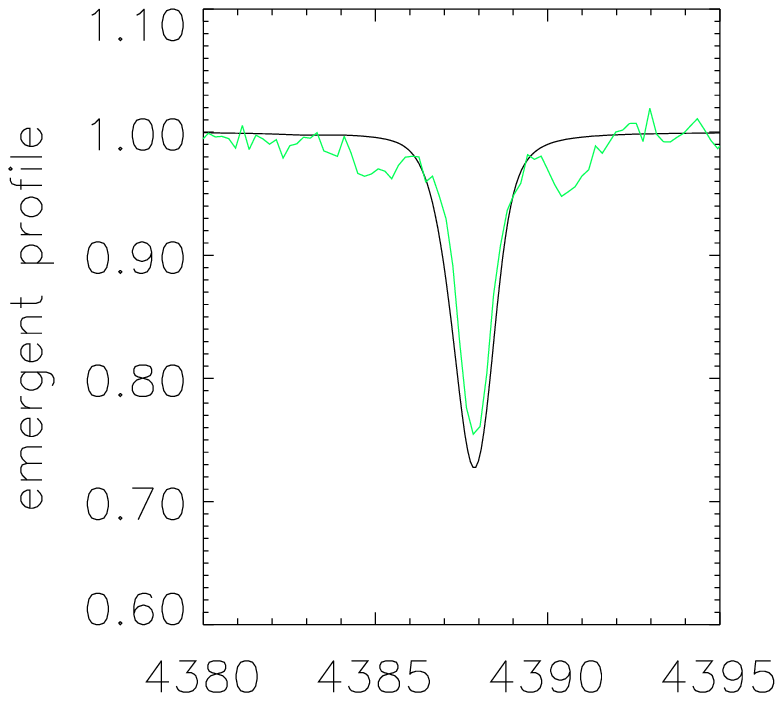}}
\end{minipage}
\\
\begin{minipage}{3.5cm}
\resizebox{\hsize}{!}
{\includegraphics{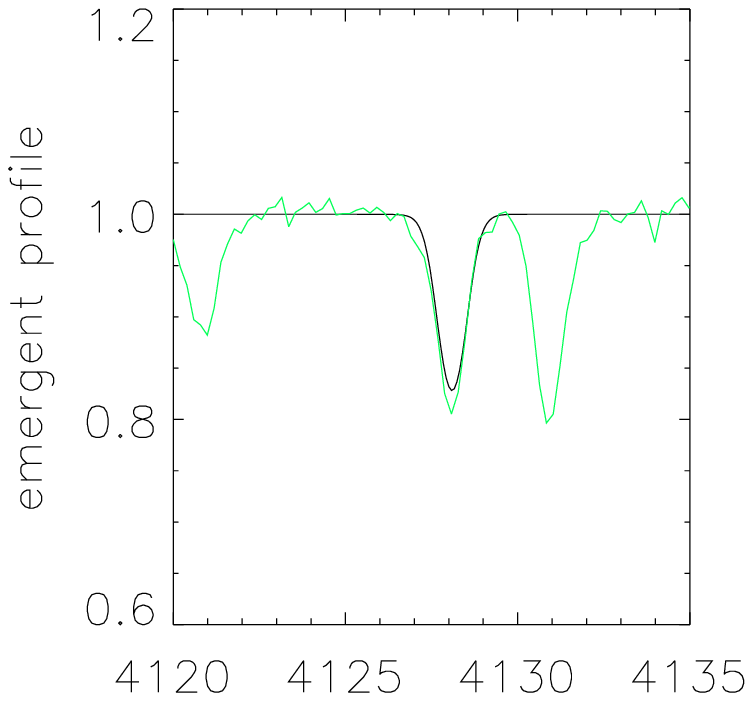}}
\end{minipage}
\hfill
\begin{minipage}{3.5cm}
\resizebox{\hsize}{!}
{\includegraphics{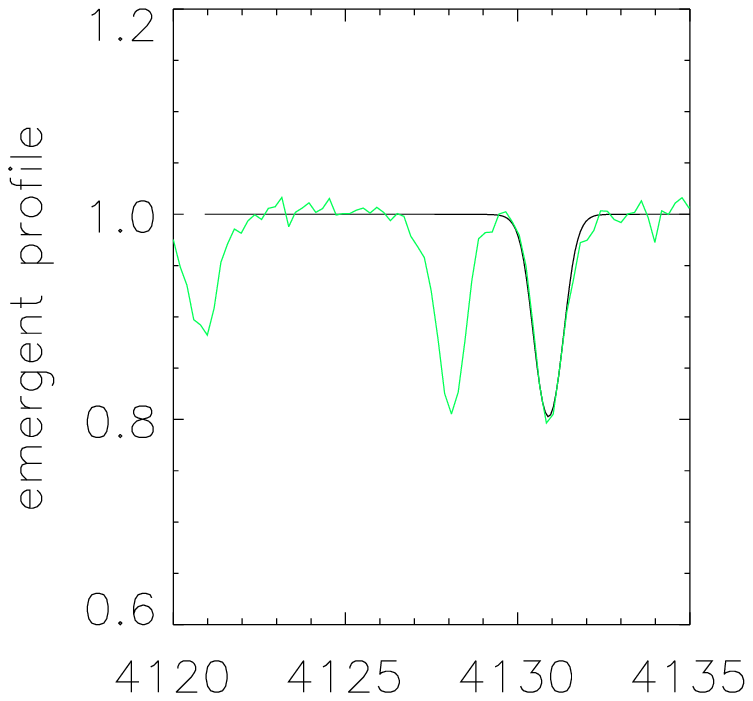}}
\end{minipage}
\hfill
\begin{minipage}{3.5cm}
\resizebox{\hsize}{!}
{\includegraphics{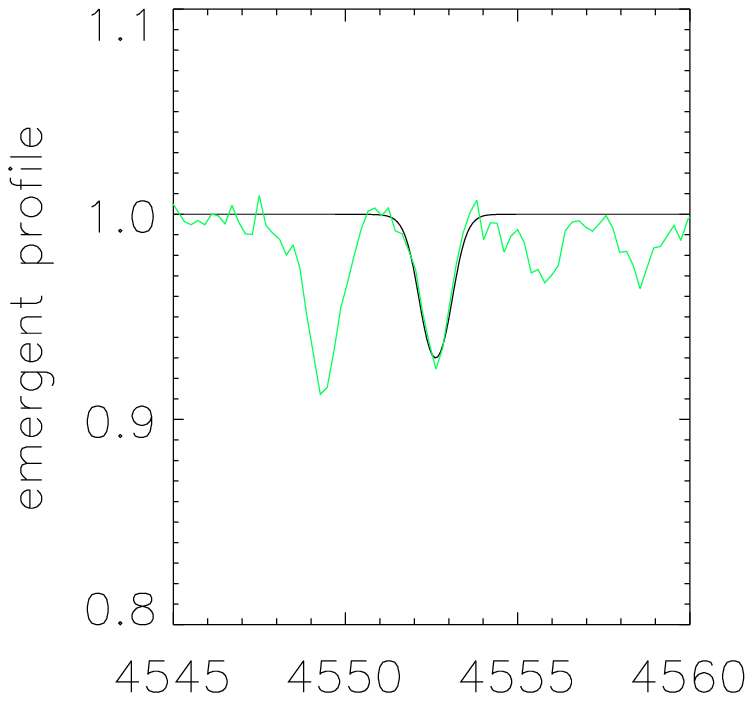}}
\end{minipage}
\hfill
\begin{minipage}{3.5cm}
\resizebox{\hsize}{!}
{\includegraphics{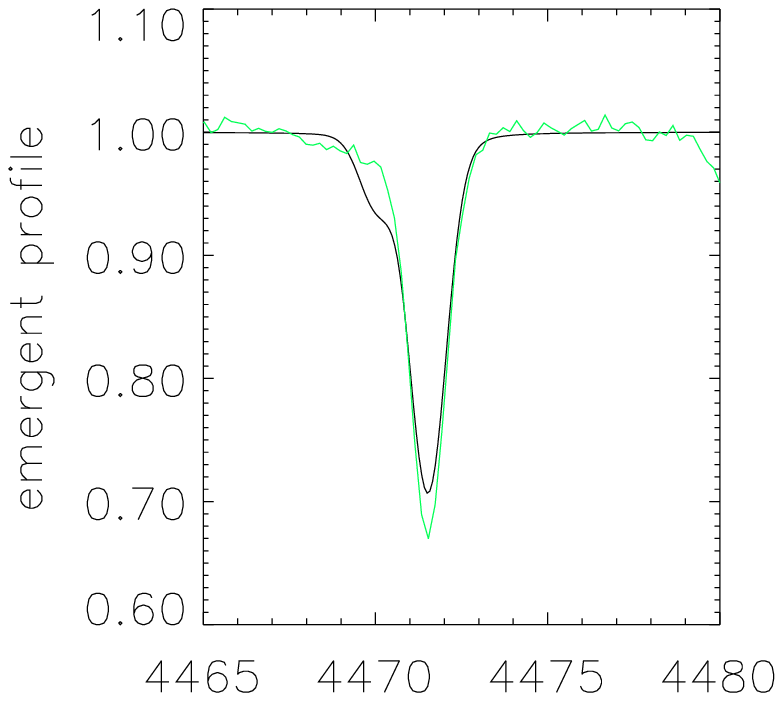}}
\end{minipage}
\hfill
\begin{minipage}{3.5cm}
\resizebox{\hsize}{!}
{\includegraphics{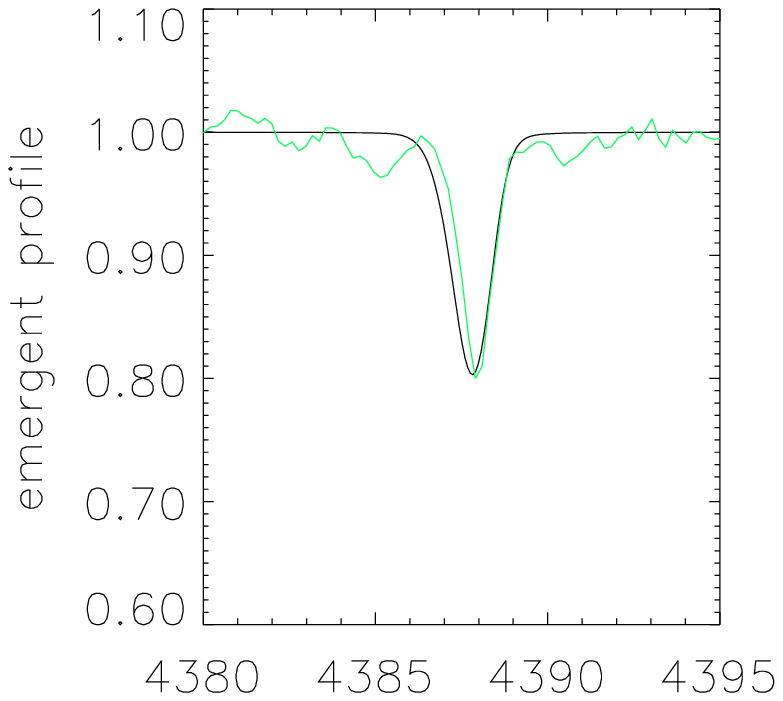}}
\end{minipage}
\\
\begin{minipage}{3.5cm}
\resizebox{\hsize}{!}
{\includegraphics{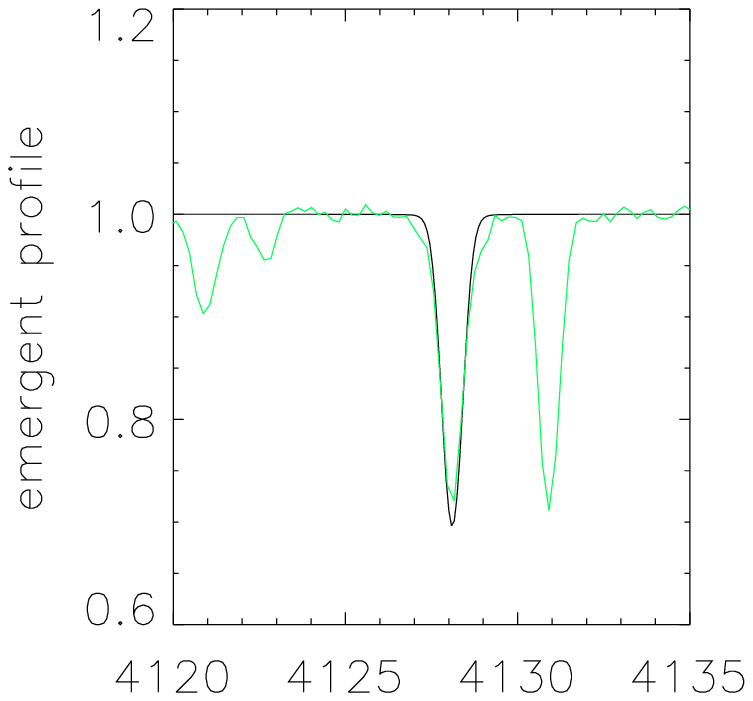}}
\end{minipage}
\hfill
\begin{minipage}{3.5cm}
\resizebox{\hsize}{!}
{\includegraphics{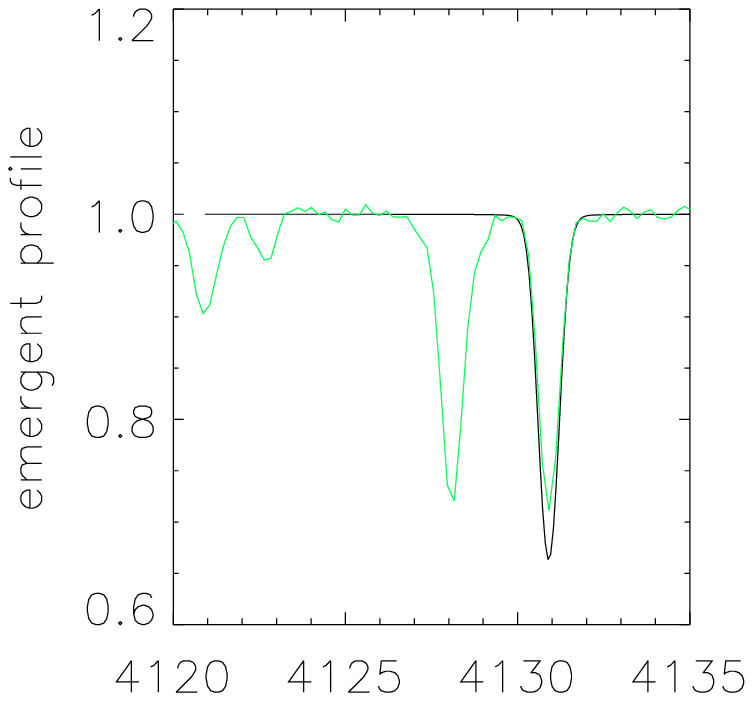}}
\end{minipage}
\hfill
\begin{minipage}{3.5cm}
\resizebox{\hsize}{!}
{\includegraphics{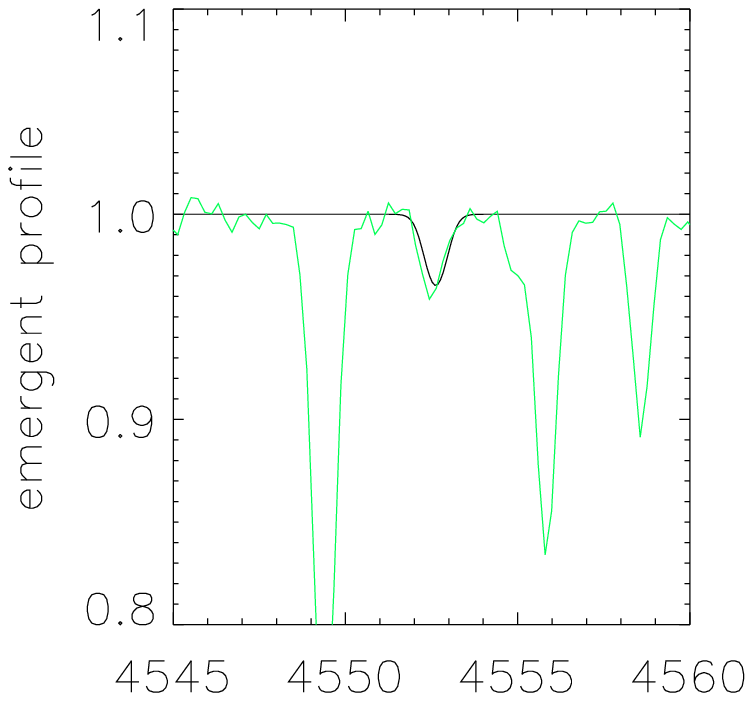}}
\end{minipage}
\hfill
\begin{minipage}{3.5cm}
\resizebox{\hsize}{!}
{\includegraphics{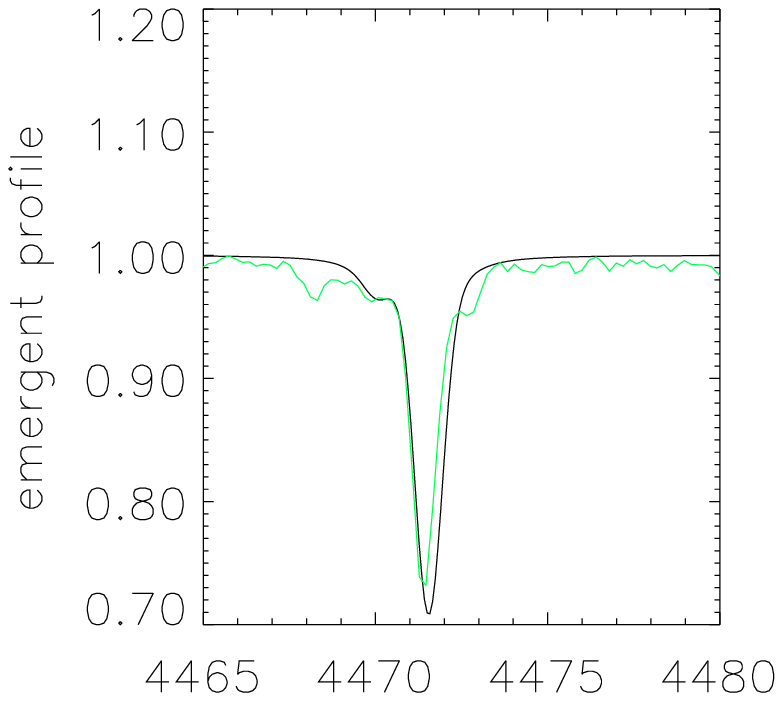}}
\end{minipage}
\hfill
\begin{minipage}{3.5cm}
\resizebox{\hsize}{!}
{\includegraphics{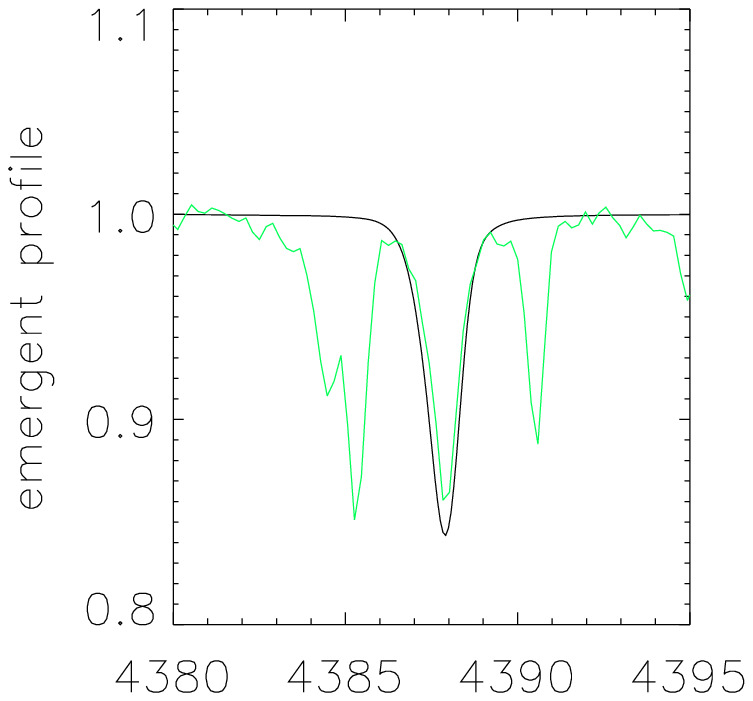}}
\end{minipage}
\\
\begin{minipage}{3.5cm}
\resizebox{\hsize}{!}
{\includegraphics{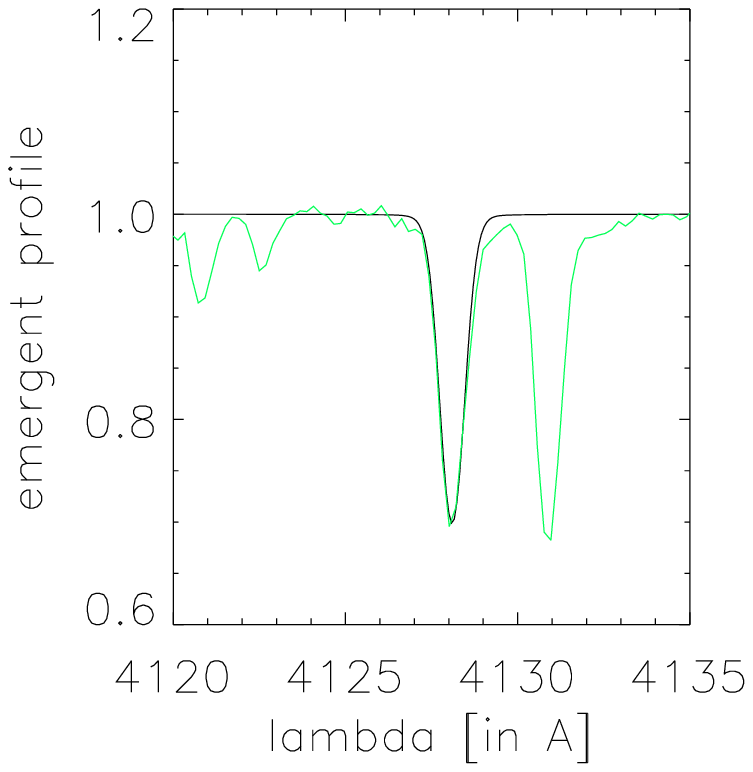}}
\end{minipage}
\hfill
\begin{minipage}{3.5cm}
\resizebox{\hsize}{!}
{\includegraphics{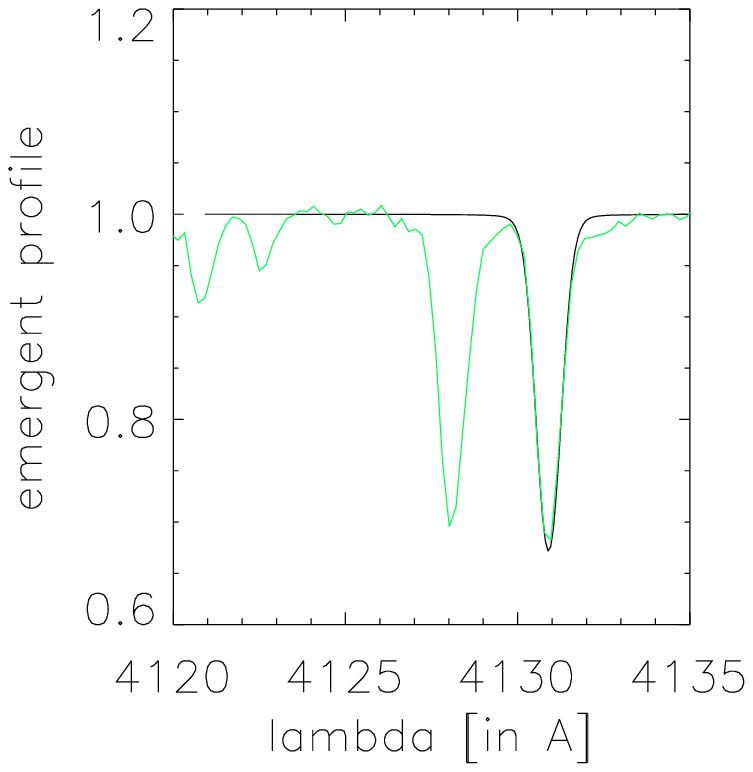}}
\end{minipage}
\hfill
\begin{minipage}{3.5cm}
\resizebox{\hsize}{!}
{\includegraphics{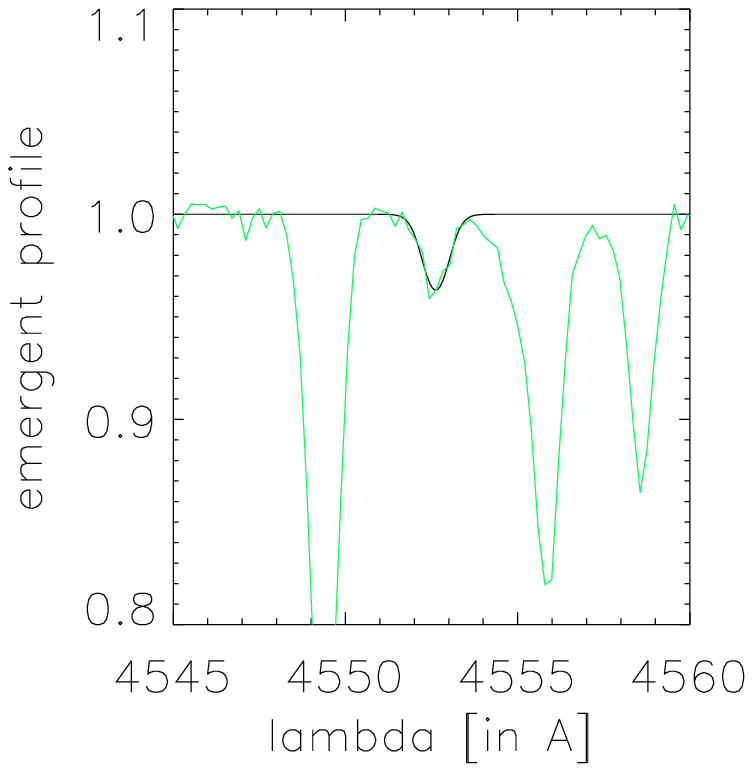}}
\end{minipage}
\hfill
\begin{minipage}{3.5cm}
\resizebox{\hsize}{!}
{\includegraphics{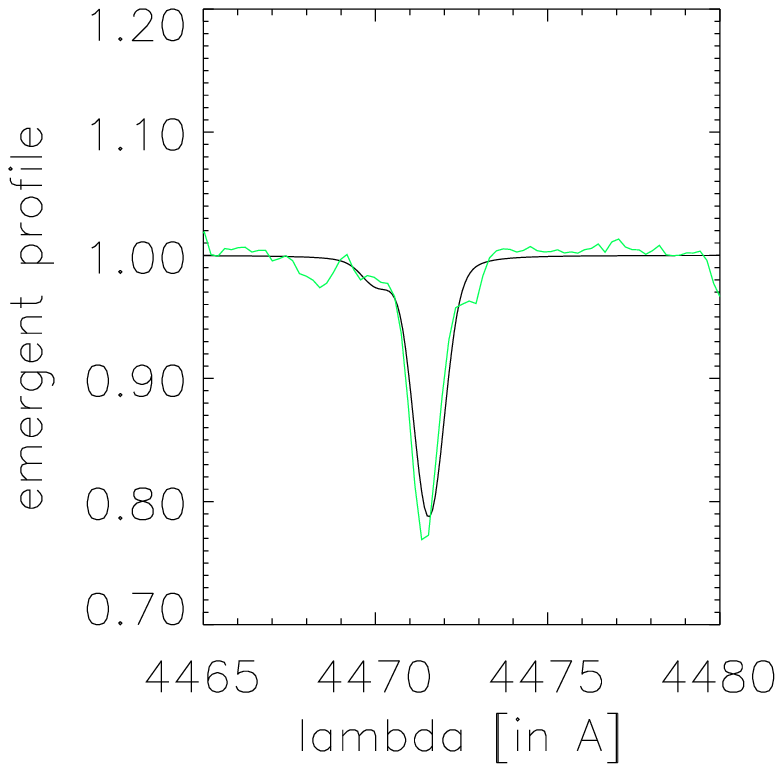}}
\end{minipage}
\hfill
\begin{minipage}{3.5cm}
\resizebox{\hsize}{!}
{\includegraphics{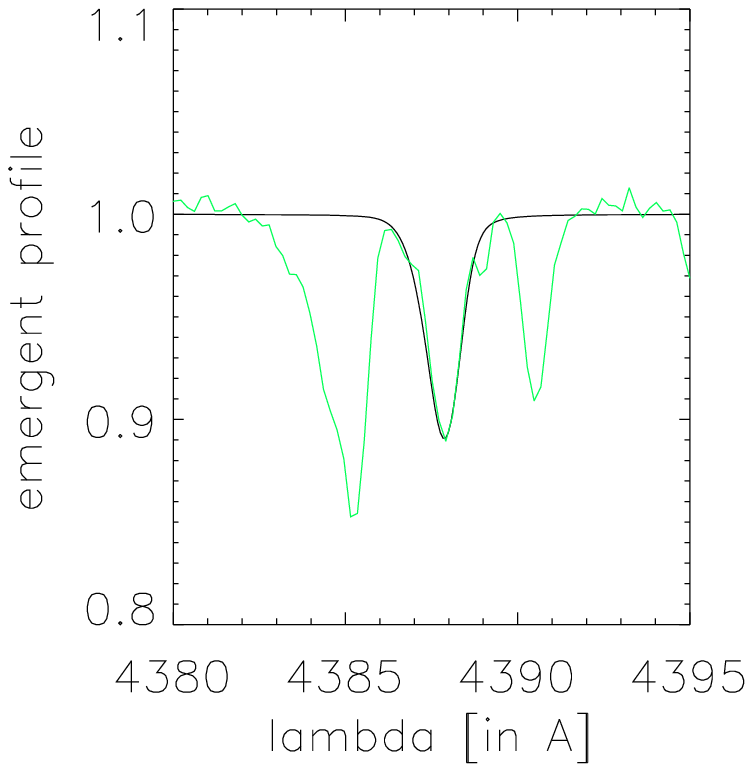}}
\end{minipage}
\caption{Final (``best'') fits to Silicon and Helium lines. From top to
bottom: HD~198\,478 (B2.5), HD~191\,243 (B5Ib), HD~199\,478 (B8ae),
HD~212\,593 (B9 Iab) and HD~202\,850 (B9 Iab). Note that the forbidden
component in the blue wing of He~I $\lambda$4471 is over-predicted in a
number of cases, and that the synthetic He~I singlet at $\lambda$4387 is
systematically too strong.}
\label{he_prof_2}
\end{figure*}

In particular, and at ``normal'' helium abundance (\Yhe=0.1$\pm$0.02), the
singlet line at $\lambda$4387 is somewhat over-predicted for all stars in
this subgroup, except for the coolest one - HD~202\,850. At the same time,
the triplet transitions at $\lambda\lambda$4471 and 4713 have been
under-predicted (HD~206\,165, B2 and HD~198\,478, B2.5), well reproduced 
(HD~191\,243, B5 and HD~199\,478, B8) or over-predicted (HD~212\,593, B9).
Additionally, in half of these stars (HD~206\,165, HD~198\,478, and
HD~199\,478) the strength of the forbidden component of He~I~4471 was
over-predicted, whereas in the other half this component was well
reproduced. In all cases, however, these discrepancies were not so large as
to prevent a globally satisfactory fit to the available He~I lines in
parallel with Silicon. Examples illustrating these facts are shown in
Fig.~\ref{he_prof_2}. Again, there are at least two principle
possibilities to solve these problems: to adapt the He abundance or/and to
use different values of \vmic, on the assumption that this parameter varies
as a function of atmospheric height (cf. Sect.~\ref{siIII}).

\paragraph{Helium abundance.} For all sample stars a ``normal'' helium
abundance, \Yhe = 0.10, was adopted as a first guess. Subsequently, this
value has been adjusted (if necessary) to improve the Helium line fits.  For
the two hottest stars with well reproduced Helium lines (and two ionisation
stages being present!), an error of only $\pm$0.02 seems to be appropriate
because of the excellent fit quality. Among those, an overabundance in
Helium (\Yhe = 0.2) was found for the hypergiant HD~190\,603, which might 
also be expected according to its evolutionary stage.

In mid and late B-type stars, on the other hand, the determination of \Yhe
was more complicated, due to problems discussed above.
Particularly for stars where the discrepancies between synthetic and
observed triplet and singlet lines were opposite to each other, no
unique solution could be obtained by varying the Helium abundance,
and we had to increase the corresponding error bars 
(HD~206\,165 and HD~198\,478).
For HD~212\,593, on the other hand (where all available singlet and triplet
lines turned out to be over-predicted), a Helium depletion by 30 to 40\%
would be required to reconcile theory with observations. 

All derived values are summarised in Column~8 of Table~\ref{para_2}, but note
that alternative fits of similar quality are possible for those
cases where an overabundance/depletion in He has been indicated, namely by
using a solar Helium content and \vmic\ being a factor of two larger/lower
than inferred from Silicon: Due to the well known dichotomy between
abundance and micro-turbulence (if only one ionisation stage is present), a
unique solution is simply not possible, accounting for the capacity of the
diagnostic tools used here.

\medskip Column~4 of Table~\ref{para_2} lists all effective temperatures as
derived in the present study. As we have seen, these estimates are
influenced by several processes and estimates of other quantities, among
which are micro- and macro-turbulence, He and Si abundances, surface gravity
and mass-loss rate (where the latter two quantities are discussed in the
following).  Nevertheless, we are quite confident that, to a large extent,
we have consistently and partly independently (regarding \vmic, \vmac and Si
abundances) accounted for these influences. Thus, the errors in our \Teff
estimates should be dominated by uncertainties in the fitting procedure,
amounting to about $\pm$500 K. Of course, these are differential errors
assuming that physics complies with all our assumptions, data and
approximations used within our atmosphere code. 
\begin{figure}
\begin{minipage}{4.2cm}
\resizebox{\hsize}{!}
{\includegraphics{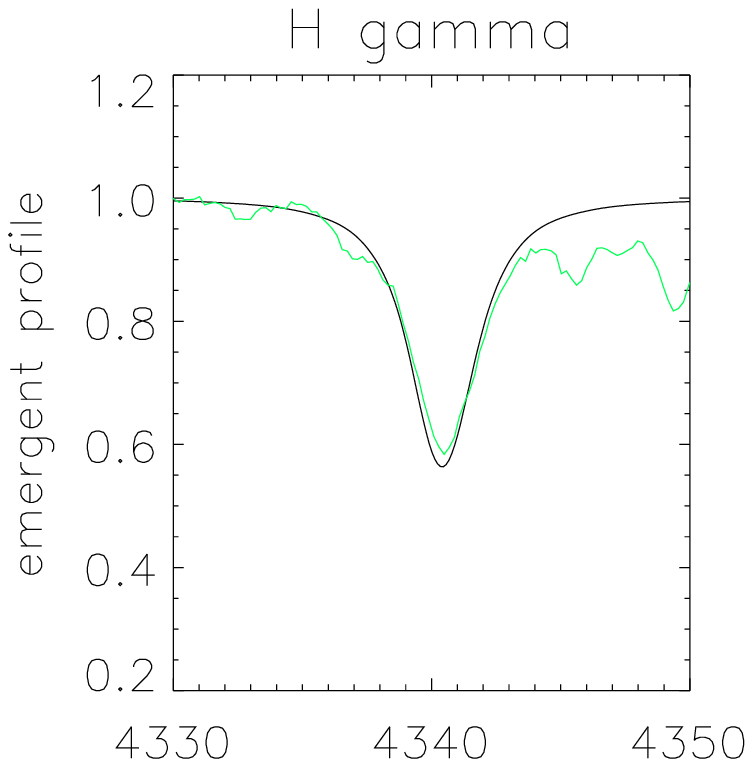}}
\end{minipage}
\hfill
\begin{minipage}{4.2cm}
\resizebox{\hsize}{!}
{\includegraphics{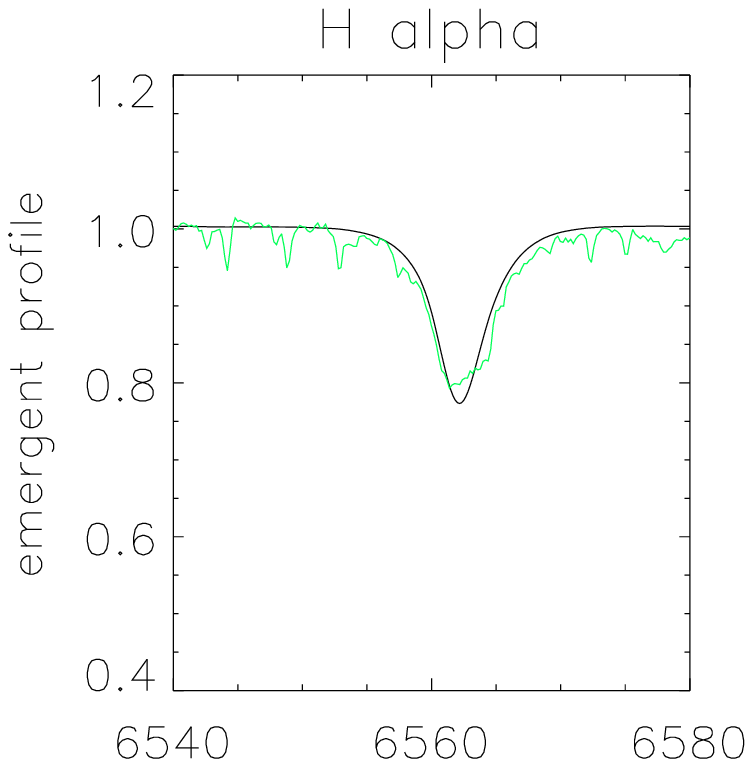}}
\end{minipage}
\\
\begin{minipage}{4.2cm}
\resizebox{\hsize}{!}
{\includegraphics{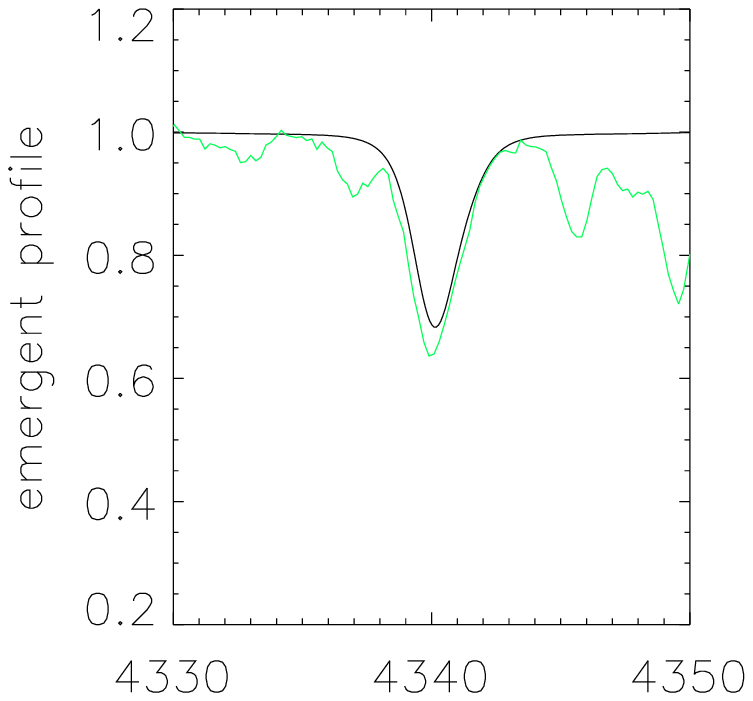}}
\end{minipage}
\hfill
\begin{minipage}{4.2cm}
\resizebox{\hsize}{!}
{\includegraphics{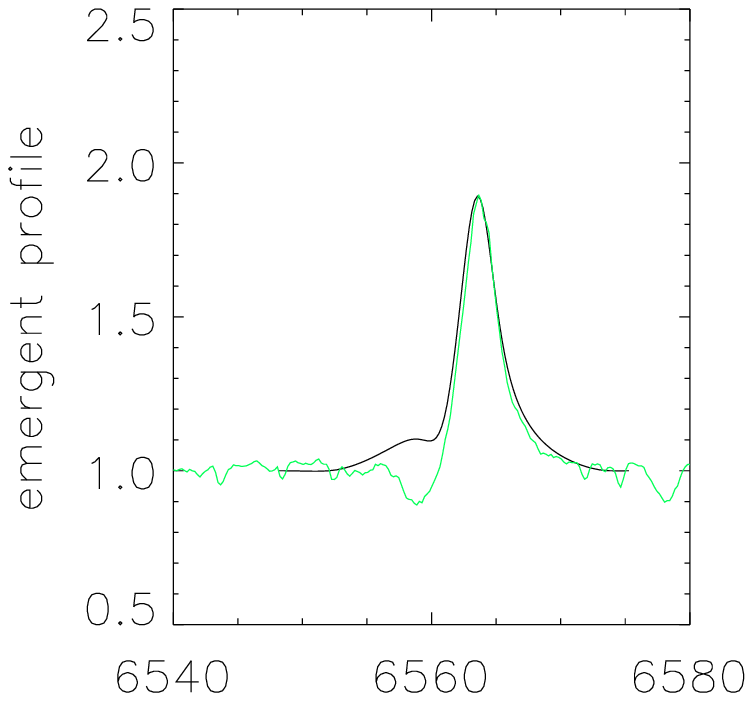}}
\end{minipage}
\\
\begin{minipage}{4.2cm}
\resizebox{\hsize}{!}
{\includegraphics{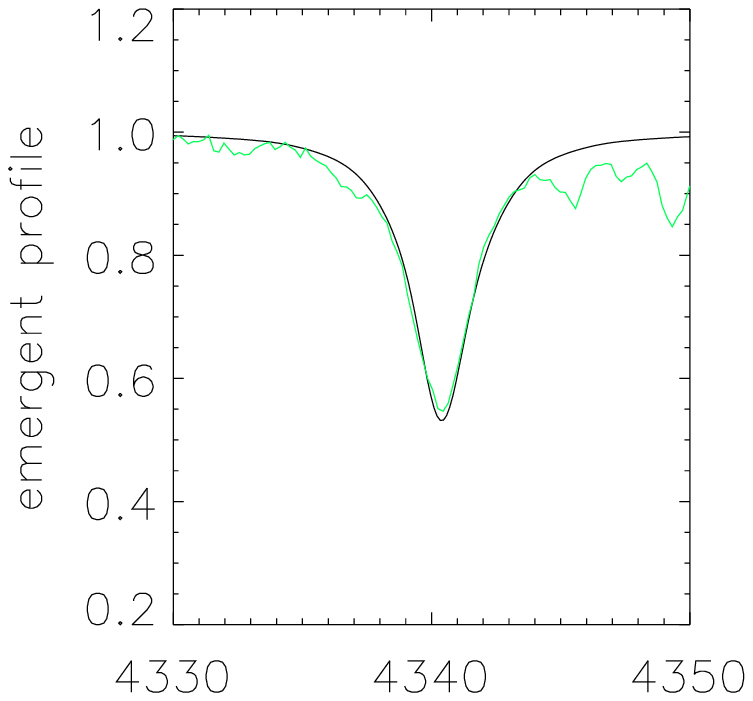}}
\end{minipage}
\hfill
\begin{minipage}{4.2cm}
\resizebox{\hsize}{!}
{\includegraphics{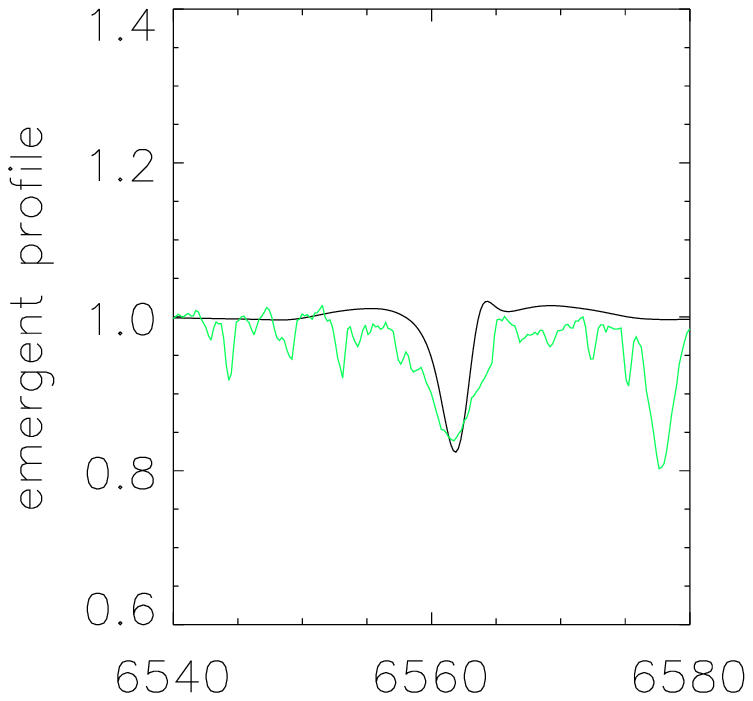}}
\end{minipage}
\\\
\begin{minipage}{4.2cm}
\resizebox{\hsize}{!}
{\includegraphics{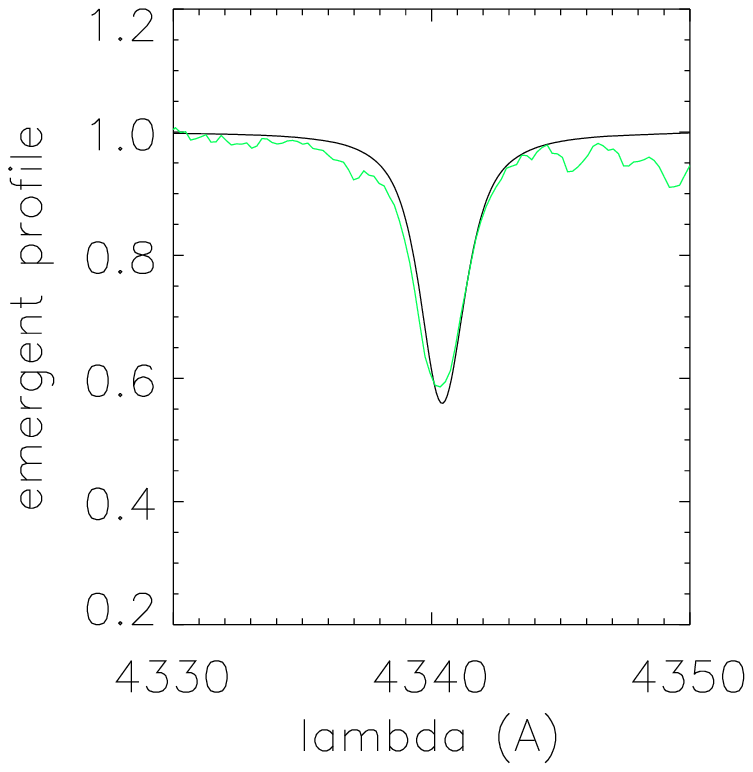}}
\end{minipage}
\hfill
\begin{minipage}{4.2cm}
\resizebox{\hsize}{!}
{\includegraphics{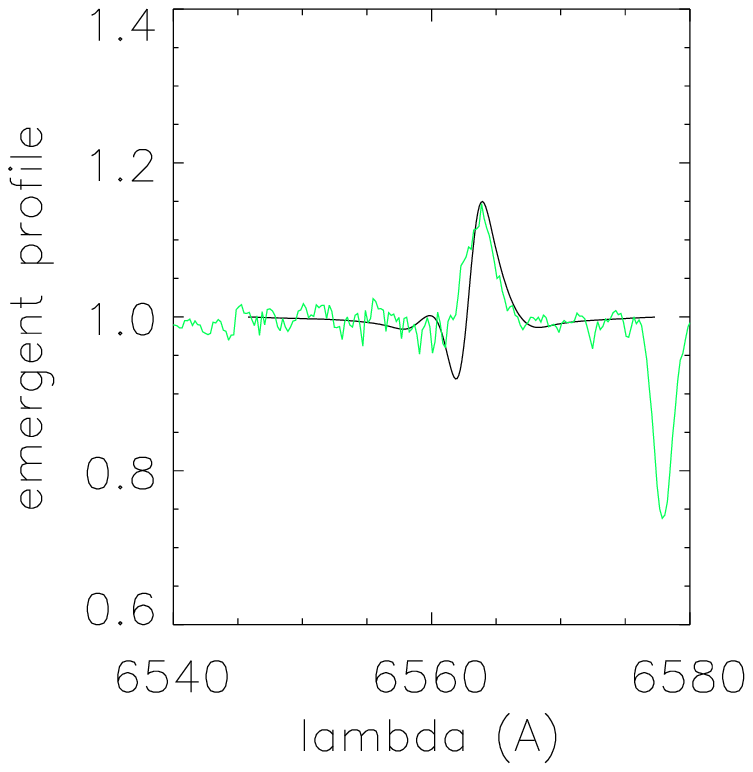}}
\end{minipage}
\caption{Fit quality for \Hgama  and H$_\alpha$,  for early 
and mid subtypes. From top to bottom: HD~185\,859 (B0.5Ia), HD~190\,603 
(B1.5Ia+), HD~206\,165 (B2Ib) and HD~198\,478 (B2.5).}
\label{Hg_Ha_1}
\end{figure}
\begin{figure}
\begin{minipage}{4.2cm}
\resizebox{\hsize}{!}
{\includegraphics{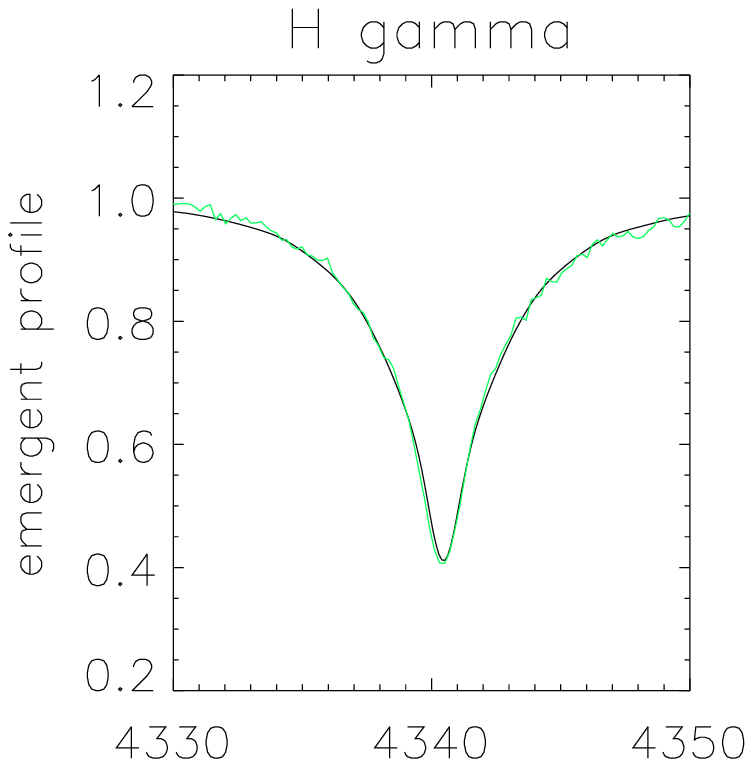}}
\end{minipage}
\hfill
\begin{minipage}{4.2cm}
\resizebox{\hsize}{!}
{\includegraphics{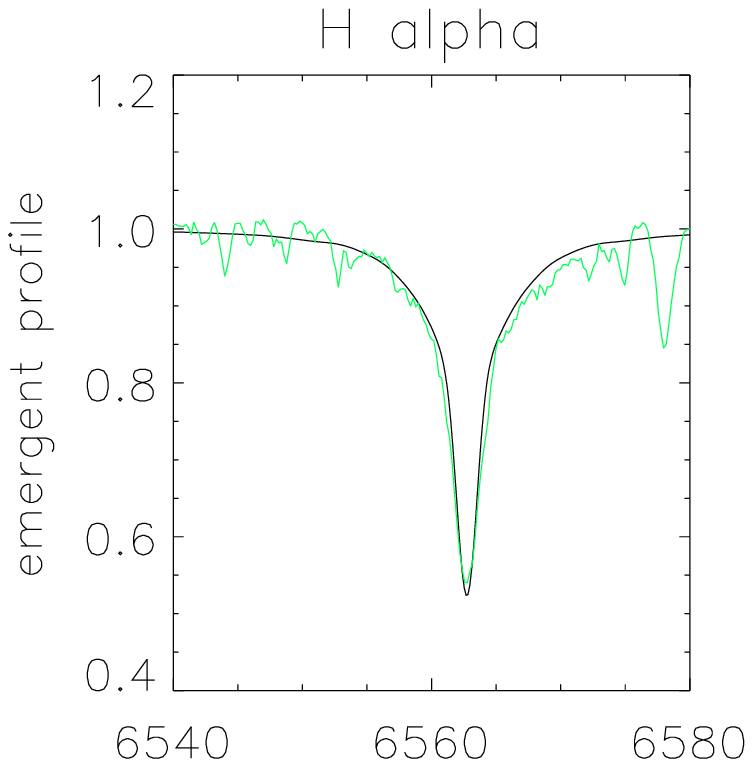}}
\end{minipage}
\\
\begin{minipage}{4.2cm}
\resizebox{\hsize}{!}
{\includegraphics{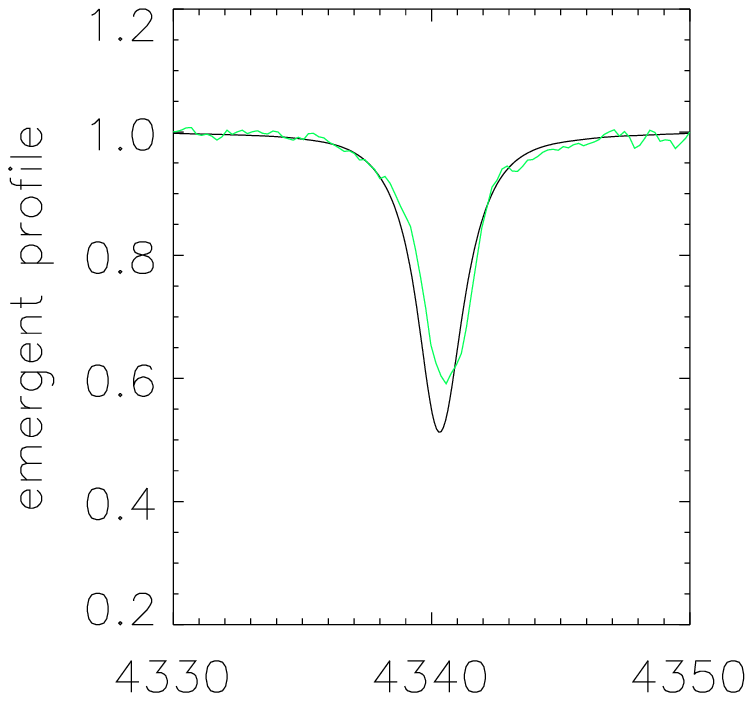}}
\end{minipage}
\hfill
\begin{minipage}{4.2cm}
\resizebox{\hsize}{!}
{\includegraphics{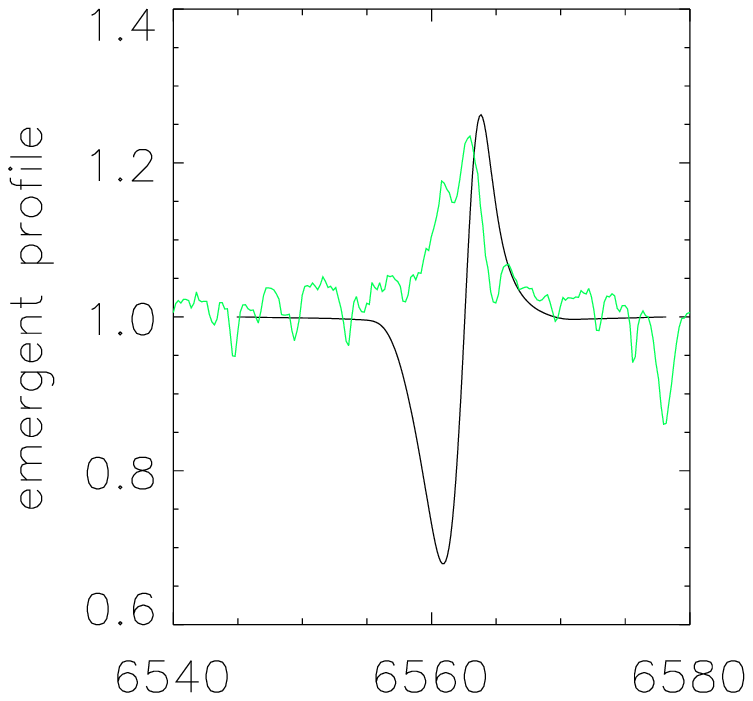}}
\end{minipage}
\\
\begin{minipage}{4.2cm}
\resizebox{\hsize}{!}
{\includegraphics{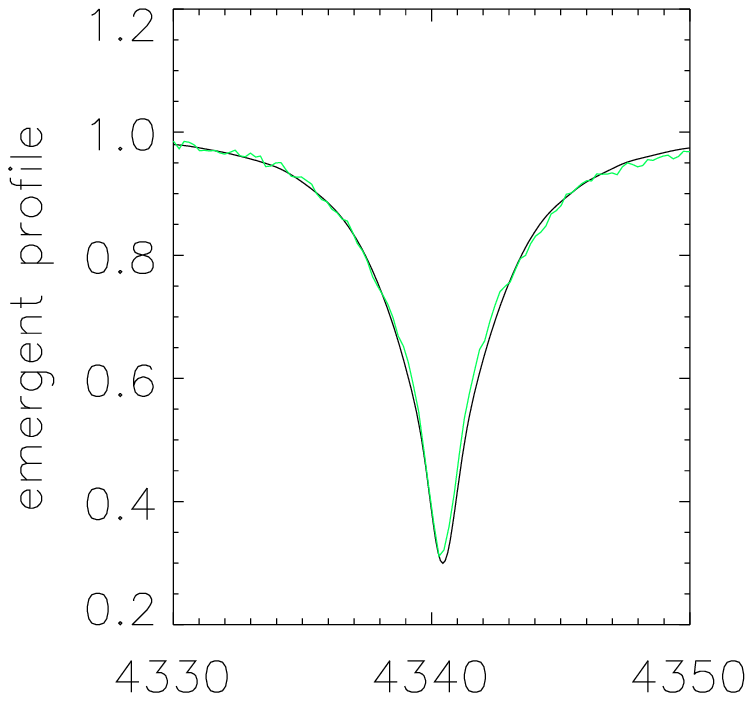}}
\end{minipage}
\hfill
\begin{minipage}{4.2cm}
\resizebox{\hsize}{!}
{\includegraphics{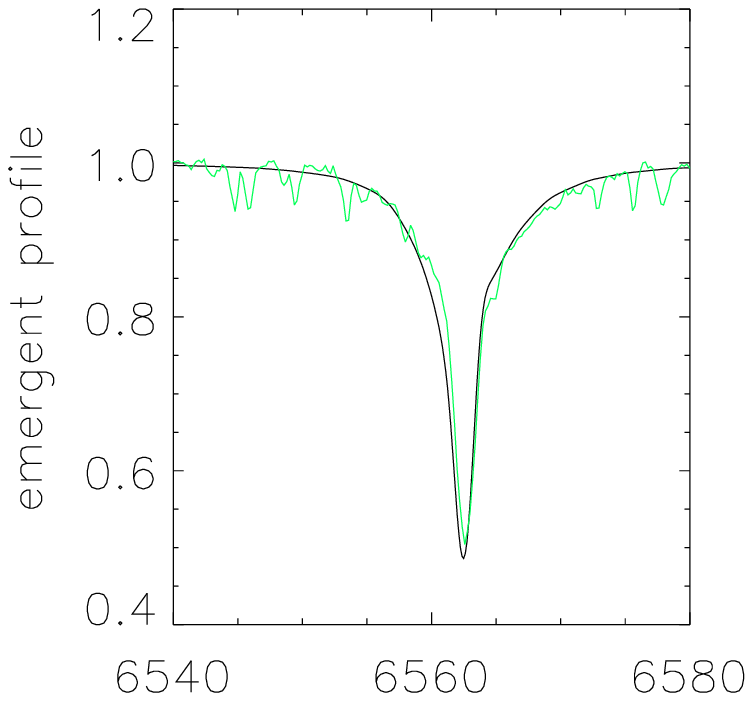}}
\end{minipage}
\\
\begin{minipage}{4.2cm}
\resizebox{\hsize}{!}
{\includegraphics{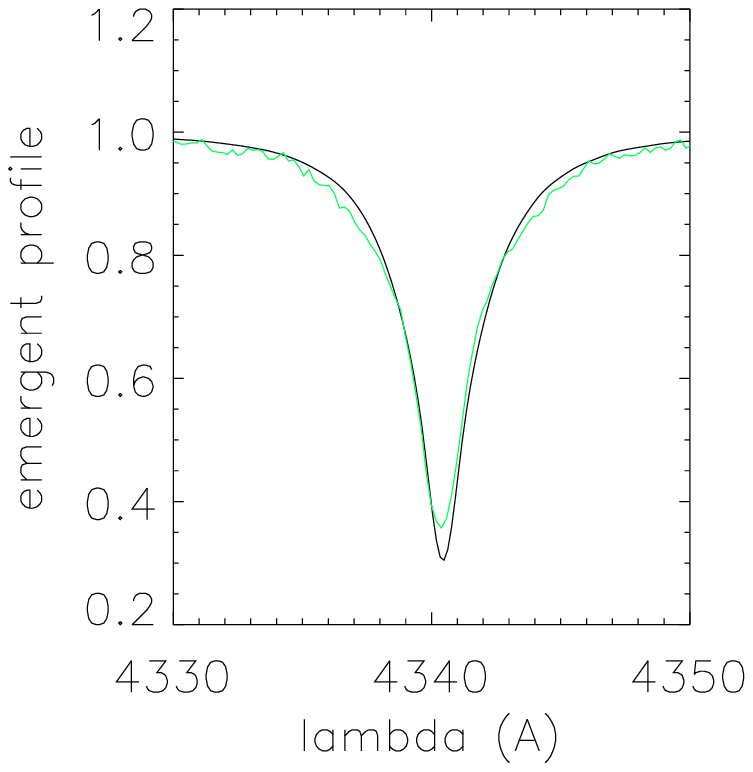}}
\end{minipage}
\hfill
\begin{minipage}{3.5cm}
\resizebox{\hsize}{!}
{\includegraphics{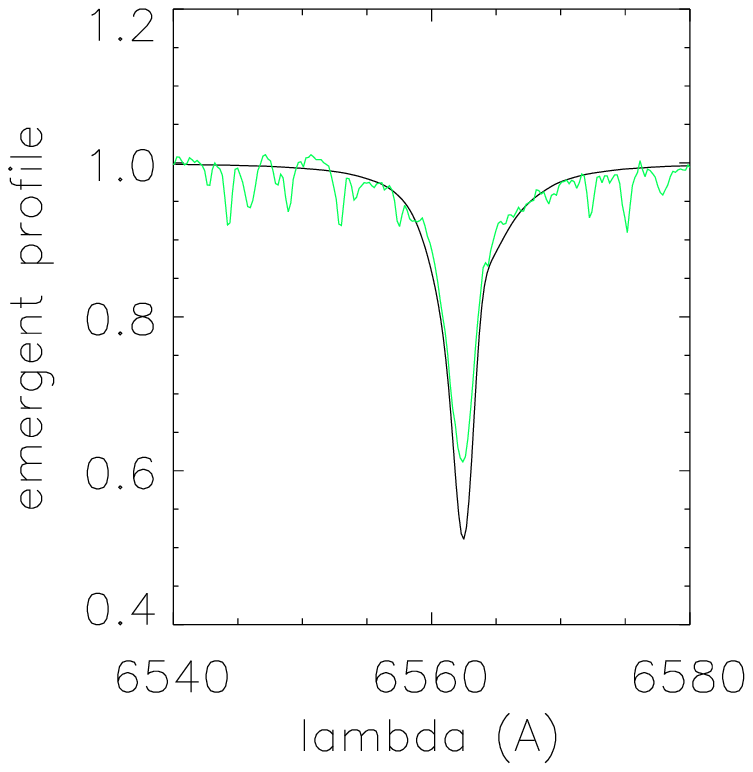}}
\end{minipage}
\caption{As for Fig.~\ref{Hg_Ha_1}, but for late B subtypes. From top to 
bottom: HD~191\,243 (B5Ib), HD~199\,478 (B8ae), HD~212\,593 (B9Iab) and 
HD~202\,850 (B9Iab).}
\label{Hg_Ha_2}
\end{figure}

\subsection{Surface gravity}

Classically, the Balmer lines wings are used to determine the surface
gravity, $\log g$, where only higher members (\Hgama and \Hdelta when
available) have been considered in the present investigation to prevent a
bias because of potential wind-emission effects in \Ha and \Hbeta. Note 
that due to stellar rotation the \logg\ values derived from such
diagnostics are only $effective$ values. To derive the {\it true} gravities,
$\log g_{\rm true}$, required to calculate masses, one has to apply a centrifugal
correction (approximated by \vsini$^2$/\Rstar), though for all
our sample stars this correction was found to be typically less than 0.03
dex.

Corresponding values for effective and corrected surface gravities are
listed in Columns 5 and 6 of Table~\ref{para_2}. The error of these 
estimates was consistently adopted as $\pm$ 0.1~dex due to the rather good 
quality of the fits and spectra (because of the small centrifugal
correction, corresponding errors can be neglected) except for HD~190\,603
and HD~199\,478 where an error of $\pm$0.15~dex was derived instead. This
point is illustrated in Figs~\ref{Hg_Ha_1} and \ref{Hg_Ha_2} where our
final (``best'') fits to the observed \Hgama profiles are shown. Note that
the relatively large discrepancies in the cores of HD~190\,603 and
HD~199\,478 might be a result of additional emission/absorption from 
large-scale structures in their winds \citep{Rivi, Mar06}, which cannot be
reproduced by our models (see also Sec.~\ref{wind} below). At least for
HD~190\,603, an alternative explanation in terms of too large a mass-loss
rate (clumping effects in \Ha) is possible as well.

\subsection{Stellar radii, luminosities and masses}

The input radii used to calculate our model grid have been drawn from 
evolutionary models. Of course, these radii are somewhat different
from the finally adopted ones (listed in Column~7 of Table~\ref{para_2})
which have been derived following the procedure introduced by
\citet{Kudritzki80} (using the de-reddened absolute magnitudes from 
Table~\ref{log_phot} and the theoretical fluxes of our models). 
With typical uncertainties of $\pm$500 K in our \Teff and of $\pm$0.3 
to 0.5 mag in \MV, the error in the stellar radius is dominated by
the uncertainty in \MV, and is of the order of $\Delta$log~\Rstar= 
$\pm$0.06...0.10, i.e., less than 26\% in $R_\star$. 

Luminosities have been calculated from the estimated effective temperatures and
stellar radii, while masses were inferred from the ``true'' surface
gravities. These estimates are given in Columns 9 and 10 of
Table~\ref{para_2}, respectively. The corresponding errors are
less then $\pm$0.21~dex in \logl and $\pm$0.16 
to 0.25~dex in log~\Mstar.

The spectroscopically estimated masses of our SG targets range from 7
to 53 \Msun.\footnote{A mass of 7 \Msun as derived for HD~198\,478
(second entry) seems to be rather low for a SG, suggesting that
the B-V colour adopted from {\it SIMBAD} is probably underestimated.}
Compared to the evolutionary masses from \citet{MM00} and apart from two
cases, our estimates are generally lower, by approximately 0.05 to 0.38~dex,
with larger differences for less luminous stars. While for some stars the
discrepancies are less than or comparable to the corresponding errors (e.g.,
HD~185\,859, HD~190\,603 first entry, HD206\,165), they are significant for
some others (mainly at lower luminosities) and might indicate a ``mass
discrepancy'', in common with previous findings 
\citep{crowther06,Trundle05}.

\subsection{Wind parameters} 
\label{wind}

\paragraph{Terminal velocities.} For the four hotter stars in our sample,
individual terminal velocities are available in the literature, determined
from UV P Cygni profiles \citep{LSL95,PBH,Howarth97}. For these stars, we
adopted the estimates from \citet{Howarth97}. Interestingly, the initially 
adopted value of 470~\kms for \vinf of HD~198\,478 did not provide a 
satisfactory fit to H$_\alpha$, which in turn required a value of about
200~km~s$^{\rm -1}$. This is at the lower limit of the ``allowed'' range, since the
photospheric escape velocity, $v_{\rm esc}$, is of the same order. Further
investigations, however, showed that at a different observational epoch the
\Ha profile of HD~198\,478 indeed has extended to about 470~\kms
\citep{crowther06}. Thus, for this object we considered a rather large
uncertainty, accounting for possible variations in $v_\infty$.

Regarding the four cooler stars, on the other hand, we were forced to 
estimate \vinf by employing the spectral type - terminal velocity
calibration provided by \citet{kud00}, since no literature values could be
found and since archival data do not show saturated P Cygni profiles which
could be used to determine \vinf. In all but one of these objects
(HD~191\,243, first entry), the calibrated \vinf-values were  lower
than the corresponding escape velocities, and we adopted \vinf = \vesc to 
avoid this problem.

The set of \vinf-values used in the present study is listed in Column 11 of
Table~\ref{para_2}. The error of these data is typically less than 100~\kms
\citep{PBH} except for the last four objects where an asymmetric error of
-25/+50\% was assumed instead, allowing for a rather large insecurity
towards higher values. 

\paragraph{Velocity exponent $\beta$.} In stars with denser winds (\Ha in
emission) $\beta$ can be derived from \Ha with relatively high precision and
in parallel with the mass loss rate, due to the strong sensitivity of the \Ha
profile shape on this parameter (but see below). On the other hand, for
stars with thin winds (\Ha in absorption) the determination of $\beta$ from
optical spectroscopy alone is (almost) impossible and a typical value of
$\beta=1$ was consistently adopted, but lower and larger values have been
additionally used to constrain the errors. Note that for two of these stars
we actually found indications for values larger than $\beta$=1.0
(explicitly stated in Table~\ref{para_2}). 

\paragraph{Mass-loss rates,} $\dot M$, have been derived from fitting the
observed \Ha profiles with model calculations. The obtained estimates are
listed in column 13 of Table~\ref{para_2}. Corresponding errors, accumulated
from the uncertainties in $Q$\footnote{We do not directly derive the mass-loss 
rate by means of \Ha, but rather the corresponding optical depth
invariant $Q$, see \citet{Markova04, repo}.}, \Rstar and
\vinf, are typically less than $\pm$0.16~dex for the three hotter stars in 
our sample  and less than  $\pm$0.26~dex for the rest, due to more insecure 
values of \vinf and $Q$. 
\begin{table*}
\caption
{Final results for our sample of Galactic B-SGs: Stellar and wind
parameters adopted ($\rightarrow$\MV) and derived using FASTWIND. \Teff in kK, 
\Rstar in \Rsun, \Mstar in \Msun, \vinf in \kms, \Mdot in $10^{-6} {\rm M_{\odot}/yr}$.  
$D_{\rm mom}$ (in cgs-units) denotes the modified wind-momentum rate.
High precision $\beta$-values are given bold-faced. For non-tabulated errors, see text.}
\label{para_2}
\tabcolsep1.15mm
\begin{tabular}{llccccclcrrlll}
\hline 
\hline
~\\
\multicolumn{1}{c}{Object}
&\multicolumn{1}{l}{Sp}
&\multicolumn{1}{r}{\MV}
&\multicolumn{1}{c}{\Teff}
&\multicolumn{1}{c}{\logg}
&\multicolumn{1}{c}{\loggtr}
&\multicolumn{1}{c}{\Rstar}
&\multicolumn{1}{c}{\Yhe}
&\multicolumn{1}{c}{\logl}
&\multicolumn{1}{r}{\Mstar}
&\multicolumn{1}{r}{\vinf}
&\multicolumn{1}{c}{$\beta$}
&\multicolumn{1}{c}{log \Mdot}
&\multicolumn{1}{l}{\logwm}
\\
\hline
HD 185\,859  &B0.5 Ia  &-7.00 &26.3 &2.95 &2.96 &35  &0.10$\pm$0.02 &5.72  &41$^{\rm +27}_{\rm -16}$ &1\,830 &{\bf1.1}$\pm$0.1$^{a)}$ &-5.82$\pm$0.13     &29.01$\pm$0.20 \\ 
HD 190\,603  &B1.5 Ia+ &-8.21 &19.5 &2.35 &2.36 &80  &0.20$\pm$0.02 &5.92  &53$^{\rm +41}_{\rm -23}$ &485    &{\bf2.9}$\pm$0.2        &-5.70$\pm$0.16     &28.73$\pm$0.22 \\  
             &         &-7.53& &     &2.36 &58&    &5.65&28$^{\rm +21}_{\rm -12}$ &      &                    &-5.91$\pm$0.16    &28.45$\pm$0.22 \\  
HD 206\,165  &B2 Ib    &-6.19 &19.3 &2.50 &2.51 &32  &0.10 - 0.20   &5.11  &12$^{\rm +7}_{\rm -4}$ &640&{\bf1.5}$^{\rm +0.2}_{\rm -0.1}$ $^{a)}$&-6.57$\pm$0.13 &27.79$\pm$0.17 \\  
HD 198\,478  &B2.5 Ia  &-6.93 &17.5 &2.10 &2.12 &49  &0.10 - 0.20   &5.31  &11$^{\rm +5}_{\rm -3}$  &200...470&{\bf1.3}$\pm$0.1      &-6.93...-6.39       &26.97...27.48 \\ 
             &         &-6.37& &     &2.12 &38  &              &5.08&7$^{\rm +3}_{\rm -2}$   &       &                &-7.00...-6.46  &26.84...27.36 \\ 
HD 191\,243  &B5 Ib    &-5.80 &14.8 &2.60 &2.61 &34  &0.09$\pm$0.02 &4.70 &17$^{\rm +9}_{\rm -6}$ &470 &0.8...1.5  &-7.52$^{\rm+0.26}_{\rm-0.20}$     &26.71$^{\rm+0.27}_{\rm-0.23}$ \\ 
             &  &-6.41 &     &     &2.60 &46  & &4.96  &31$^{\rm +17}_{\rm -11}$ & &   &-7.30$^{\rm+0.25}_{\rm-0.17}$       &27.00$^{\rm+0.25}_{\rm-0.21}$\\  
HD 199\,478  &B8 Iae   &-7.00 &13.0 &1.70 &1.73 &68  &0.10$\pm$0.02 &5.08  &9$^{\rm +5}_{\rm -3}$   &230    &0.8...1.5       &-6.73...-6.18               &27.33...27.88 \\ 
HD 212\,593  &B9 Iab   &-6.50 &11.8 &2.18 &2.19 &59  &0.06 - 0.10   &4.79  &19$^{\rm +13}_{\rm -8}$  &350    &0.8...1.5       &-7.04$^{\rm+0.25}_{\rm-0.19}$      &27.18$^{\rm+0.28}_{\rm-0.24}$\\
HD 202\,850  &B9 Iab   &-6.18 &11.0 &1.85 &1.87 &54  &0.09$\pm$0.02 &4.59  &8$^{\rm +4}_{\rm -3}$   &240    &0.8...1.8       &-7.22$^{\rm+0.25}_{\rm-0.17}$      &26.82$^{\rm+0.25}_{\rm-0.20}$\\
\end{tabular}
\\
\\
\footnotesize{$^{a)}$ \Ha (though in absorption) indicates $\beta > 1$.} 
\normalsize
\end{table*}

The errors in $Q$ itself have been determined from the fit-quality to \Ha
and from the uncertainty in $\beta$ (for stars of thin winds), while the
contribution from the small errors in \Teff have been neglected. Since we
assume an unclumped wind, the {\it actual} mass-loss rates of our sample
stars might, of course, be lower. In case of small-scale clumping, this
reduction would be inversely proportional to the square root of the
effective clumping factor being present in the \Ha forming region (e.g.,
\citealt{puls06} and references therein).
In Figures~\ref{Hg_Ha_1} and \ref{Hg_Ha_2} we present our final (``best'')
\Ha fits for all sample stars. Apparent problems are:

\smallskip
\noindent
$\bullet$
The P~Cygni profile of \Ha seen in  HD~190\,603 (B1.5 Ia+) is not 
well fitted. The model fails to reproduce the depth and the 
width of the absorption trough. This (minor) discrepancy, however, has no 
effect on the derived \Mdot and $\beta$, because these parameters are mainly 
determined by the emission peak and the red emission wing of the line which 
are well reproduced. 

\noindent
$\bullet$
In two sample stars, HD~198\,478 and HD~199\,478, the observations 
show \Ha in emission whilst the models predict profiles of P Cygni-type. At
least for HD~198\,478, a satisfactory fit to the red wing and the emission
peak became possible, since the observed profile is symmetric with respect
to rest wavelength, thus allowing us to estimate $\beta$ and \Mdot 
within our assumption of homogeneous and spherically
symmetric winds (see below). 
For $\dot M$, we provide lower and upper limits in Table~\ref{para_2},  
corresponding to the lower and upper limits of the adopted \vinf
(see also Sect.~\ref{wlr_comp1}). For HD~199\,478, on the other hand, 
$\beta$ is more insecure due to the strong asymmetry in the profile
shape, leading to a larger range in possible mass loss rates and wind momenta.

\noindent
$\bullet$
The \Ha profile of HD~202\,850, the coolest star in our sample, 
is not well reproduced: the model predicts more absorption in the core 
than is actually observed.

\smallskip
\noindent
The most likely reason for our failure to reproduce certain profile shapes
for stars with {\it dense} winds is our assumption of smooth and spherically
symmetric atmospheres. Besides the open question of small-scale clumping
(which can change the \Ha morphology quite substantially\footnote{and
might introduce a certain ambiguity between $\beta$ and the run of the
clumping factor, if the latter quantity is radially stratified}, cf.
\citealt{puls06}), the fact that in two of the four problematic cases
convincing evidence has been found for the presence of time-dependent 
large-scale structure (HD~190\,603, \citealt{Rivi}) or deviations from spherical
symmetry and homogeneity (HD~199\,478, \citealt{Mar06}) seems to support
such a possibility. Note that similar problems in reproducing certain \Ha
profile shapes in B-SGs have been reported by \citet[ from here on
KPL99]{Kud99}, \citet{Trundle04}, \citet{crowther06} and \citet{Lefever}.

\begin{figure*}
\begin{minipage}{8.8cm}
\resizebox{\hsize}{!}
{\includegraphics{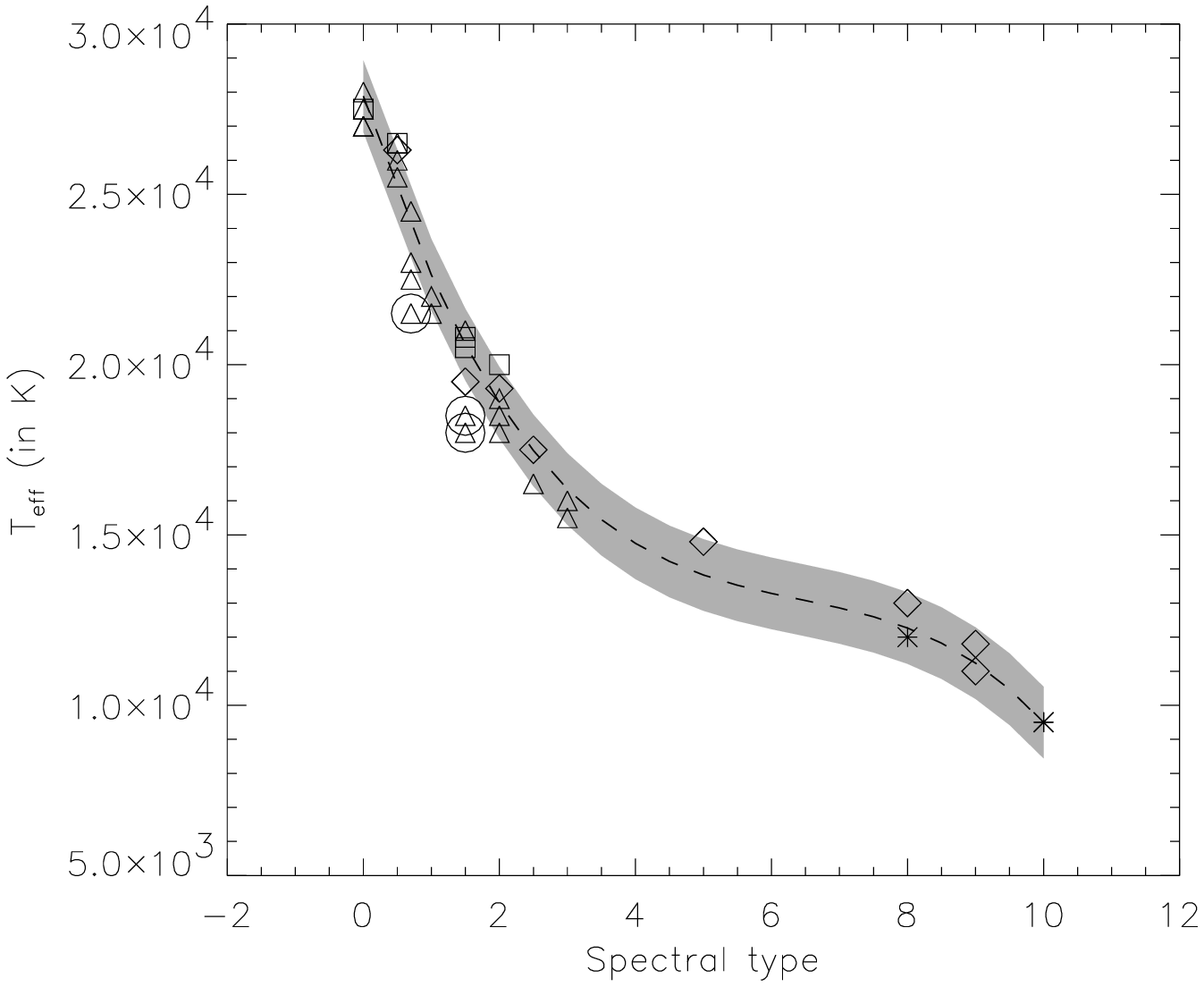}}
\end{minipage}
\hfill
\begin{minipage}{8.8cm}
\resizebox{\hsize}{!}
{\includegraphics{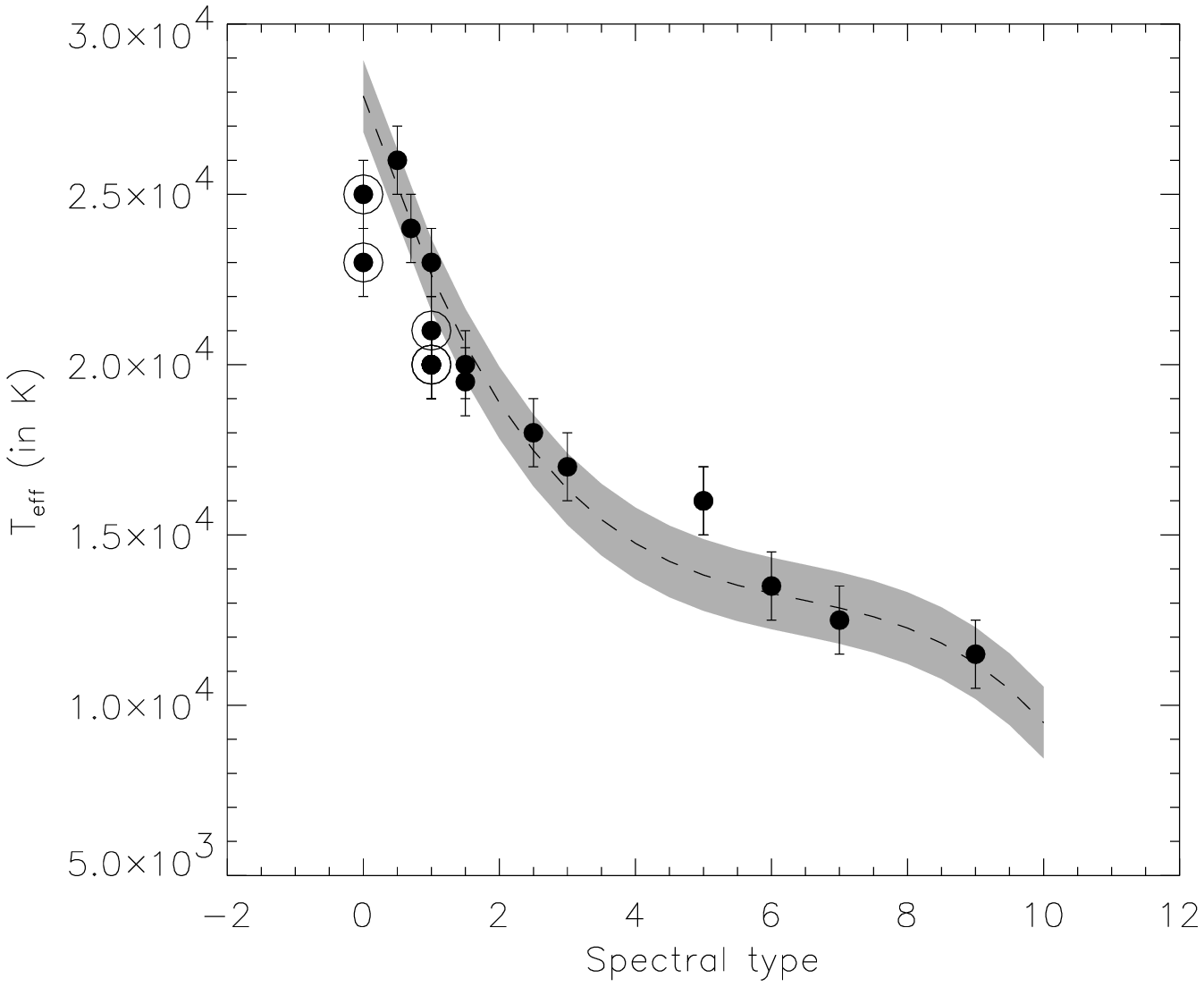}}
\end{minipage}

\caption{{\bf Left:} Comparison of effective temperatures as derived in the 
present study with data from similar investigations. Diamonds - our data;
triangles - data from \citet{crowther06}; squares - \citet{Urb};
asterisks - \citet{Przb06}. Large circles mark the three 
objects with strongest winds, which simultaneously show the largest
deviation in \Teff from the mean (all from the sample by Crowther et al.).
The dashed line represents a 3rd order polynomial fit to the data accounting
for the individual errors in \Teff, and the grey-shaded area denotes the
corresponding standard deviation. Spectral types refer to B-stars (i.e.,
``-1'' corresponds to O9, and ``10'' to A0).  \newline {\bf Right:} \Teff
estimates for GROUP~I stars from \citet{Lefever}, as a function of
spectral type. The error bars correspond to $\pm$1\,000~K, and the dashed
line/shaded area refer to the fit on the left. Large circles mark
data points which deviate significantly from this regression (see text).} 
\label{teff_comp}
\end{figure*}

A comparison of present results with such from previous studies 
\citep{crowther06, BC} for three stars in common indicates that the
parameters derived by \citet{crowther06} for HD~190\,603 and HD~198\,478 are
similar to ours (accounting for the fact that higher \Teff and \MV result 
in larger \logg and $\dot M$, respectively, and vice versa). The mass-loss rates
from \citet{BC} (derived from the IR-excess!) for HD~198\,478 and
HD~202\,850, on the other hand, are significantly larger than ours and those
from Crowther et al., a problem already faced by \citet{Kud99} in a
similar (though more simplified) investigation. This might be either due to
certain inconsistencies in the different approaches, or might point to the
possibility that the IR-forming region of these stars is more heavily
clumped than the \Ha forming one.

\section{The \Teff\ scale for B-SGs -- comparison with other
investigations} \label{teffscale}

\subsection{Line-blanketed analyses}  

Besides the present study, two other investigations have determined the
effective temperatures of {\it Galactic} B-type SGs by methods
similar to ours, namely from Silicon and Helium (when possible) ionisation
balances, employing state of the art techniques of quantitative spectroscopy
on top of high resolution spectra covering all strategic lines.
\citet{crowther06} have used the non-LTE, line blanketed code CMFGEN
\citep{hil98} to determine \Teff of 24 supergiants (luminosity classes Ia,
Ib, Iab, Ia+) of spectral type B0-B3 with an accuracy of $\pm$1\,000 K, 
while \citet{Urb} employed FASTWIND (as done here) and determined effective 
temperatures of five early B (B2 and earlier) stars of luminosity classes
Ia/Ib with an (internal) accuracy of $\pm$500~K. In addition, \citet{Przb06}
have recently published very precise temperatures (typical error of $\pm$200
K) of four BA SGs (among which one B8 and two A0 stars), again
derived by means of a line-blanketed non-LTE code, in this case in
plane-parallel geometry neglecting wind effects.

Motivated by the good correspondence between data from FASTWIND 
and CMFGEN (which has also been  noted by \citealt{crowther06}), we plotted
the effective temperatures of all four investigations, as a function of
spectral type (left panel of Fig.~\ref{teff_comp}). Overplotted (dashed
line) is a 3rd order polynomial regression to these data, accounting for the
individual errors in \Teff, as provided by the different investigations. The
grey-shaded area denotes the standard deviation of the regression.
Obviously, the correspondence between the different datasets is (more than)
satisfactory: for a given spectral subtype, the dispersion of the data does
not exceed $\pm$1000 K. There are only three stars (marked with large circles),
all from the sample of Crowther et al., which make an exception, showing
significantly lower temperatures: HD~190\,603, HD~152\,236 and HD~2\,905.
Given their strong P~Cygni profiles seen in \Ha and their high luminosities
- note that the first two stars are actually hypergiants - this result
should not be a surprise though (higher luminosity $\rightarrow$ denser wind
$\rightarrow$ stronger wind blanketing $\rightarrow$ lower \Teff). 

Very recently, \citet{Lefever} published a study with the goal to test
whether the variability of a sample of 28 periodically pulsating, Galactic
B-type SGs is compatible with opacity driven non-radial pulsations.
To this end, they analysed this sample plus 12 comparison objects, also by
means of FASTWIND, thus providing additional stellar and wind parameters of
such objects. In contrast to both our investigation and those mentioned
above, Lefever et al. could not use the Si (He) ionisation {\it balance} to
estimate \Teff, but had to rely on the analysis of one ionisation stage
alone, {\it either} Si~II {\it or} Si~III, plus two more He~I lines
($\lambda$4471 and $\lambda$6678). The reason for doing so was the (very)
limited spectral coverage of their sample (though at very high resolution),
with only one representative Silicon ionisation stage observed per object.

Given the problems we faced during our analysis (which only appeared
because we had a much larger number of lines at our disposal) and the fact
that Lefever et al. were not able to {\it independently} estimate Si
abundances and \vmic of their sample stars (as we have done here), the
results derived during this investigation are certainly prone to larger
error bars than those obtained by methods where {\it all} strategic lines 
could be included (see Lefever et al. for more details).  

The right-hand panel of Fig.~\ref{teff_comp} displays their temperature 
estimates for stars from the so-called GROUP~I (most precise parameters), 
overplotted by {\it our} regression from the left panel. The error bars
correspond to $\pm$1\,000~K quoted by the authors as a nominal error. 
While most of their data are consistent (within their errors) with our
regression, there are also objects (marked again with large circles) which
deviate significantly. 

Interestingly, all outliers situated below the regression are stars of early
subtypes (B0/B1), which furthermore show P~Cygni profiles with relatively
strong emission components in \Ha (except for HD~15\,043 which exhibits \Ha
in absorption), a situation that is quite similar to the one observed on the
left of this figure. (We return to this point in the next section.)

On the other hand, there are two stars of B5-type with same \Teff, which lie
above the regression, i.e., seem to show ``overestimated'' temperatures. The
positions of these stars within the \Teff-spectral type plane have been
extensively discussed by \citet{Lefever} who suggested that the presence of
a radially stratified micro-turbulent velocity (as also discussed by us) or
a Si abundance being lower than adopted (solar) might explain the
overestimate (if so) of their temperatures. Note, however, that the
surface gravity of HD~108\,659 (=2.3), one of the Lefever et al. B5 targets,
seems to be somewhat large for a SG but appropriate for a bright
giant. Thus, it might still be that the ``overestimated'' temperature of
HD~108\,659 is a result of its misclassification as a SG whilst
actually it is a bright giant. This possibility, however, cannot be applied to
the other B5 target, HD~102\,997, which has \logg of 2.0 (and \MV of -7.0),
i.e., is consistent with its classification as a supergiant. 

Interestingly, the surface gravity of ``our'' B5 star, HD~191\,243
(\logg=2.6), appears also to be larger than what is typical for a 
supergiant of B5 subtype.
With a distance modulus of 2.2~kps \citep{Humphreys78}, the absolute
magnitude of HD~191\,243 would be more consistent with a supergiant
classification, but with d=1.75~kpc \citep{GS} a luminosity class II is more
appropriate. Thus, this star also seems to be misclassified. \footnote{Note
that already \citet{Lennon92} suggested that HD~191\,243 
{\bf is likely} a bright giant, but their argumentation was based more on
qualitative rather than on quantitative evidence.} 

\begin{figure*}
\begin{minipage}{8.8cm}
\resizebox{\hsize}{!}
{\includegraphics{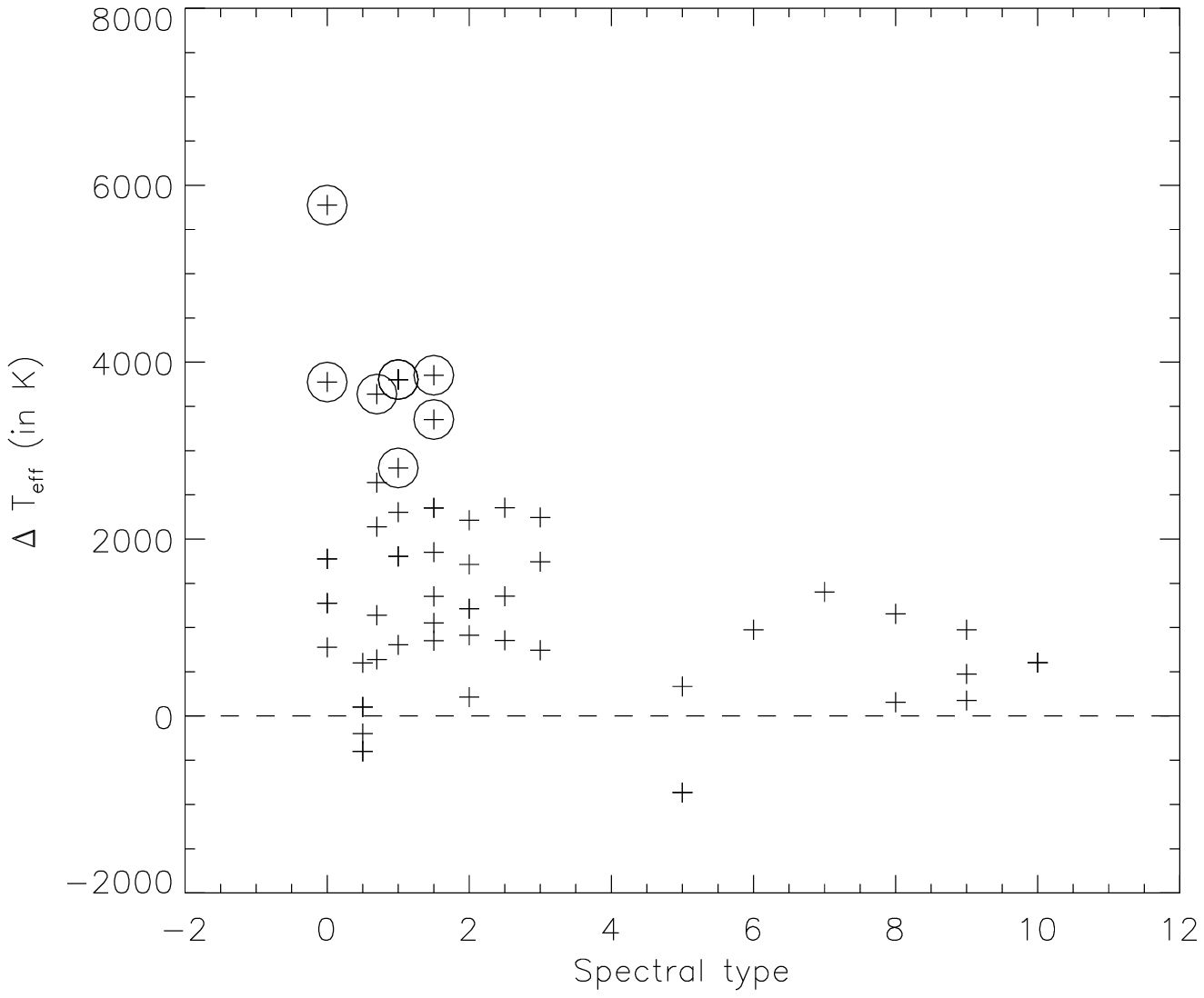}}
\end{minipage}
\begin{minipage}{8.8cm}
\resizebox{\hsize}{!}
{\includegraphics{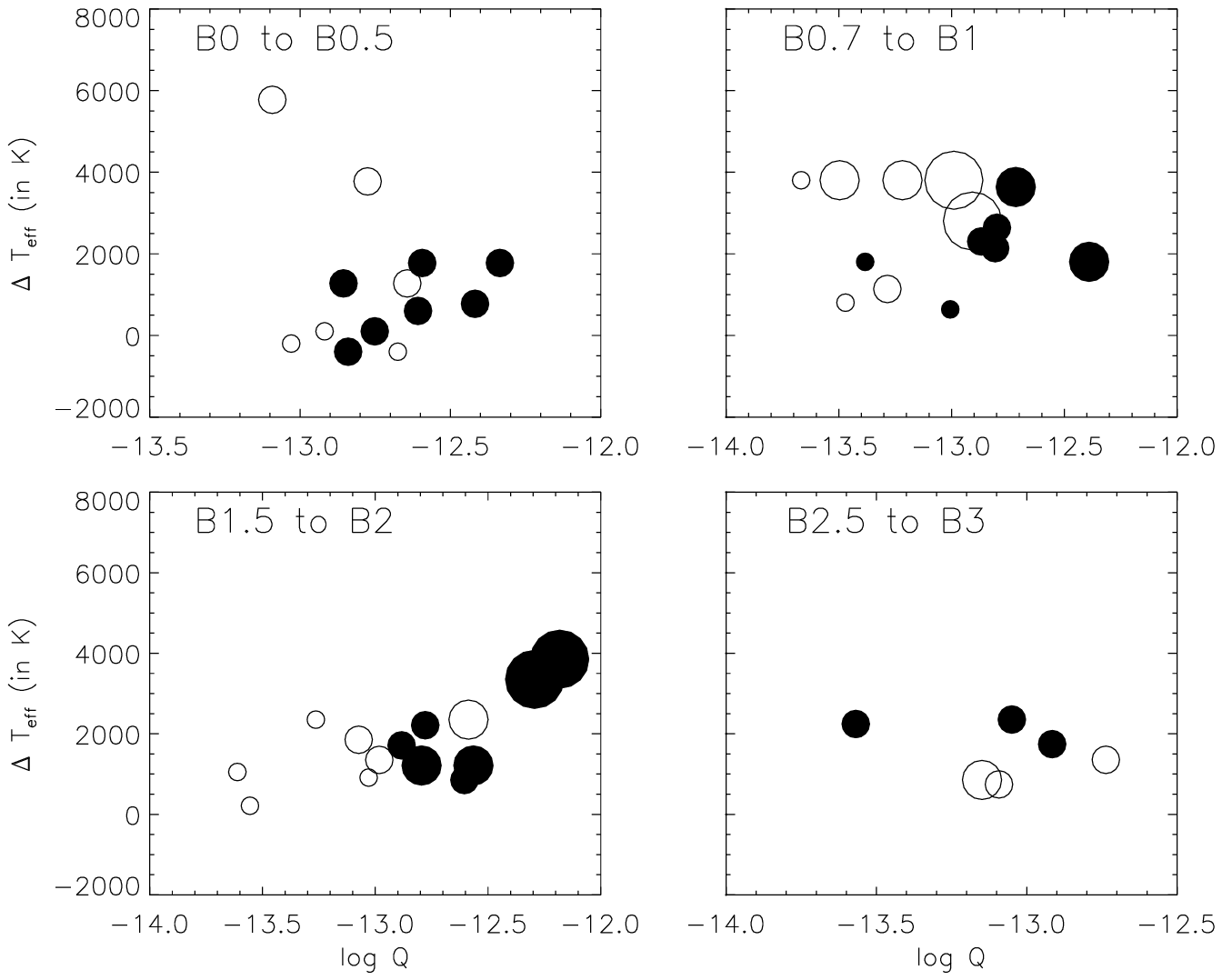}}
\end{minipage}
\caption{
Differences between ``unblanketed'' and ``blanketed'' effective temperatures
for the combined sample (this work, \citealt{Urb}, \citealt{crowther06},
\citealt{Przb06} and GROUP~I objects from \citealt{Lefever}), as a
function of spectral type (left panel) and as a function of $\log Q$ for
individual subtypes, B0 to B3 (right panel).  Unblanketed \Teff\ are from
\citet{McE99}.\newline 
{\bf Left}: Large circles denote the same objects as in
Fig.~\ref{teff_comp}.  The data point indicating a significant {\it
negative} temperature difference corresponds to the {\it two} B5 stars (at
same temperature) from the Lefever et al.  sample. 
\newline
{\bf Right}: The size of the symbols corresponds to the size of the 
peak emission seen in \Ha. Filled circles mark data from CMFGEN, open 
circles those from FASTWIND.}
\label{teff_dev}
\end{figure*}

\subsection{Temperature revisions due to 
line-blanketing and wind effects}
\label{line block}

In order to estimate now the effects of line-blocking/blanketing together
with wind effects in the B supergiant domain (as has been done previously
for the O-star domain, e.g., \citealt{Markova04, repo, martins05}), we have
combined the different datasets as discussed above into one sample, keeping
in mind the encountered problems. 

Figure~\ref{teff_dev}, left panel, displays the differences between 
``unblanketed'' and ``blanketed'' effective temperatures for this combined
sample, as a function of spectral type. The ``unblanketed'' temperatures
have been estimated using the \Teff-Spectral type calibration provided by
\citet{McE99}, based on unblanketed, plane-parallel, NLTE model atmosphere
analyses. Objects enclosed by large circles are the same as in
Fig.~\ref{teff_comp}, i.e., three from the analysis by Crowther et al.,
and seven from the sample by Lefever et al.\footnote{Four B1 stars from 
the Lefever et al. sample have the same \Teff and thus appear as one 
data point in Figs 8 (right) and 9.} As to be expected and as noted by previous 
authors on the basis of smaller samples (e.g.,
\citealt{crowther06, Lefever}), the ``blanketed'' temperatures of Galactic 
B-SGs are systematically lower than the ``unblanketed'' ones. The
differences range from about zero to roughly 6\,000~K, with a tendency to
decrease towards later subtypes (see below for further discussion).

The most remarkable feature in Figure~\ref{teff_dev} is the large dispersion
in $\Delta$\Teff for stars of early subtypes, B0-B3. Since the largest
differences are seen for stars showing P~Cygni profiles with a relatively 
strong emission component in \Ha, we suggest that most of this dispersion is
related to wind effects. 

To investigate this possibility, we have plotted the distribution of the
$\Delta$\Teff-values of the B0-B3 object as a function of the
distant-invariant optical depth parameter $\log Q$ (cf.
Sect.~\ref{results}). Since the \Ha emission strength does not depend on 
$Q$ alone but also on \Teff - for same $Q$-values cooler objects have more 
emission due to lower ionisation - stars with individual subclasses were
studied separately to diminish this effect. The right-hand panel of 
Fig.~\ref{teff_dev} illustrates our results, where the size of the circles
corresponds to the strength of the emission peak of the line. Filled 
symbols mark data from CMFGEN, and open ones data from FASTWIND.
Inspection of these data indicates that objects with stronger \Ha emission
tend to show larger log~$Q$-values and subsequently higher $\Delta$\Teff -
 a finding that is model independent. This tendency is particularly
evident in the case of B1 and B2 objects. On the other hand, there are at
least three objects which appear  to deviate from this rule, but this might
still be due to the fact that the temperature dependence of $Q$ has not been
completely removed (of course, uncertainties in $\beta$, \Teff and \logg can
also contribute).  All three stars (HD~89\,767,  HD~94\,909
(both B0) and HD~154\,043 (B1)) are from the Lefever et al. sample
and do not exhibit strong \Ha emission but nevertheless the highest
$\Delta$\Teff among the individual subclasses.

In summary, we suggest that the dispersion in the derived effective
temperature scale of early B-SGs is physically real and originates from wind
effects. Moreover, there are three stars from the Lefever et al. GROUP
I sample (spectral types B0 to B1) whose temperatures seem to be 
significantly underestimated, probably due to insufficient diagnostics.  In
our follow-up analysis with respect to wind-properties, we will discard
these ``problematic'' objects to remain on the ``conservative'' side.
\begin{figure}
\begin{minipage}{8.8cm}
\resizebox{\hsize}{!}
{\includegraphics{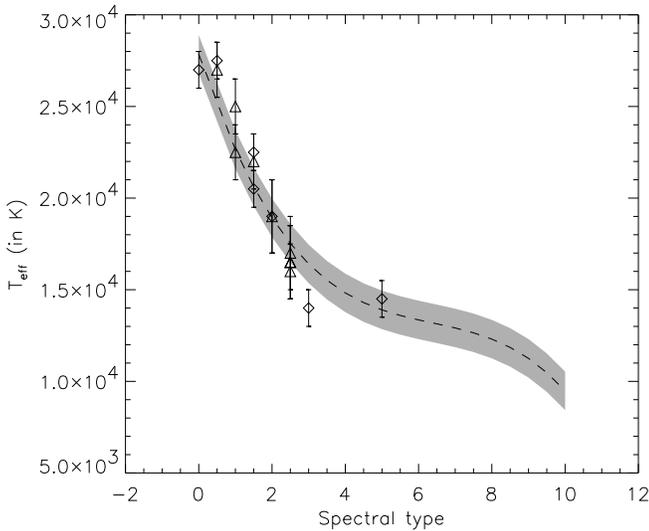}}
\end{minipage}

\caption{Temperature scale for Galactic B-SG as derived in the present study
(dashed, see text), compared to \Teff estimates for similar stars in the SMC
(from \citealt{Trundle04} (diamonds) and \citealt{Trundle05} (triangles)).
The grey area denotes the standard deviation of the regression for Galactic
objects. Spectral types account for metallicity effects (from 
\citealt{lennon97}), see text.  
} 
\label{teff_metal}
\end{figure}

\subsection{Comparison of the temperature scales 
of Galactic and SMC B supergiants}

Wanted to obtain an impression of the influence of
metallicity on the temperature scale for B-type SGs, by comparing Galactic
with SMC data. To this end, we derived a \Teff-Spectral type calibration for
Galactic B-SGs on basis of the five datasets discussed above, discarding
only those (seven) objects from the Lefever et al. sample where the temperatures 
might be particularly affected by strong winds or other uncertainties 
(marked by large circles in Fig,~\ref{teff_comp}, right). Accounting
for the errors in $T_{\rm eff}$, we obtain the following regression (for a
precision of three significant digits) 
\beq
\Teffe\,=\,27\,800\,-\,6\,000\,{\rm SP}\,+\,878\,{\rm SP}^2\,-\,45.9\,
{\rm SP}^3,
\eeq
where ``SP'' (0-9) gives the spectral type (from B0 to B9), and the
standard deviation is $\pm$1040 K.  This regression was then compared to
\Teff estimates obtained by \citet{Trundle04} and \citet{Trundle05} for
B-SGs in the SMC.  

We decided to compare with these two studies {\it only}, because Trundle et
al.  have used a similar (2004) or identical (2005) version of FASTWIND as
we did here, i.e., systematic, model dependent differences between different
datasets can be excluded and because the metallicity of the SMC is
significantly lower than in the Galaxy, so that metallicity dependent
effects should be maximised.

The outcome of our comparison is illustrated in 
Fig.~\ref{teff_metal}: In contrast to the O-star case (cf. 
\citealt{massey04, massey05,mokiem06}), the data for the SMC stars are,
within their errors, consistent with the temperature scale for their
Galactic counterparts. This result might be interpreted as an indication of
small or even negligible metallicity effects (both directly, via
line-blanketing, and indirectly, via weaker winds) in the temperature regime
of B-SGs, at least for metallicities in between solar and SMC (about
0.2 solar) values. Such an interpretation would somewhat contradict our
findings about the strong influence of line-blanketing in the
Galactic case (given that these effects should be lower in the SMC), but
might be misleading since Trundle et al. (2004, 2005) have used the spectral
classification from \citet{lennon97}, which already accounts for the lower
metallicity in the SMC. To check the influence of this re-classification, we
recovered the original (MK) spectral types of the SMC targets using data
provided by \citet[Table 2]{lennon97}, and subsequently compared them to our
results for Galactic B-SGs. Unexpectedly, SMC objects still do not show any
systematic deviation from the Galactic scale but are, instead, distributed
quite randomly around the Galactic mean. Most plausibly, this outcome
results from the large uncertainty in spectral types as determined by
\citet{AZO75}\footnote{using low quality objective prism spectra in
combination with MK classification criteria, both of which contribute to
the uncertainty.}, such that metallicity effects cannot become apparent for
the SMC objects considered here. Nevertheless, we can also conclude that the
classification by \citet{lennon97} has been done {\bf in a perfect way}, namely
that Galactic and SMC stars of similar spectral type also have similar
physical parameters, as expected.
\begin{figure*}
\begin{minipage}{8.8cm}
\resizebox{\hsize}{!}
{\includegraphics{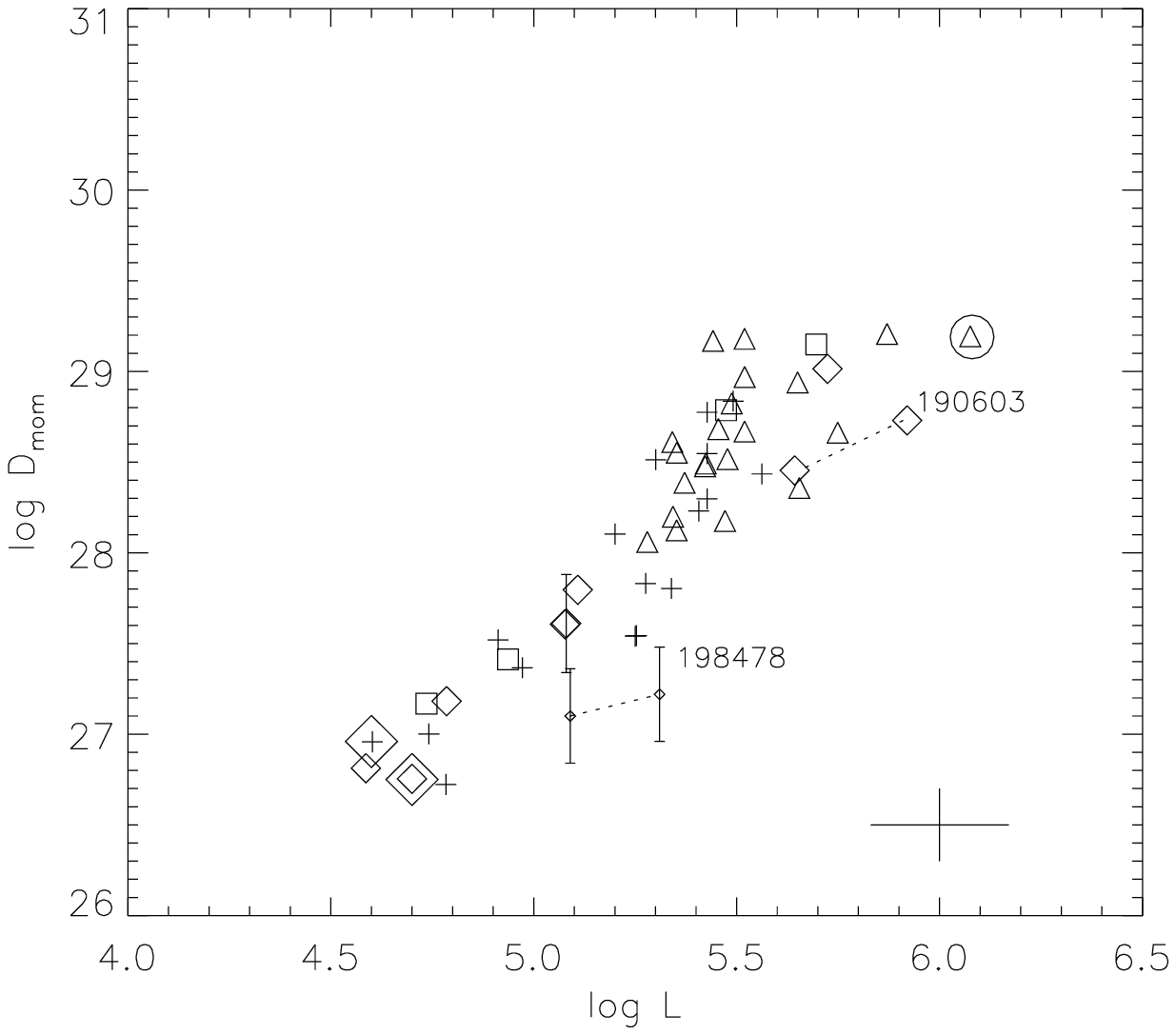}}
\end{minipage}
\begin{minipage}{8.8cm}
\resizebox{\hsize}{!}
{\includegraphics{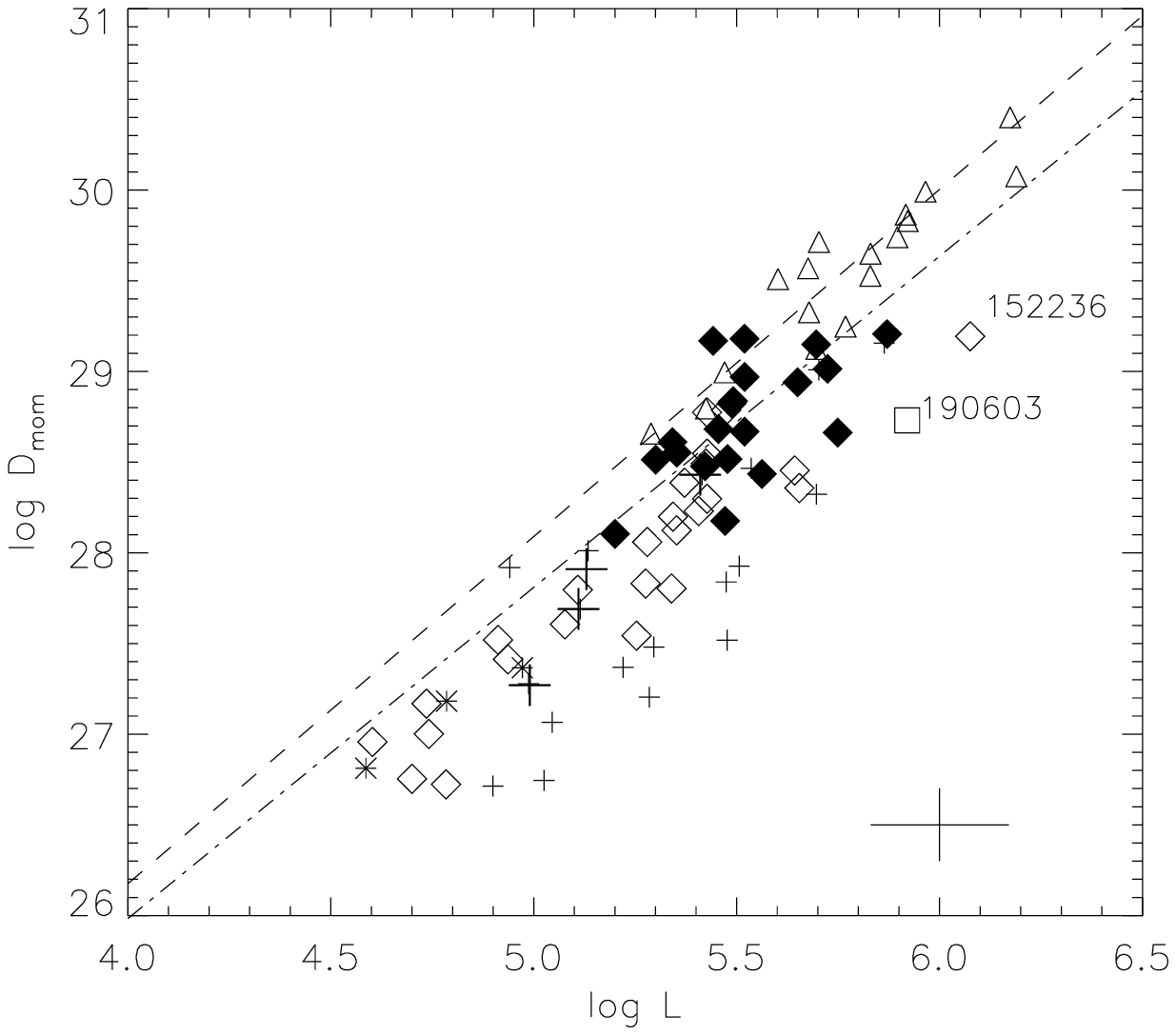}}
\end{minipage}
\caption{ 
{\bf Left}: WLR for Galactic
B-SGs, based on data from the combined sample: diamonds - our data;
triangles - data from \citet{crowther06}; squares - data from \citet{Urb}
and plus signs - data from \citet{Lefever} (GROUP~I, without the three 
``problematic'' objects). The two pairs of symbols connected with dashed
lines correspond to the two entries for HD~190\,603 and HD~198\,478 as
listed in Table~\ref{para_1}. For the latter object and for HD~199\,478, we
also provide error bars indicating the rather large uncertainty in their
wind-momenta. Specially marked objects are discussed in the text.
\newline
{\bf Right:} Wind-momenta of B-SGs from the left (diamonds and asterisks) are
compared with similar data for O-SGs (triangles). Filled diamonds indicate
B-type objects with \Teff $\geq$21\,000~K, asterisks such with \Teff
$\leq$12\,500~K and open diamonds B-types with temperatures in between
12\,500 and 21\,000~K. The high luminosity solution for HD~190\,603 is
indicated by a square. Overplotted are the early/mid B -  (small plus-signs)
and A SG (large plus-signs) data derived by KPL99 and the theoretical
predictions from \citet{Vink00} for Galactic SGs with
27\,500$\leq$\Teff$\leq$50\,000 (dashed-dotted) and with
12\,500$\leq$\Teff$\leq$ 22\,500 (dashed) \newline Error bars provided in
the lower-right corner of each panel represent the typical errors in \logl
and \logwm for data from our sample. Maximum errors in \logwm are about 50\%
larger.
} 
\label{wlr}
\end{figure*}

\section{Wind-momentum luminosity relationship}
\label{wlr_comp}

\subsection{Comparison with results from similar studies} 
\label{wlr_comp1}

Using the stellar and wind parameters, the modified wind momenta can be
calculated (Table~\ref{para_2}, column 14), and the wind-momentum 
luminosity diagram constructed. The results for the combined sample (to
improve the statistics, but without the ``problematic'' stars from the
Lefever et al. GROUP~I sample) are shown on the left of Figure~\ref{wlr}.  Data
from different sources are indicated by different symbols.  For HD~190\,603 
and HD~198\,478, both alternative entries (from Table~\ref{para_1}) are
indicated and connected by a dashed line.  Before we consider the global
behaviour, we first comment on few particular objects.

\smallskip
\noindent
$\bullet$
The position of HD~190\,603 corresponding to $B-V$=0.540 (lower 
luminosity) appears to be more consistent with the distribution of the 
other data points than the alternative position with $B-V$=0.760.
In the following, we give more weight to the former solution.

\noindent
$\bullet$
The positions of the two B5 stars suggested as being misclassified 
(HD~191\,243  and HD~108\,659, large diamonds) fit well the global 
trend of the data, implying that these bright giants do not behave differently 
from supergiants.

\noindent
$\bullet$
The minimum values for the wind momentum of HD~198\,478 
(with \vinf=200~\kms) deviate strongly from the global trend, whereas the 
maximum ones (\vinf=470~\kms) are roughly consistent with this trend. 
For our follow-up analysis, we discard this object because of the very
unclear situation. 

\noindent
$\bullet$
HD~152\,236 (from the sample of Crowther et al., marked with a
large circle) is a hypergiant
with a very dense wind, for which the authors adopted \Rstar = 112 \Rsun,
which makes this object the brightest one in the sample.  

\paragraph{Global features.} From the left of Figure~\ref{wlr}, we see that
the lower luminosity B-supergiants seem to follow a systematically lower 
WLR than their higher luminosity counterparts, with a steep transition 
between both regimes located in between \logl=5.3 and \logl= 5.6.
(Admittedly, most of the early type (high L) objects are Ia's, whereas the
later types concentrate around Iab's with few Ia/Ib's.) 
This finding becomes even more apparent when the WLR is extended towards
higher luminosities by including Galactic O supergiants (from
\citealt{repo, Markova04, Herrero02}), as done on the right of the same
figure.

KPL99 were the first to point out that the offsets in the corresponding WLR 
of OBA-supergiants depend on spectral type, being strongest for O-SGs, 
decreasing from B0-B1 to B1.5-B3 and {\it increasing} again towards A 
supergiants. While some of these results have been confirmed by recent 
studies, others have not \citep{crowther06, Lefever}. 

To investigate this issue in more detail and based on the large sample
available now, we have highlighted the early objects (B0-B1.5,
21\,000$\leq$\Teff$\leq$27\,500~K) in the right-hand panel of
Figure~\ref{wlr} using filled diamonds. (Very) Late objects with \Teff$\leq$
12\,500~K have been indicated by asterisks, and intermediate temperature
objects by open diamonds. Triangles denote O-SGs. Additionally, the 
theoretical predictions by \citet{Vink00} are provided via dashed-dotted and 
dashed lines, corresponding to the temperature regimes of O and B-supergiants,
respectively (from here on referred to as ``higher'' and ``lower''
temperature predictions). Indeed,

\smallskip
\noindent
$\bullet$
O-SGs show the strongest wind momenta, determining a
different relationship than the majority of B-SGs (see below). 

\noindent
$\bullet$
the wind momenta of B0-B1.5 subtypes are larger than those 
of B1.5-3, and both follow a different relationship. However, a direct 
comparison with KPL99 reveals a large discrepancy for mid B1.5-B3 subtypes
($\Delta$\logwm about 0.5 dex), while for B0-B1.5 subtypes their results
are consistent with those from our combined dataset.

\noindent
$\bullet$
Late B4-B9 stars follow the same relationship as mid subtypes.

\smallskip \noindent 
Thus, the only apparent disagreement with earlier
findings relates to the KPL99 mid-B types, previously pointed out by \citet{crowther06}, and suggested to be a result of
line blocking/blanketing effects not accounted for in the KPL99
analysis.\footnote{These authors have employed the {\it unblanketed} 
version of FASTWIND \citep{Santolaya} to determine wind parameters/gravities
while effective temperatures were adopted using the unblanketed,
plane-parallel temperature scale of \citet{McE99}.} After a detailed
investigation of this issue for one proto-typical object from the KPL99
sample (HD~42\,087), we are convinced  that the neglect of line
blocking/blanketing cannot solely account for such lower wind momenta. Other
effects must also contribute, e.g., overestimated $\beta$-values, though at
least the latter effect still leaves a considerable discrepancy.

On the other hand, a direct comparison of the KPL99 A-supergiant dataset 
(marked with large plus-signs on the right of Fig.~\ref{wlr}) with data 
from the combined sample shows that their wind momenta seem to be quite 
similar to those of mid and late B subtypes. Further investigations based 
on better statistics are required to clarify this issue.

\subsection{Comparison with theoretical predictions and the bistability jump}

According to the theoretical predictions by \citet{Vink00}, Galactic
supergiants with effective temperatures between 12\,500 and 22\,500~K
(spectral types B1 to B9) should follow a WLR different from that of hotter
stars (O-types and early B subtypes), with wind momenta being systematically
{\it larger}. From Figure~\ref{wlr} (right), however, it is obvious that the
observed behaviour does not follow these predictions. Instead, the majority
of O-SGs (triangles -- actually those with \Ha in emission, see below)
follow the low-temperature predictions (dashed line), while most of the
early B0-B1.5 subtypes (filled diamonds) are consistent with the
high-temperature predictions (dashed-dotted), and later subtypes (from B2
on, open diamonds) lie below (!), by about 0.3 dex. Only few early B-types
are located in between both predictions or close to the low-temperature one.

The offset between both theoretical WLRs has been explained by
\citet{Vink00} due to the {\it increase} in mass-loss rate at the
bi-stability jump (more lines from lower iron ionization stages available to
accelerate the wind), which is only partly compensated by a drop in terminal
velocity. The size of the jump in $\dot M$, about a factor of five, was
determined requiring a drop in \vinf by a factor of two, as extracted from
earlier observations \citep{LSL95}. 

However, more recent investigations (\citealt{crowther06}, see also
\citealt{Evans2}) have questioned the presence of such a ``jump'' in \vinf,
and argued in favour of a gradual decrease in \vinf/\vesc, from $\sim$3.4
above 24 kK to $\sim$1.9 below 20 kK. 

In the following, we comment on our findings regarding this problem
in some detail,  
(i) because of the significant increase in data (also at lower \Teff), (ii)
we will tackle the problem by a somewhat modified approach and 
(iii) recently a new investigation of the bistability jump by means of {\it radio}
mass-loss rates has been published \citep{benaglia07} which gives additional
impact and allows for further comparison/conclusions.

\begin{figure}
\resizebox{\hsize}{!} {\includegraphics{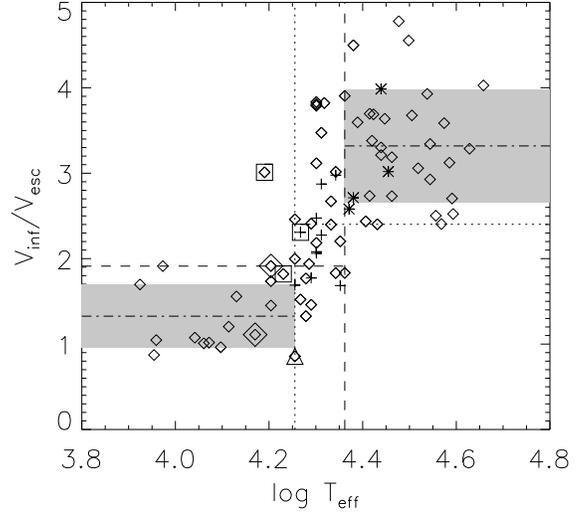}}
\caption{Ratio of \vinf/\vesc as a function of log \Teff. Diamonds: combined
OBA-supergiant sample as in Fig.~\ref{wlr}, right panel.
Asterisks/plus-signs refer to the early/mid B supergiants from the sample by
KPL99. Large diamonds again correspond to the potentially misclassified
B5-objects HD~191\,243 and HD~108\,659. The triangle denotes the hypergiant
HD~152\,236 and the three squares - the various positions of HD~53\,138 as
derived by \citet{crowther06}, \citet{Lefever} and KPL99
(references ordered by increasing \Teff). Individual errors extend
from 33\% to 43\%. See text.}
\label{jump}
\end{figure}

First, let us define the ``position'' of the jump by means of the
\vinf/\vesc-ratios from the OBA-supergiant sample as defined in the previous
section (excluding the ``uncertain'' object HD~198\,478).  In
Figure~\ref{jump}, two temperature regimes with considerably different
values of such ratios have been identified, connected by a transition zone.
In the high temperature regime (\Teff $>$ 23~kK), our sample provides
$\vinfe/\vesce \approx 3.3 \pm 0.7$, whereas in the low temperature one
(\Teff $<$ 18~kK), we find $\vinfe/\vesce \approx 1.3 \pm 0.4$. (Warning:
The latter estimate has to be considered cautiously, due to the large
uncertainties at the lower end where $\vinfe = \vesce$ has been adopted for
few stars due to missing diagnostics.) Note that the individual errors
for $\vinfe/\vesce$ are fairly similar, of the order of 33\% (for $\Delta$\MV = 0.3, 
$\Delta$\logg = 0.15 and $\Delta$\vinf/\vinf = 0.25) to 43\% (in the most
pessimistic case $\Delta$\MV = 1.0), similar to the
corresponding Fig.~8 by Crowther et al..

In the {\it transition zone}, a variety of ratios are present, thus
supporting the findings discussed above. Obviously though, large ratios 
typical for the high temperature region are no longer present from the
centre of the transition region on, so we can define a ``jump
temperature'' of \Teff $\approx$ 20,000~K. Nevertheless, we have shifted the
border of the high-temperature regime to \Teff = 23kK, since at least low
ratios are present until then (note the dashed vertical and horizontal lines
in Fig.~\ref{jump}). The low temperature border has been defined
analogously, as the coolest location with ratios $> 2$ (dotted lines).

By comparing our (rather conservative) numbers with those from the
publications as cited above, we find a satisfactory agreement, both with
respect to the borders of the transition zone as well as with the average
ratios of \vinf/\vesc. In particular, our high temperature value is almost
identical to that derived by \citet{crowther06}; \citealt{kud00} provide
an average ratio of 2.65 for \Teff $>$ 21 kK), whereas in the low
temperature regime we are consistent with the latter investigation
(Kudritzki \& Puls: 1.4). The somewhat larger value found by Crowther et
al. results from missing latest spectral subtypes. 

Having defined the behaviour of $v_\infty$, we investigate
the behaviour of \Mdot, which is predicted to increase more strongly than \vinf
decreases. As we have already seen from the WLR, this most probably is {\it
not} the case for a statistically representative sample of ``normal''
B-SGs, but more definite statements become difficult for two
reasons. First, both the independent ($\log L$) and the dependent (\logwm)
variable depend on \Rstar (remember, the fit quantity is not \Mdot but $Q$),
which is problematic for Galactic objects. Second, the wind-momentum rate is
a function of $L$ but not of \Teff alone, such that a division of different
regimes becomes difficult. To avoid these problems, let us firstly
recapitulate the derivation of the WLR, to see the differences compared to
our alternative approach formulated below. 

From the scaling relations of line-driven wind theory, we have
\beqa
\Mdote &\propto& k^\ao L^\ao (\Mstare(1-\Gamma))^{1-\ao} \label{mdotscal}\\
\vinfe &=& \Cinf \vesce, \qquad \vesce \propto (\Mstare(1-\Gamma))^\half
\eeqa
where $\alpha'$ is the difference between the line force multipliers $\alpha -
\delta$ (corresponding to the slope of the line-strength distribution
function and the ionisation parameter; for details, see \citealt{Puls00}),
$k$ - the force-multiplier parameter proportional to the effective number of
driving lines and $\Gamma$ - the (distance independent!) 
Eddington parameter.  Note that the relation for \Mdot is
problematic because of its mass-dependence, and that \Mdot itself depends
on distance. By multiplying with \vinf and (\Rstar/\Rsun)$^{1/2}$, we obtain
the well-known expression for the (modified) wind-momentum rate,
\beqa
\Dmom &=& \Mdote \vinfe (\Rstare/\Rsune)^\half \propto k^\ao \Cinf L^\ao
(\Mstare(1-\Gamma))^{-\varepsilon} \\
\varepsilon&=&\ao - \threehalf \\
\log \Dmom &\approx& \ao \log L + \Do\\
\Do &=& \ao \log k + \log \Cinf + {\rm const} 
\eeqa
where we have explicitly included here those quantities which are dependent
on spectral type (and metallicity). Remember that this derivation assumes
the winds to be unclumped, and that $\varepsilon$ is small, which is true 
at least for O-supergiants \citep{Puls00}. 

Investigating  various possibilities, it turned out that the
(predicted) scaling relation for a quantity defined similarly as the
optical-depth invariant is particularly advantageous:
\beqa
Q'&=:&\frac{\Mdote}{\Rstare^\threehalf} \frac{\geff}{\vinfe} \propto 
\frac {k^\ao}{\Cinf} \Teffe^\af \geff^{-\varepsilon}, \qquad \geff \propto
\frac{\Mstare(1-\Gamma)}{\Rstare^2} \label{lqp}\\
\log Q' &\approx& \af \log \Teffe + \Do'\\
\Do' &=& \ao \log k - \log \Cinf + {\rm const'} 
\eeqa

\begin{figure}
\resizebox{\hsize}{!}
{\includegraphics{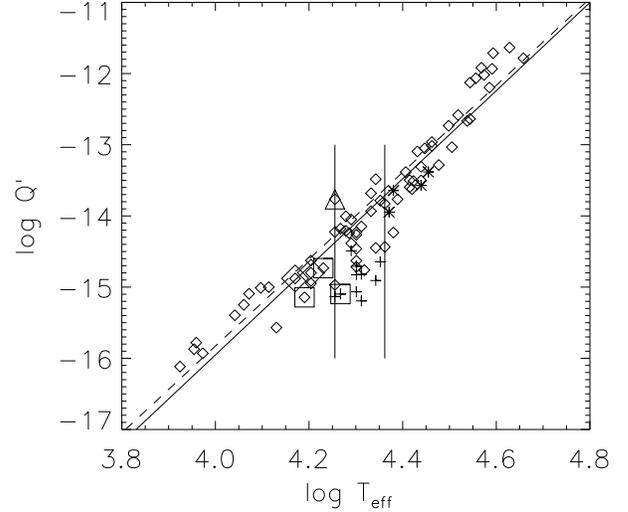}}
\caption{
Modified optical depth invariant, $\log Q'$ (Eq.~\ref{lqp}, as a function of
log \Teff. Symbols as in Fig.~\ref{jump}, with vertical bars indicating the
``transition'' region between 18 and 23 kK. Note that the mid B-type
supergiants from the KPL99 sample (plus-signs) deviate from the trend
displayed by the other objects. Regression for the complete sample
(excluding HD~53\,138, HD~152\,236 and the KPL99 B-supergiants) overplotted
in solid; dashed regression similar, but additionally excluding the objects
in the transition region (see text).} 
\label{teffqp}
\end{figure}

This relation for the {\it distance independent} quantity $Q'$ becomes a
function of log \Teff and $\Do'$ alone if $\alpha'$ were exactly 2/3, i.e.,
under the same circumstances as the WLR. Obviously, this relation has all
the features we are interested in, and we will investigate the temperature
behaviour of \Mdot by plotting log~$Q'$ vs. log \Teff.  We believe that the 
factor $\vinfe/\geff$ is a  monotonic
function on both sides of and through the transition zone, as is also the
case for $\geff$ itself. Thus, the $\log Q' - \log \Teffe$ relation should
react only on differences in the effective number of driving lines and on
the different ratio \vinf/\vesc on both sides of the transition region.

Fig.~\ref{teffqp} displays our final result. At first glance, there is almost
no difference between the relation on both sides of the ``jump'', whereas
inside the transition zone there is a large scatter, even if not accounting
for the (questionable) mid-B star data from KPL99.

Initially, we calculated the average slope of this relation by linear
regression, and then the corresponding slope by additionally excluding the
objects inside the transition region (dashed). Both regressions give similar
results, interpreted in terms of $\alpha'$ with values of 0.65 and 0.66 (!),
respectively\footnote{slope of regression should correspond to 4/$\alpha'$,
{\it if} the relations were unique.}, and with standard errors regarding
$\log Q'$ of $\pm 0.33$ and $\pm 0.28$ dex. 

If the relations indeed were identical on both sides of the jump, 
we would also have to conclude that the offset, 
$\Do'$, is identical on both sides of the transition region. In this case,
the decrease in \vinf/\vesc within the transition zone has to be more or
less exactly balanced by the {\it same} amount of a decrease in $k^\ao$,
i.e., both \Mdot and \vinf are decreasing in parallel, in complete
contradiction to the prediction by Vink et al.

A closer inspection of Fig.~\ref{teffqp} (in combination with the
corresponding WLR of Fig.~\ref{wlr}) implies an alternative
interpretation. At the hottest (high luminosity) end, we find the typical
division of supergiants with \Ha in emission and absorption, where the
former display an offset of a factor 2{\ldots}3 above the mean relation, a
fact which has been interpreted to be related to wind-clumping previously.

Proceeding towards lower temperatures, the $Q'$ relation becomes well
defined between roughly 31kK and the hot side of the transition zone (in
contrast to the WLR, which shows more scatter, presumably due to uncertain
\Rstar). Inside the transition zone and also in the WLR around \logl
$\approx$ 5.45, a large scatter is present, followed by an apparent steep
decrease in $\log Q'$ and wind-momentum rate, where the former is located
just at the ``jump temperature'' of 20~kK. Note that the mid-B type objects
of the KPL99 sample are located just in this region. From then on, $Q'$
appears to remain almost constant until 14~kK, whereas the WLR is 
rather flat between 5.1 $<$ \logl $<$ 5.4, in agreement with the findings by 
\citet[ their Fig.~8]{benaglia07}. At the lowest temperatures/luminosities, 
both $Q'$ and the WLR decrease again, with a similar slope as in the hot 
star domain. This offers a  possibility of a discontinuous 
behaviour, but, again, in contradiction to what is predicted. 

We now quantify the behaviour of the mass-loss rate in the low
temperature region (compared to the high temperature one), in a more
conservative manner than estimated above, by using both the $\log Q'$
relation {\it and} the WLR. Accounting for the fact that the corresponding
slopes are rather similar on both sides of the transition zone, we define a
difference of offsets,
\beqa
\Delta \Do  & \approx & \ao \Delta \log k + \Delta \log \Cinf \\
\Delta \Do'  & \approx & \ao \Delta \log k - \Delta \log \Cinf,
\eeqa
evaluated with respect to ``low'' minus ``high''. From the WLR, we have
$\Delta \Do < 0$, whereas the $Q'$ relation implies $\Delta \Do' \ge 0$, to
be cautious. Thus, the change in $\ao \Delta \log k$ (which expresses the
difference in log \Mdot on both sides of the jump, cf. Eq.~\ref{mdotscal})
is constrained by
\beq
\Delta \log \Cinf \le \ao \Delta \log k < -\Delta \log \Cinf.
\eeq
To be cautious again, we note that $\Delta \log \Cinf$ should lie in the
range log(1.9/2.4) {\ldots} log(1.3/3.3) = -0.1 {\ldots} -0.4, accounting
for the worst and the average situation (cf. Fig.~\ref{jump}).

Thus, the scaling factors of mass-loss rates on both sides of the jump
(cool vs. hot) differ by
\beq
0.4{\ldots} 0.8 \le k^{\ao} < 1.25{\ldots} 2.5
\label{mdotfactor}
\eeq
i.e., \Mdot either decreases in parallel with \vinf/\vesc or increases {\it
marginally}. 

\begin{figure}
\resizebox{\hsize}{!} {\includegraphics{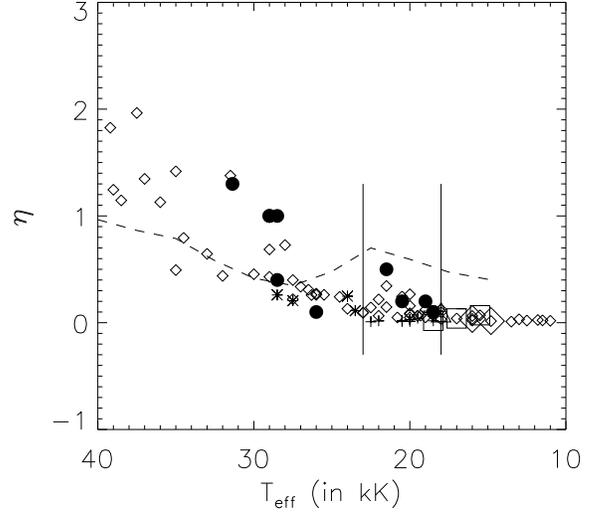}}
\caption{Wind-efficiency, $\eta$, as a function of \Teff, for our combined
sample (symbols as in Fig.~\ref{jump}) and objects as analysed by
\citealt{benaglia07} (filled dots, mass-loss rates from radio excess). Eight
objects in common have been discarded from the {\it latter} sample.
Overplotted (dashed) are their theoretical predictions, based on the models
by \citet{Vink00}.}
\label{eta}
\end{figure}

\paragraph{Wind efficiencies.} Before discussing the implications of these
findings, let us come back to the investigations by \citet{benaglia07} who
recently reported evidence of the possible presence of a local maximum in
the wind efficiency, $\eta$ = \Mdot \vinf /(L/c), around 21\,000~K, which
would be at least in {\it qualitative} agreement with theoretical
predictions. In Figure~\ref{eta}, we compare the wind-efficiencies as
derived for our combined sample (from \Ha) to corresponding data from their
radio measurements (filled dots). The dashed line in the figure displays the
theoretical predictions, which, again, are based on the models by
\citet{Vink00}.

There are eight stars in common with our sample for which we display the \Ha
results only, not to artificially increase the statistics.  At least for
five of those, all of spectral type B0 to B2, a direct comparison of the \Ha
and radio results is possible, since same values of \Teff, \Rstar and \vinf
have been used to derive the corresponding wind efficiencies. In all but one
of these stars\footnote{HD~41\,117, with \Mdot from \Ha being 0.37~dex lower
than from the radio excess.}, radio and optical mass-loss rates agree within
0.2~dex, which is comparable to the typical uncertainty of the optical data.
Translated to potential wind-clumping, this would mean that the outer and
inner wind-regions were affected by similar clumping
factors, in analogy to the findings for {\it thin} O-star winds
\citep{puls06}. From Fig.~\ref{eta} now, several issues are apparent: 

\smallskip
\noindent
$\bullet$
As for the wind-momenta and mass-loss rates, also the wind-efficiencies of
OB-supergiants do not behave as predicted, at least globally.\footnote{In
contrast to $Q'$, $\eta$ is not completely radius-independent, but includes
a dependence $\propto \Rstare^{-0.5}$, both if \Mdot is measured by \Ha and
by the radio excess.} Instead, they follow a different trend (for
$\alpha'=2/3$, one expects $\eta \propto \Teffe^2 \, (\Rstare^{0.5} k^{1.5}
\Cinf)$, i.e., a parabola with spectral type dependent offset), where, as we
have already seen, the offset at the cool side of the jump is much lower
than in the simulations by Vink et al. Actually, this is true for almost
the complete B-SGs domain (between 27 and 10 kK).

\noindent
$\bullet$
Similar as in the {\it observed} wind-momentum luminosity diagram
(Fig.~\ref{wlr}, right panel), some of the O-supergiants do follow the
predictions, while others show wind-efficiencies which are larger by up to a
factor of two. Note that this result is supported by {\it both} \Ha {\it
and} radio diagnostics. If this discrepancy were interpreted in terms of 
small-scale clumping, we would have to conclude that the winds of these
objects are moderately clumped, even at large distances from the stellar
surface.

\noindent
$\bullet$
Within the transition zone, 
a large scatter towards higher values of $\eta$ is observed, which, if not
due to systematic errors in the adopted parameters, 
indeed might indicate the presence of a local maximum, thus supporting the
findings of \citet{benaglia07}. From a careful investigation of the
distribution of stellar radii, terminal velocities and mass-loss rates, we
believe that this local bump does not seem to be
strongly biased by such uncertainties, but is instead due to a real increase
in \Mdot.

\subsection{Discussion}

Regarding a comparison with theoretical models, the major conclusion to be
drawn from the previous section is as follows. 
In addition to the well-known factor of two discrepancies for dense O-SG winds, the
most notable disagreement (discarding local effects within the transition
zone for the moment) is found in the low \Teff/low $L$ B-SG domain, confirming
the analysis by \citet{crowther06}. The predictions by Vink et al. clearly
require the decrease in \vinf to be {\it over}compensated by an increase in
\Mdot throughout the complete mid/late B-star regime, whereas our analysis
has shown that this is not the case. At best, \Mdot increases by the same
amount as \vinf decreases, though a reduction of \Mdot seems to be more
likely, accounting for the fact that the upper limit in Eq.~\ref{mdotfactor}
is a rather conservative estimate.

Since the calculation of {\it absolute} mass-loss rates and wind-momenta is a
difficult task and depends on a number of uncertainties (see below), let us
firstly consider the possibility that at least the predictions regarding the
{\it relative} change in \Mdot (from hot to cool objects) are correct, and
that clumping affects this prediction only marginally.

In this case, the most simple explanation for the detected discrepancy is
that cooler objects are less clumped than hotter ones. Since Vink et al.
predict an increase in \Mdot of a factor of five, this would imply 
that the clumping factors for hotter objects are
larger by factors of 4 (most optimistic case) to 156 (worst case) compared
to those of cooler ones.\footnote{from Eq.~\ref{mdotfactor} with ratios
of $(5/2.5)^2$ and $(5/0.4)^2$} Given  our present knowledge (see
\citealt{FMP06}, \citealt{puls06} and references therein), this is not
impossible, but raises the question about the physical origin of such a
difference.
This hypothesis would also imply that {\it all} B-SG mass-loss rates are
overpredicted, though to a lesser extent for cooler subtypes.

In the alternative, and maybe more reasonable scenario that the clumping
properties of OBA supergiants were not too different, we would have to
conclude that at least the low temperature predictions suffer from unknown 
defects. 
Note, however, that a potential ``failure'' of these predictions 
does not invalidate the radiation driven wind theory itself. The actual
mass-loss rates depend on the effective number of driving lines, and, at
least in principle, this number should {\it decrease} towards lower \Teff,
due to an increasing mismatch between the position of these lines and the
flux maximum (e.g., \citealt{Puls00}). In Vink's models, it increases
instead because Fe~{\sc iii} has many more lines than Fe~{\sc iv}, and because
these lines are distributed over a significant spectral range. 
The absolute number of these lines and their strengths, however, depend on
details of the available data (not  forgeting the elemental abundances,
\citealt{KK07}), a consistent description of the 
ionisation/excitation equilibrium and also on other, complicating effects
(e.g., the diffuse radiation field diminishing the line acceleration in the
lower wind, \citealt{OP99}, and the potential influence of microturbulence,
\citealt{Lucy07}), which makes quantitative predictions fairly
ambiguous. Moreover, if the winds were clumped, this would influence
the hydrodynamical simulations, due to a modified ionisation structure. 

{\it That} there is an effect which is most probably related to the
principal bistability mechanism \citep{PP90} remains undisputed, and is
evident from the more or less sudden decrease in \vinf/\vesc. Additionally, 
there is a large probability that at least inside the transition zone a
``local'' increase of (\Mdot \vinf) is present, which would partly support
the arguments by Vink et al., though not on a global scale. Furthermore, the
scatter of $Q'$ (and wind-momentum rate) turned out to be {\it much} larger
in the transition region than somewhere else. This might be explained by the
fact that hydrogen begins to recombine in the wind just in this region,
whereas the degree of recombination depends on a multitude of parameters,
thus leading to the observed variety of mass-loss rates and terminal
velocities. Finally, note that at least the observed hypergiant
seems to be consistent with the bistability scenario, which, after all, has
been originally ``invented'' for these kind of objects. \bigskip

\section{Summary and future work}
\label{summary}

In this study, we have presented a detailed investigation of the optical
spectra of a small sample of Galactic B supergiants, from B0 to B9. Stellar
and wind parameters have been obtained by employing the NLTE, unified model
atmosphere code FASTWIND \citep{Puls05}, assuming unclumped winds.
The major findings of our analysis can be summarised as follows.

\smallskip
\noindent
1. We confirm recent results \citep{Ryans,dft06, simon} of the
presence of a (symmetric) line-broadening mechanism in addition to stellar
rotation, denoted as ``macro-turbulence''. The derived values of \vmac are 
highly supersonic, decreasing from $\approx$ 60~\kms at B0 to
$\approx$ 30~\kms at B9. 

\smallskip \noindent 2.  We determined the Si abundances of our sample stars
in parallel with their corresponding micro-turbulent velocities.\\ 
 (i) For all but one star, the estimated Si abundances {\bf were}
consistent with the corresponding solar
value (within $\pm$0.1 dex), in agreement with similar
studies \citep{GL,Roll,Urb,Przb06}. For HD~202\,850, on the other hand, an
overabundance of about 0.4~dex has been derived, suggesting that this late-B
supergiant might be a silicon star.\\ 
(ii) The micro-turbulent velocities tend to decrease towards later B
subtypes, from 15 to 20~\kms at B0 (similar to the situation in
O-supergiants) to 7~\kms at B9, which is also a typical value for
A-SGs. \\
(iii) The effect of micro-turbulence on the derived
effective temperature was negligible as long as Si lines from
the two major ions are used to determine it.  

\smallskip 
\noindent 
3. Based on our \Teff estimates and incorporating data from similar
investigations \citep{crowther06,Urb,Przb06,Lefever}, we confirm
previous results (e.g., \citealt{crowther06}) 
on a 10\% downwards revision of the effective temperature scale of
early B-SGs, required after incorporating the 
effects of line blocking/blanketing. Furthermore, we suggest a similar
correction for mid and late subtypes. When strong
winds are present, this reduction can become  a factor of two larger, 
similar to the situation encountered in O-SGs \citep{Crowther02}.

\smallskip 
\noindent 
4. To our surprise, a comparison with data from similar SMC objects
\citep{Trundle04, Trundle05} did not reveal any systematic difference
between the two temperature scales. This result is interpreted as an 
indication that  the re-classification scheme as developed by
\citet{lennon97} to account for lower metal line strengths in SMC
B-SGs also removes the effects of different degrees of line
blanketing.

\smallskip 
\noindent 
5. Investigating the wind properties of a statistically significant sample
of supergiants  
with \Teff between 10 and 45 kK, we identified a number of discrepancies 
between theoretical predictions \citep{Vink00} and observations.
In fair accordance with recent results \citep{Evans2, crowther06}, our sample
indicates a gradual decrease in \vinf in the bi-stability
(``transition'') region, which is located at lower temperatures than
predicted: 18 to 23~kK (present study) against 22.5 to 27 kK.

By means of a $newly$ defined, distance independent quantity, $Q'=
\Mdote/\Rstare^{1.5} \, \geff/\vinfe)$ we have investigated the behaviour of
\Mdot as a function of \Teff.  Whereas inside the transition zone a large
scatter is present (coupled with a potential {\it local} maximum in wind
efficiency around 21~kK), $Q'$ remains a well defined function with low
scatter in the hot and cool temperature region outside the transition zone.
Combining the behaviour of $Q'$ and the modified wind-momentum rate, the
change in \Mdot over the bi-stability jump (from hot to cool) could be
constrained to lie within the factors 0.4 to 2.5, to be conservative. Thus,
\Mdot either decreases in parallel with \vinf/\vesc (more probable), or, at
most, the decrease in \vinf is just balanced by a corresponding increase in
\Mdot (less probable). This finding contradicts the predictions by
\citet{Vink00} that the decrease in \vinf should be {\it over-compensated}
by an increase in \Mdot, i.e., that the wind-momenta should increase over
the jump. Considering potential clumping effects, we have argued that 
such effects will not change our basic result, unless hotter objects turn
out to be substantially more strongly clumped than cooler ones. In any case, at
least in the low temperature region present theoretical predictions for
\Mdot are too large! 

\medskip \noindent This finding is somewhat similar to the recent
``weak-wind problem'' for late O-dwarfs\footnote{A detailed UV-analysis by
\citet{martins04} showed the mass-loss rates of young late-O dwarfs in N81
(SMC) to be significantly smaller (factors 10 to 100) than theory predicts
(see also \citealt{bouret03}). In the Galaxy, the same dilemma applies to
the O9V 10~Lac(\citealt{Herrero02}) and maybe also for $\tau$ Sco (B0.2V),
which show very low mass-loss rates.}, though probably to a lesser extent. 
Thus, it might be that our understanding of radiation driven winds is
not as complete as thought only a few years ago. Thus, it is of
extreme importance to continue the effort of constructing
sophisticated wind models, including the aforementioned effects
(wind-clumping, diffuse radiation field, micro-turbulence), both in terms of
stationary and time-dependent simulations. With respect to the objects of
the present study, a re-analysis of the ``peculiar'' mid-type B-supergiants
from the KPL99 sample is urgently required as well. Finally, let us (once
more) point to the unresolved problem of macro-turbulence, which implies the
presence of rather deep-seated, statistically distributed and highly
supersonic velocity fields. How can we explain such an effect within our
present-day atmospheric models of hot, massive stars?
   
\acknowledgements{Many thanks to our anonymous referee for very
constructive suggestions on the original manuscript. We like to thank
Jorick Vink for providing us with his theoretical predictions for the
behaviour of wind-efficiency $\eta$ vs. \Teff. This investigation was
supported in part both by a NATO CLG No. PST/CLG 980007 and
by a Bulgarian NSF grand No. 1407/04. J.P. gratefully acknowledges 
travel support by the Spanish MEC through project AYA2004-08271-CO2.}

\end{document}